\documentclass[11pt]{article}

\usepackage{makeidx}
\makeindex

\usepackage{newclude}
\usepackage{versions}
\usepackage{graphicx}

\usepackage[most]{tcolorbox} 

\usepackage{fancyvrb}

\usepackage[english]{babel}
\usepackage{amsfonts}
\usepackage{amstext}
\usepackage{amsmath}
\usepackage{amssymb,dsfont,mathtools}
\usepackage{bbm} 
\usepackage{amsthm} 
\usepackage{epsfig, subfigure}
\usepackage{natbib}

\usepackage{color}

\usepackage[colorlinks,citecolor=blue,urlcolor=blue]{hyperref}
\RequirePackage{hypernat}

\usepackage{url}

\usepackage{multirow}
\usepackage{booktabs}

\let\code=\texttt

\usepackage{xparse} 
\usepackage{etoolbox} 

\newtheorem{theorem}{Theorem}[section]
\newtheorem{proposition}{Proposition}[section]

\newtheorem{lemma}{Lemma}[section]
\newtheorem{remark}{Remark}[section]

\newtheorem*{assumption*}{\assumptionnumber}
\providecommand{\assumptionnumber}{}
\makeatletter 
\newenvironment{assumption}[1]
{%
  \renewcommand{\assumptionnumber}{Assumption #1}%
  \begin{assumption*}%
    \protected@edef\@currentlabel{#1}%
  }
  {%
  \end{assumption*}
}

\NewDocumentCommand\Nb{O{}}{
  \ifstrempty{#1}{
    N_{[\, ]}
  }{
    N_{[\, ]}\left( {#1} \right)
  }
}

\NewDocumentCommand\Nc{O{}}{
  \ifstrempty{#1}{
    N
  }{
    N\left( {#1} \right)
  }
}

\NewDocumentCommand\cC{O{}}{
  \ifstrempty{#1}{
    \mc C
  }{
    \mc C \left( {#1} \right)
  }
}

\global\long\def\inv#1{\frac{1}{#1}}

\DeclareMathOperator*{\Var}{Var}

\newcommand{\lp}{\left(} 
  \newcommand{\rp}{\right)}
\newcommand{\lb}{\left\{} 
  \newcommand{\rb}{\right\}}
\newcommand{\ls}{\left[} 
  \newcommand{\rs}{\right]}
\newcommand{\la}{\left\langle}
\newcommand{\ra}{\right\rangle}

\newcommand{\wh}{\widehat}
\newcommand{\wt}{\widetilde}

\newcommand{\bs}[1]{\boldsymbol{#1}}

\newcommand{\PP}{{\mathbb P}}
\newcommand{\DD}{{\mathbb D}}

\newcommand{\GG}{{\mathbb G}}

\newcommand{\RR}{\mathbb{R}}

\newcommand{\EE}{\mathbb{E}}

\newcommand{\mono}{\mathcal{I}}

\newcommand{\mcFpi}{\mathcal{F}_g^{-1}}

\newcommand{\one}{\mathbbm{1}}
\newcommand{\One}[1]{\mathbbm{1}_{\{#1\}}}
\newcommand{\mc}[1]{\mathcal {#1}}

\newcommand{\ord}[1]{_{(#1)}} 

\newcommand{\argmin}{\mathrm{argmin}}

\newcommand{\sign}{\operatorname{sign}}
\newcommand{\GCM}{\operatorname{GCM}}

 \newcommand{\mle}{\widehat{\theta}_n}
\newcommand{\mlevr}{\widehat{\theta}^0_n} 
 \newcommand{\mlereg}{\widehat{r}_n} 

\newcommand{\ur}[1]{\widehat{#1}_n} 
\newcommand{\vr}[1]{\widehat{#1}_n^0} 

\newcommand{\densA}{f} 

\newcommand{\remCRD}[1]{{\noindent  \color{red}{(#1 --Charles)}}
  \index{CRD}}
\newcommand{\modCRD}[1]{{\noindent  \color{purple}#1} \index{CRD}} 

\renewcommand{\remCRD}[1]{}
\renewcommand{\modCRD}[1]{{#1} \index{CRD}} 

\newenvironment{longform}{%
  \par \vspace{.2cm} \color{green} { \bf  Notes to self:}%
}{%
  $\blacktriangle$  \vspace{.1cm}
}

\newcommand{\tkappa}{\breve{\kappa}}
\newcommand{\cc}{c_0(a_0)}

\excludeversion{mylongform}

\excludeversion{myscribbles}

\excludeversion{mysolution}

\excludeversion{supplement-file}

\excludeversion{main-file}

\includeversion{all-in-one-file}
\newcommand{\mydoctitle}{Doubly robust 
  pointwise confidence intervals for a monotonic continuous treatment effect curve}

\begin{supplement-file}
  \title{Supplementary Material for ``\mydoctitle''}
\end{supplement-file}

\begin{main-file}
  \title{\mydoctitle}
\end{main-file}

\begin{all-in-one-file}
  \title{\mydoctitle}
\end{all-in-one-file}

\author{Charles R. Doss \thanks{Charles R. Doss, 224 Church St.\ SE,
     Minneapolis, MN 55455. Email: \url{cdoss@stat.umn.edu}} \\
   School of Statistics, University of Minnesota}

\date{}

\begin{document}


\maketitle

\begin{all-in-one-file}
  \begin{abstract}
We study nonparametric inference for the causal dose-response (or treatment effect) curve when the treatment variable is continuous rather than binary or discrete.  We do this by developing  doubly robust
confidence intervals for the continuous treatment effect curve (at a fixed point) under the assumption that it is monotonic, based on inverting a likelihood ratio-type test.  Monotonicity of the treatment effect curve is often a very natural assumption, and this assumption removes the need to choose a smoothing or tuning parameter for the nonparametrically estimated curve.  The likelihood ratio procedure is effective because it allows us to avoid estimating the curve's unknown bias, which is challenging to do.
\modCRD{The test statistic is ``doubly robust'' in that a remainder term is the product of errors for the two so-called nuisance functions that naturally arise (the outcome regression and generalized propensity score functions), which allows one nuisance to be estimated poorly if the other is estimated well.}
Furthermore, we propose a version of our test or confidence interval that is adaptive to a range of the unknown curve's flatness level. We present versions with and without cross fitting.  We illustrate the new methods via simulations and a study of a dataset relating the effect of nurse staffing hours on hospital performance.


  \end{abstract}
\end{all-in-one-file}

\begin{main-file}
  \begin{abstract}
    
  \end{abstract}
\end{main-file}

\begin{supplement-file}
  \begin{abstract}
    This supplement contains proofs and other technical material for
``\mydoctitle''.  Reference numbers to equations, theorems, or other
statements in that document have a unified numbering system with this
document.  The bibliography is shared and found at the end of the main
document.

  \end{abstract}
\end{supplement-file}







\tableofcontents

\section{Introduction}

We are interested in testing hypotheses and forming confidence intervals for the value of a continuous causal treatment effect curve, denoted $\theta_0(\cdot)$, at a fixed point based on observational data.
Much of the classical literature for developing valid causal inference is focused on the case of binary or discrete treatments, but recently there has been a renewed focus on developing methods for the case of continuous treatments.  Performing honest causal inference with observational data requires
accounting for confounding variables,  which are
variables related to both the outcome and the treatment.
Under the  ``no unmeasured
confounders'' assumption, one can adjust for the observed confounding variables in order to perform valid inference for the causal effect curve.
Adjustments can be made through the so-called (generalized) propensity score function and the outcome regression function.
Causal inference procedures for an average treatment effect estimand use the propensity score \citep{Hirano:2004in, Imai:2004gd, Galvao:2015ju}, or the outcome regression function \citep{Imbens:2004cz, Hill:2011bn}, or combine both \citep{Scharfstein_Rotnitzky_Robins_1999,  robins2001comment,VanDerLaan:2003uu, Bang_Robins_2005}.
In the framework of semiparametric statistics, the outcome regression function and the propensity score function
are often referred to as ``nuisance parameters''
(possibly infinite-dimensional); and in problems where the semiparametric efficiency bound is well defined, one needs to use both of these nuisance parameters to attain that efficiency bound.
These methods generally have the feature that they can be consistent for the causal estimand even if one of the nuisance parameters is model misspecified, which is where the term 
``double robustness'' arises.
Put another way,  methods which make use of both the propensity score and outcome regression nuisance parameters  are less  susceptible to the curse of dimensionality than methods that use just one of the nuisance parameters; in the latter type of approach,  the theoretical rate of convergence of the estimator of
the causal parameter %
is the same as that of the estimator of the nuisance parameter, which may be high dimensional and so have a slow rate of convergence.
On the other hand, in the doubly robust approach,
the rate of the leading error term in estimating the causal estimand is determined by the
rate of the product of the two nuisance parameters' error terms, 
which may be much faster than either individual rate is.

Somewhat recently, a nonparametric doubly robust estimation method has been proposed
(\cite{Kennedy:2017cq}),
allowing the  flexibility to use nonparametric machine learning methods for modeling the nuisance parameters.
Further work has now developed doubly robust estimation methods and limit distribution theory for the  causal effect curve based on the assumption that the curve is monotonic
\citep{Westling_Gilbert_Carone_2020,Westling_Carone_2020},
which is a very natural assumption in the setting of causal inference where, for instance, a given treatment may be believed a priori to either be beneficial or to be neutral but  to be unlikely to have a negative effect.  In fact, in some cases if the estimated treatment curve is non-monotone (for a reasonable range of treatment), this might be considered a sign that not all confounders have been captured. Besides the improved efficiency that monotonicity provides, a benefit to making use of this shape constraint is that it allows to avoid the selection of tuning parameters.  For smoothness-based nonparametric  methods, selecting tuning parameters (e.g., bandwidths, penalty parameters, etc.) is often an important aspect of estimation and their correct selection is often a necessary but sometimes complicated step for estimation and inference.  The monotonicity asssumption allows us to avoid tuning parameter selection entirely (or, put another way, monotonicity-based estimators often automatically select locally optimal tuning parameters).
For further motivation for the monotonicity assumption, see examples in \cite{Westling_Gilbert_Carone_2020,Westling_Carone_2020}.

In the present paper, we work under the monotonicity assumption on the treatment effect curve $\theta_0(\cdot)$, and we develop doubly robust %
pointwise confidence intervals for the treatment effect curve $\theta_0(a_0)$ at a fixed treatment value $a_0$.  The intervals are developed based on a likelihood ratio (LR) statistic.
The authors of \cite{Westling_Gilbert_Carone_2020} develop a Wald-type of confidence interval for a monotone treatment curve which requires estimating unknown curvature parameters (i.e., $\theta_0'(a_0)$) of the treatment curve and plugging those in to the limit distribution.  The downside to this approach is that estimating the unknown curvature parameters can be difficult and create problems for inference
(see, for example, discussion in the introduction of \cite{doss2019concave} about a different problem but with the same general concerns), and the efficacy of the procedure depends strongly on knowing a priori the order of smoothness and flatness of the unknown curve.
\cite{takatsu2024debiased} also consider inference for a continuous treatment curve via debiasing, but with a smoothness rather than a monotonicity constraint.
Another approach for forming confidence intervals is the bootstrap.
However, in general, nonparametric (``pairs'' or ``residuals'') regression bootstrap procedures are not expected to work automatically for inference for a nonparametrically estimated regression function.
But \cite{cattaneo2023bootstrap}  %
have recently developed a ``bootstrap-assisted'' approach that successfully performs inference in ``generalized Grenander models'' which include the problem we consider here.
Instead of using a bootstrap, we develop here a LR approach that has been very successful in the (non-causal) monotone regression setting
\citep{Banerjee:2001jy, Banerjee:2005gw, Banerjee:2005im,banerjee2007:LRTmonoresponse, Groeneboom_Jongbloed_2015}.
The main benefit to the  LR approach is that it avoids the estimation of some unknown nuisance
parameters that the other approaches need to estimate.
\modCRD{Here, the ``nuisance parameters'' are different than those from above [the outcome regression and propensity score]. The main nuisance parameter that creates well-known difficulties is the derivative $\theta_0'$, which is needed (assuming it exists) for plug-in type confidence intervals. LRTs avoid estimating $\theta_0'$.}
This allows the LRT  to be quite efficient, as we demonstrate in our simulation studies.  Perhaps more importantly, it allows a procedure to be developed that is {\it adaptive} to a broad range of models (different levels of flatness, see Assumption~\ref{assm:model} below) without having prior knowledge of what model is true. 
Although \cite{cattaneo2023bootstrap} develop a procedure that is adaptive over a range of flatness regimes, the procedure still involves implicit estimation of the relevant curvature.

\modCRD{That adaptation is possible for estimation or inference for a (nonparametrically estimated) monotone function is well known at this point and has been studied in a variety of settings;
  see the review \cite{guntuboyina2018nonparametric}. 
  For instance, when the true function being estimated is constant, it is often possible to estimate  and form confidence intervals for it at the parametric rate $n^{-1/2}$, up to poly-log factors, rather than the usual much slower nonparametric rates. In the setting of the present paper, the dose response function being constant is the very important situation in which there is a null treatment effect, and so being able to estimate at fast rates of convergence is particularly beneficial.
  Thus, adaptation to no-treatment-effect is another very significant benefit to using monotonicity in the context of the dose response curve. 
}

Our approach is based on developing a doubly robust ``likelihood ratio test''\footnote{Actually the statistic is based on a residual sum of squares criterion (and we do not make any Gaussianity assumption) but ``likelihood ratio test'' is common terminology so it is what we use.} procedure for testing the null hypothesis $H_0 \colon \theta_0(a_0) = t_0$ against $H_1 \colon \theta_0(a_0) \ne t_0$ for fixed treatment and outcome values $a_0, t_0$.  
Basic limit theory for a monotone dose response estimator $\wh \theta_n$ has been developed already
\citep{Westling_Gilbert_Carone_2020,Westling_Carone_2020}; here we must develop and study a null hypothesis  monotone estimator $\wh \theta^0_n$ constrained to satisfy $\wh \theta_n^0(a_0) = t_0$, and then use that to form a likelihood ratio statistic (LRS).
We also extend the results for $\wh \theta_n$ to a broader set of model assumptions, so that we can consider adaptive behavior. 
The main benefit to the LR approach is that at least in parametric problems, LRS's are (under regularity) {\it asymptotically pivotal} (or satisfy the Wilks phenomenon) meaning that the limit distribution is universal (a chi-squared in regular parametric problems) and there are no unknown nuisance parameters to estimate.  This is practically quite valuable since it can simplify inference, avoid extraneous model assumptions, and lead to a robust procedure.
In our nonparametric setting, it is not quite true that our LRS is asymptotically pivotal,  as there remains a ``variance'' nuisance parameter (which in turn  depends on the outcome regression and propensity function nuisance parameters), but it can be doubly robustly estimated {without requiring any further assumptions beyond what is originally needed for our inference procedure.}
A confidence interval can be formed by  inverting the likelihood ratio test (LRT).  (Computationally, this can be implemented by a simple grid search which is feasible for our univariate confidence intervals.)

To summarize, our contributions are as follows.  Our main contribution is 
a new doubly robust test (and corresponding CI) procedure.
We show it to be consistent under the null hypothesis,
as long as at least one of the two nuisance parameters is specified correctly (under
some assumptions  on the rate(s) of nuisance estimator convergence).
The procedure requires only nonparametric assumptions on the treatment effect curve,
unlike
\cite{Robins:2000gv} and \cite{Neugebauer:2007fg}.
We develop our procedure either under entropy conditions on the nuisance parameters or under a sample splitting regime that avoids such entropy conditions. 
The test/CI is very efficient as demonstrated by simulation studies.
Next, we go beyond CI's that depend on knowing the unknown flatness or smoothness of the truth, and we are crucially able to  develop a procedure that is {adaptive} to different flatness levels.
\begin{mylongform}
  \begin{longform}
    Code that implements our doubly robust dose response test and CI is
    available in the R \citep{R-language} package \code{DRDRmonoLRT},
    available on the author's
    website\footnote{\url{http://users.stat.umn.edu/~cdoss/packages/DRDRmonoLRT.tar.gz}}.
  \end{longform}
\end{mylongform}

It is worth commenting that although flexible machine learning methods can (and often should)
be used to alleviate model misspecification, they should not be viewed as  entirely removing the issue.  Machine learning methods still require some structural assumptions
(e.g., sparsity, additive structure, only low order interactions) on the underlying model without which they may effectively be considered misspecified (have very slow rates of convergence). 
Additionally, their practical implementations  often require multiple tuning parameters (the poor choice of which could again be considered analogous to model misspecification).

The rest of the paper is organized as follows. %
In Section~\ref{sec:notat-probl-setup} we introduce notation, the problem setup along with causal assumptions, and present an introduction to the causal methodology on which our procedure is based.  Then in Section~\ref{sec:caus-isot-regr} we %
develop our procedure and theoretical results.  %
Section~\ref{sec:simulations} contains simulation results and in Section~\ref{sec:data-example} we present analysis on a data example relating nurse staffing to hospital effectiveness.
Our main interest in this paper is in the causal setting, but along the path to studying that setting we need to also study the non-causal (classical) monotone regression setting.
We do this in Appendix~\ref{sec:non-causal-isotonic-regression}.
(So some readers may prefer to warm up by reading Appendix~\ref{sec:non-causal-isotonic-regression} before
proceeding to Section~\ref{sec:notat-probl-setup} and onwards.) 
In the rest of the current section we review the literature on continuous causal treatment effect estimation and inference.  
\begin{all-in-one-file}
  Most proofs, with a few exceptions, are given in the Appendices.
  We also present  in Appendix Section~\ref{sec:sample-splitting}
  a sample splitting (cross fitting) variation of our procedure to remove complexity conditions on nuisance estimators.
\end{all-in-one-file}
\begin{main-file}
  Most proofs, with a few exceptions, 
  are provided in the supplementary material \citep{DRDRlocalmono-supp}.
  We also present there in Appendix Section~\ref{sec:sample-splitting}
  a sample splitting (cross fitting) variation of our procedure to remove complexity conditions on nuisance estimators.
  {Section references that begin with a capital letter (``A'', ``B'', etc.) refer to the supplementary document.}
\end{main-file}

\subsection{Literature on continuous treatment effects}
\label{sec:disc-recent-liter}

For many years, the causal inference literature focused more heavily on binary or discrete treatments, but recently there has been
renewed interest in the setting of continuous treatment variables.
The recent literature on doubly robust methods for the dose-response curve starts with 
\cite{Kennedy:2017cq}, on which other works, including the present paper,
build.
\cite{Kennedy:2017cq} have developed a method for efficient doubly
robust estimation of the treatment effect curve.  
Denote the outcome
regression function by $\mu$ with true value $\mu_0$, and denote the propensity score function
by $\pi$ with true value $\pi_0$.
Their method is
based on a pseudo-outcome $\xi \equiv \xi(\bs Z; \pi, \mu)$,
which depends on the sample point $\bs Z$, and on the nuisance functions
$\pi$, $\mu$.  The pseudo-outcome $\xi$ has the key double robustness property that if {\it either} $\pi = \pi_0$ or $\mu = \mu_0$, then $\EE( \xi( \bs Z; \pi, \mu) | A = a)$ is equal to $\theta_0(a)$.
The general estimation  procedure of   \cite{Kennedy:2017cq} is then a natural two-step procedure: (1) estimate the nuisance functions $(\pi_0, \mu_0)$ by some estimators $(\wh \pi, \wh \mu)$,
which the user can choose as they wish, %
and
construct (observable) pseudo-outcomes $\wh \xi_i$
(which approximate $\xi_i$ and depend on $\wh \pi,$ $\wh \mu$), and (2)
regress the pseudo-outcomes on $A$ using some nonparametric method (e.g., local linear regression).  As we described above, the error term from the nuisance parameter estimation is given by the product of the error term for estimating $\pi_0$ and for estimating $\mu_0$, so is smaller than either, partially alleviating the curse of dimensionality.

Several works have now made use of the pseudo-outcome approach of
\cite{Kennedy:2017cq}, or similar approaches.
\cite{Westling_Gilbert_Carone_2020, Semenova:2017uc} use the pseudo-outcomes
of \cite{Kennedy:2017cq} with alternative estimation techniques, and \cite{Colangelo:2020tt, su2019non} use similar
pseudo-outcomes (and study particular nuisance estimators).
The authors of \cite{Westling_Gilbert_Carone_2020}  develop a doubly robust estimator of a continuous treatment effect curve; they develop a procedure
based on the assumption that the true effect curve satisfies the shape constraint of monotonicity,
and we build on their work in this paper.
\cite{Colangelo:2020tt} provide an alternative motivation for a related pseudo-outcome to that of 
\cite{Kennedy:2017cq}, study a sample-splitting variation of the
estimation methodology of
\cite{Kennedy:2017cq}, and also consider estimating the gradient of
the treatment curve.
\cite{chen2016personalized, Kallus:2018up, schulz2021doubly, pmlr-v151-chen22c} go beyond estimation/inference for the dose-response curve and consider the setting of optimal treatment regimes,
and there are a large number of scientific areas
where continuous treatments arise (e.g., \cite{Kreif:2015cp, coulombe2021estimating} in the health sciences).

\remCRD{ placement of this paragraph?}

\modCRD{The ``double robust'' terminology has multiple meanings, depending on the context.
  Here we show that estimator and test statistic limit distributions hold under Condition \ref{assumption:estimators} below which requires  a second order (product) remainder term to be smaller than the (nonparametric) rate of convergence of our estimators. Since our estimand can only  be estimated at slower than root-$n$ rates, this allows one nuisance to be fully misspecified if the other is estimated quickly enough, so that the limit distributions and test are doubly robust. 

  We note that the smoothness/complexity of $\mu_0$ would generally imply a bound on the smoothness/complexity of $\theta_0$.  %
  From a theoretical standpoint, one might complain that if the estimator for $\mu_0$ converges faster than that for $\theta_0$, then the practitioner has made a poor choice of some model.  However, one of the benefits of the  pseudo-outcome framework is that it allows the user to separate out the model for the nuisances from the model for $\theta_0$. It may be reasonable to use models that don't quite match in many practical scenarios, e.g.\  for the outcome regression use a parametric model (which is reasonable but may be slightly misspecified) but a more flexible model for the target of interest for which one wants to make minimal assumptions and avoid all possibility of model misspecification. 
  It is possible then to get the parametric model correct and have that nuisance estimated more quickly than $\theta_0$ is (allowing for the other nuisance to be fully  misspecified), justifying the ``doubly robust'' terminology which is common in the literature (on
  inference) that relies on the
pseudo-outcomes of \cite{Kennedy:2017cq}. }

\section{Causal notation and problem setup}
\label{sec:notat-probl-setup}

In this section we introduce the notation, problem setup, estimand, and lay out the building blocks for our method.

\subsection{Notation}
\label{sec:notation}

We observe $n$ i.i.d.\ copies $\bs W_1, \ldots, \bs W_n$ of $\bs W = (\bs L, A, Y)$ with support $\mc W := \mc L \times \mc A \times \mc Y$ where
$\mc{A}$ is bounded, from a distribution $\PP_0$ which has density $p_0$ (with respect to some dominating measure $\nu$).
Here, $\bs L \in \RR^d$ are the observed covariates/confounder variables, $A \in \RR$ is the continuous univariate treatment variable, and $Y \in \RR$ is the outcome/response variable.  We use $\EE(\cdot)$ and $\PP(\cdot)$ for generic expectation and probability statements when the random variables and data generating processes have been defined.  For a function $h$ we let $\| h \|^p_p := \int h^p d\PP_0$ (when this quantity is well defined). For a measure $Q$ on $\bs x \in \mc{X}$ and an integrable function $f$ (which could be itself random), we use the operator notation $Qf(\bs X) := \int_{\mc X} f(\bs x) dQ(\bs x)$ which we may abbreviate as $Q f$ when there is no ambiguity about the variables over which we integrate, and we use $Q(\cdot)$ for probability statements according to a probability measure $Q$.  We use $\| f(\bs X) \|_2^2$ to denote $ \int f(\bs x)^2 \, d\PP_0(\bs x)$, the (squared) $L^2(\PP_0)$ norm over the variable $\bs X$.  We let $L^\infty[-K,K]$ for %
$K>0$ denote the Lebesgue $L^\infty$ function space on $[-K,K]$.
 
We use the subscript of ``$0$'' to refer to true parameters generally.
For instance, we let $\mu_0(\bs l, a) := \EE(Y | \bs L = \bs l, A = a)$ denote the true outcome regression function, we let $\pi_0(a | \bs l) : = \frac{\partial}{\partial a} \PP_0(A \le a | \bs L = \bs l)$ denote the true generalized propensity score, we let
$F_0(a) := \PP_0(A \le a)$ be the true cumulative distribution function of $A$ and 
$\densA_0(a) :=\frac{d}{d a} F_0(a)$ be the true marginal density of $A$.  Let $g_0(a, \bs l) := \pi_0(a | \bs l) / \densA_0(a)$ be the normalized propensity function.  We use the symbols $\mu$, $\pi$, $\densA$, and $g$ for generic versions of these quantities, and we let $\eta := (\mu, g)$ be the combined nuisance parameter(s).  We let $\PP_n$ denote the empirical distribution of the data.
We will let GCM denote the so-called greatest convex minorant, discussed in further detail below. 
When they are well defined, we let  $\partial f(\cdot +)$ and $\partial f(\cdot -) $ denote the right- and left-derivatives of a function $f$.
An %
isotonic %
estimator is generally formed by taking the (left) derivative of the GCM; for a function $X(\cdot)$ we denote this isotonization by %
\begin{equation}
  \label{eq:defn-M}
  \mono(X) :=    \partial \GCM(X)(\cdot -)
\end{equation}
or  by $\mono(X)(u)$ for the value at a fixed point $u$.  If the isonization %
is restricted to a given interval $I$ we write $\mono_I(X)$ for $\partial \GCM_I(X)(\cdot -)$.
We use
``$\stackrel{d}{=}$'' or ``$=_d$'' to denote equality in distribution and ``$\to_d$'' to denote convergence in distribution.
Our target parameter of interest is the {\it G-computed regression function,}
\begin{equation}
  \label{eq:2}
  \theta_0(a) :=
  \EE(  \EE( Y | A=a, \bs L)).
\end{equation}
This quantity is related to the so-called causal dose-response curve under identifying assumptions.

\subsubsection{Limit distribution notation}
\label{sec:limit-distr-notat}

The following notation will be used when we present asymptotic limit distribution results; we present it here for ease of reference. The parameter(s) $\beta_0$ (and also $\rho_0(a_0)$) will be defined in Assumption~\ref{assm:model} below. We let
\begin{equation}
  \label{eq:defn:tn}
  t_n := n^{-1 / (2\beta_0 + 1)},
\end{equation}
which is the local scale to `zoom in' around $a_0$. 
The limit distributions will depend on a standard Brownian motion $W$ on $\RR$ with $W(0)=0$.  Then we define  $X(t) := W(t) + |t|^{\beta_0}$. We need to ``isotonize'' $X$;
let
(notationally suppressing dependence on $\beta_0$)
\begin{equation}
  \label{eq:defn-M-m0}
  M(t)  := W(t) + |t|^{\beta_0+1}
  \quad \text{and} \quad
  M^0(t)  := M(t) + \Lambda \one_{(0,\infty)}(t),
\end{equation}
where $\Lambda \equiv \Lambda_{\beta_0}$ is a random variable described in equation \eqref{eq:Lagrangemult-constants}
\begin{all-in-one-file}
  in  Appendix~\ref{app:sec:proofs-caus-estim-lemm};
\end{all-in-one-file}
\begin{main-file}
  in Appendix~\ref{app:sec:proofs-caus-estim-lemm} in 
  \cite{DRDRlocalmono-supp};
\end{main-file}
$M^0$ is defined so that the corresponding (limit) ``estimator'' based on $M^0$ satisfies the null constraint. Thus let
\begin{equation}
  \label{eq:defn:theta-theta0}
  \wh \theta :=
  \mono(M)
  \quad   \text{ and } \quad
  \wh \theta^0 := %
  \mono(M^0).
\end{equation}
It is true (from the proofs of Theorem \ref{thm:causal-fullestimator-limit} and  \ref{thm:causal-nullestimator-limit})
that $\wh \theta^0$ satisfies the limit version of the null constraint, that is $\wh \theta^0(0) = 0$.

We now introduce some  constants that arise in our limit distributions. The constants involve $\mu_\infty$ and $g_\infty$ which are
the limits of our nuisance estimators; see Assumption~\ref{assm:E-II_double-robust-setup} below for the formal definitions. 
Define 
$\kappa_0(a_0)$ and $\tkappa_0(a_0)$ by
\begin{equation}
  \label{eq:defn:kappa}
  \kappa_0(a_0) :=
  \EE_0 \lp \EE_0 \ls \delta_\infty(\bs W)^2 \vert A=a_0, \bs L \rs g_0(a_0, \bs L) \rp
\end{equation}
and $  \tkappa_0(a_0) := \kappa_0(a_0) \densA_0(a_0),$
where $\delta_\infty(\bs W) := \frac{Y - \mu_\infty(A, \bs L)}{g_\infty(A, \bs L)} + \theta_\infty(A) - \theta_0(a_0)$ and 
$\theta_\infty(b) := \int \mu_\infty(b, \bs w) d\PP_0(\bs w)$.
Let (recall that $\beta_0$ and $\rho_0(a_0)$ will be defined in Assumption~\ref{assm:model} below)
\begin{equation}
  \label{eq:defn:limit-const}
  \cc^{2\beta_0 + 1}
  :=
  \frac{\tkappa_0(a_0)^{\beta_0} \rho_0(a_0) }{(\beta_0+1) f_0^{2 \beta_0}(a_0)}
  =
  \frac{ \kappa_0(a_0)^{\beta_0} \rho_0(a_0) }{(\beta_0+1) f_0^{ \beta_0}(a_0)}.
\end{equation}

\subsection{Causal assumptions}

Formally, we choose to define our target estimand to simply be the G-computed regression function
given in \eqref{eq:2},
and regardless of whether causal identifiability assumptions hold, all of our results will apply to this parameter %
which is an identifiable statistical parameter. %
This choice slightly simplifies statements of theorems.
This parameter may also be of interest even in cases where causal assumptions do not hold, as discussed after Assumption~\ref{assm:identifiability} below.
But the setting where $\theta_0(a)$ is most interesting is when it is a causal parameter, so for completeness we will introduce the causal setup and assumptions.  As mentioned earlier, we let $Y^a$ denote the counterfactual/potential outcome corresponding to treatment level $a \in \mc A$, and then we assume that $Y = Y^A$.
Identifiability assumptions such that $\theta_0(a)$ equals $E(Y^a)$  are as follows.
\begin{assumption}{I}
  \label{assm:identifiability}
  $\phantom{blah}$
  \begin{enumerate}
  \item  Consistency/SUTVA: Assume $Y=Y^A$, and each unit's potential outcomes are independent of all other units' exposures;
  \item Positivity: There exists $\epsilon_0 > 0$ such that
    almost surely $ \pi_0(a | \bs L) \ge \epsilon_0$ for all $a \in \mc{A}$.
  \item Ignorability/unconfoundedness: We have $\EE(Y^a | \bs L, A) = \EE(Y^a | \bs L)$ almost surely
    for all $a \in \mc A$. 
  \end{enumerate}
\end{assumption}

\noindent The above assumptions for identifying the causal estimand are the standard ones in the context of observational studies with no unmeasured confounding \citep{robins1986new,gill2001causal}.  However, they are generally of course highly nontrivial, and this is particularly true in the present context of continuous treatment. Unconfoundedness is always a strong assumption.  The positivity assumption that every  treatment level may possibly be received for every (possibly high-dimensional) covariate value is  a stronger assumption when treatment is continuous than when it is binary.

On the other hand, even if the identifiability assumptions are not fully met, the adjusted regression function may still be a useful target parameter.  The causal inference assumptions can be thought of as conditions that ensure that the study population corresponds to a global/external population (where  treatment and confounders become independent).  If those assumptions do not hold, we can still interpret the target parameter as describing the effect of treatment in the study population itself.  This may be of interest, and it may indeed be more useful and interesting than the unadjusted regression function $a \mapsto \EE(Y| A=a)$, and a more succinct (univariate) representation than the perhaps high-dimensional regression function $(a, \bs l) \mapsto \mu_0(a, \bs l)$.

\subsection{Method setup}
\label{sec:method-setup}
For estimating $\theta_0(a)$, the regression of $Y$ on $A$ is generally biased, but
we can adjust for the bias (i.e., for confounding) by defining the following
``pseudo-outcomes.'' %
We let
\begin{equation}
  \label{eq:1}
  \xi(\bs W; \eta) :=
  \frac{Y - \mu(\bs L, A)}{g(A, \bs L)}  +
  \int_{\mc L} \mu(\bs l, A) d\PP_0(\bs l).
\end{equation}
This pseudo-outcome is shown in \cite{Kennedy:2017cq} to be ``doubly robust'' in that it satisfies $\EE (\xi(\bs W; \eta) | A = a) = \theta(a)$ whenever $\mu$ or $g$ is specified correctly (to be equal to the true $\mu_0$ or $g_0$).  Of course the nuisance parameters are not known so we have to estimate them.  We allow generic black-box estimators to be used that the user can specify and which we denote by $\wh \eta := (\wh \mu, \wh g).$
(We will place some conditions on the estimators later.)  We define $\wh \xi(\bs W; \eta)
\equiv \wh \xi_n(\bs W; \eta)
:=    %
(Y - \mu(\bs L, A))/g(A, \bs L)  +
\int_{\mc L} \mu(\bs l, A) d\PP_n(\bs l)$ (replacing $\PP_0$ by $\PP_n$) and then
define an observable version of the pseudo-outcome by
\begin{equation}
  \label{eq:pseudo-outcome-hat}
  \wh \xi(\bs W; \wh \eta) :=
  \frac{Y - \wh \mu(\bs L, A)}{\wh g(A, \bs L)}  +
  \int_{\mc L} \wh \mu(\bs l, A) d\PP_n(\bs l).
\end{equation}
The general idea proposed in
\cite{Kennedy:2017cq} is to use $(A_i, \wh \xi_i)$  in place of $(A_i, Y_i)$ as inputs to regression procedures
with $\wh \xi_i$, $i=1,\ldots, n$, pseudo-outcomes defined based on i.i.d.\ observations $\bs W_1, \ldots, \bs W_n$ (after sorting; full definition  given below).
In the present paper,
we consider the isotonic regression
$\mle$
of $(\wh \xi_i)_{i=1}^n$ on  $(A_i)_{i=1}^n$.  Our goal is to perform inference at a fixed point $a_0 \in \mc A$, via a likelihood ratio type of test.  (Here, ``likelihood'' will be based on residual sum of squares, meaning based on a Gaussian model for the errors, although that assumption is only for defining the test statistic and we do {\it not} require the Gaussianity model assumption to hold in our theorems.)  To
form a likelihood ratio test for $H_0: \theta_0(a_0) = t_0$, for a fixed
point $a_0$ we also consider the isotonic regression subject to the
(further) constraint/restriction that $\theta_0(a_0) = t_0$, which we refer to as $\wh \theta_n^0$.
Both estimates are (shape) ``constrained'' to be monotonic.
We will refer to $\mle$ as the ``full (model/hypothesis)'' estimator and to $\mlevr$ as the ``null (model/hypothesis)'' estimator.
To be more formal, recall $\mc M$ is the set of nondecreasing functions $\{ \theta(\cdot) : \theta(x) \le \theta(y), \text{ if } x \le y \}$ and let $\mc M_n := \{ (\theta(A_{(i)}))_{i=1}^n : \theta \in \mc{M} \}$
(with a very minor overloading of $\mc{M}_n$, used also in the previous section)
where $A_{(1)} \le \cdots A_{(i)} \le \cdots \le A_{(n)}$ are the (sorted) order statistics of  $\{ A_1, \ldots, A_n \}$.
Define $k_0$ to be the index such that $a_0 \in [A_{(k_0)}, A_{(k_0+1)})$. 
We let
$\wh \xi_i := \wh \xi(\bs W_{(i)} ; \wh \eta)$ where $W_{(i)}$ corresponds to $A_{(i)}$; that is for convenience we sort the data according to $\{A_i\}$ and, for instance, $\wh \xi_1$ is the pseudo-outcome for the smallest $A_i$ value.
Then we let $\mc{M}^0 := \{ \theta(\cdot) : \theta(A_{(k_0)}) = t_0, \theta \in \mc{M}\}$ and 
$\mc {M}_n^0 := \{ (\theta(A_{(1)}), \ldots, \theta(A_{(n)})))_{i=1}^{n}: \theta \in \mc{M}^0 \}$.
Here, as in Section~\ref{sec:non-causal-isotonic-regression}, rather than forcing $\theta_0(a_0) = t_0$ we require $\theta_0(A_{(k_0)}) = t_0$ where $A_{(k_0)}$ is the nearest treatment less than or equal to $a_0$. As mentioned in Section~\ref{sec:non-causal-isotonic-regression}, this difference makes no change asymptotically. %

\section{Causal isotonic regression}
\label{sec:caus-isot-regr}

We now proceed to develop our method for the causal setting (based on the notation and general methodology described in Section~\ref{sec:notat-probl-setup}).
\subsection{Modeling assumptions}
\label{sec:caus-model-assumpt}

We present some needed %
assumptions here.
We start with assumptions on our Nuisance parameter estimators. %
In addition to the rates of convergence conditions that we make below for the nuisance parameter estimation, we can rely on nuisance parameter family entropy (complexity) conditions that allow us to use the entire sample one time for the nuisance estimators, or we can avoid such conditions by using sample splitting / cross fitting.  We discuss cross fitting in %
Appendix~\ref{sec:sample-splitting}.
Here we state the entropy conditions (which are used for both global and local asymptotics).

\remCRD{Note: verify that the 'in probability' containment in the classes is OK. } 
\begin{assumption}{N1}\label{assm:EC-I}
  Assume    that $\wh \mu_n$ and $\wh g_n$ are elements of classes $\mc{F}_\mu$ and $\mc{F}_g$ respectively with probability converging to $1$ as $n \to \infty$.
  Assume there are constants $C, \epsilon_0, K_0, K_1, K_2 \in (0,\infty)$ and $V \in [0,2)$ such that
  \begin{enumerate}
  \item \label{enum:item:1}
    $|\mu| \le K_0$ for all $\mu \in \mc{F}_\mu$ and $K_1 \le g \le K_2$ for all $g \in \mc{F}_g$,
    and
  \item \label{enum:item:2}
    $\log( \sup_Q N(\epsilon, \mc{F}_\mu, L_2(Q))) \le C \epsilon^{-V/2}$ and
    $\log( \sup_Q N(\epsilon, \mc{F}_g, L_2(Q))) \le C \epsilon^{-V}$ for all
    $0 < \epsilon \le \epsilon_0$ where the suprema are over all probability measures $Q$.
  \end{enumerate}
\end{assumption}

\noindent
\noindent The following two conditions are sometimes described as ``double robustness conditions'' on the Nuisance estimators (because of the product that arises in Assumption~\ref{assumption:estimators}). Here (as in \cite{Westling_Gilbert_Carone_2020}) we require only that at least one of $\wh \mu_n$ or $\wh g_n$ is consistent.
\begin{assumption}{N2}\label{assm:E-II_double-robust-setup}
  Assume that $\mc{A}$ is bounded and that there exist functions
  $\mu_\infty \in \mc{F}_\mu$ and $g_\infty \in \mc{F}_g$ such that
  $\PP_0( \wh \mu_n - \mu_\infty)^2 \to_p 0$ and
  $\PP_0( \wh g_n - g_\infty)^2 \to_p 0$ as $n \to \infty$ and the set where
  $\mu_\infty = \mu_0$ or $g_\infty = g_0$ has $\PP_0$-probability one.
\end{assumption}

\begin{assumption}{N3}
  \label{assumption:estimators}
  Let $\beta_0$  be as given in Assumption~\ref{assm:model}.
  For $M>0$, let
  \begin{equation*}
    \begin{split}
      s_{n,M}
      & := \sup_{|s-a_0| \le M n^{-1/(2 \beta_0 + 1)}} 
      \| \wh \mu(s, \bs L) - \mu_0(s, \bs L) \|_2 \\
      r_{n,M}
      & := \sup_{|s-a_0| \le M n^{-1/(2 \beta_0 + 1)}} 
      \| \wh g(s, \bs L) - g_0(s, \bs L) \|_2     .
    \end{split}
  \end{equation*}
  For any $M > 0$ we assume
  $s_{n,M} r_{n,M} = o_p(n^{-\beta_0 / (2\beta_0 + 1)})$.
\end{assumption}

\remCRD{ I don't think uniform in M is needed; at least in lemma 4.2.  Just for each fixed M. This hasn't been }

\begin{remark}
  A condition that implies the estimator rate
  Assumption~\ref{assumption:estimators} is given by replacing
  $M n^{-1/(2\beta_0 + 1)}$ by some $\epsilon_0 >0$, i.e.\ defining
  $s_n := \sup_{|s-a_0| \le \epsilon_0} \| \wh \mu(s, \bs L) - \mu_0(s, \bs
  L) \|_2 $ and similarly for $r_n$ and then assuming/checking that
  $s_n r_n = o_p(n^{-\beta_0 / (2\beta_0 + 1)}).$
\end{remark}

Now we present further assumptions, on the Causal Model.
Our focus is on pointwise asymptotics; but in addition to pointwise conditions, we need conditions and results ensuring global consistency, without which we cannot be sure to have local consistency (because of the global nature of the monotonicity constraint wherein behavior at distant points is related/dependent).

\begin{assumption}{CM1} \label{assm:setup}
  The data generating setup is as described in Subsection~\ref{sec:notation}.
  We assume $\mc{A}$ is bounded.
  We assume   $\sigma^2_0(a) := \Var(Y | A = a)$ is uniformly bounded over all $a \in \mc{A}$.
\end{assumption}

\begin{assumption}{CM2}\label{assm:cts-diffable}
  We assume that (i) $F_0$ and
  $\sigma_0^2$ are continuously differentiable and that (ii) $\mu_0$, $\mu_\infty$, $g_0$, and $g_\infty$ are uniformly continuous in a neighborhood of $a_0$
  uniformly in $\bs l \in \bs L$.
\end{assumption}
\smallskip \noindent Note that \cite{Westling_Gilbert_Carone_2020} assume that $\mu_0$, $\mu_\infty$, $g_0$, and $g_\infty$ are continuously differentiable rather than just uniformly continuous. {This stronger assumption is unnecessary. }

To do so, we make the following basic (Monotonicity) Model assumptions. %
For $a \in \RR$ we define $\sign(a)$  to be $1$ if $a > 0$ and $-1$ if $a < 0$ and $0$ if $a=0$. 
\begin{assumption}{M1}
  \label{assm:model}
  For a monotone function $f$, assume for  some $\beta_0, \rho_0(a_0)$ that $f$ satisfies
  $f(a) - f(a_0) = $ 
  sign$(a-a_0) \rho_0(a_0) |a - a_0|^{\beta_0} + o(|a - a_0|^{\beta_0})$ as $a \to a_0$.
\end{assumption}
\noindent  Functions $f$ that satisfy Assumption~\ref{assm:model} 
are locally shaped like odd-powered monomials (the monotonicity restricts the possibilities for the functional form, so for instance if $\beta_0$ is an odd integer then in fact a monotone decreasing function $a \mapsto$ sign$(a-a_0) \rho_0(a_0) |a - a_0|^{\beta_0} = \rho_0(a_0) (a-a_0)^{\beta_0}$ is $\beta_0$-times differentiable). 
We refer to Assumption~\ref{assm:model}
as a ``flatness'' assumption. It is not just a smoothness assumption, although it does entail  a local $\beta_0$-H\"older continuity assumption  at $a_0$; but it also additionally enforces an  assumption of flatness 
(for instance, if $\beta_0$ is an odd integer then the assumption not only implies that  $f$ is $\beta_0$-times differentiable but also implies that all derivatives smaller than the $\beta_0$th derivative are zero). If $\beta_0  = 1$, then Assumption~\ref{assm:model} is just the standard differentiability assumption and $\rho_0(a_0)$ is the derivative of $f$ at $a_0$. 
We require $\beta_0 > 0$, which means $\theta_0$ must be continuous at $a_0$. 
\begin{assumption}{M2}
  \label{assm:model2}
  For the variable $A$, assume $A$ has a density function on $\mc{A}$ and
  assume that density function is bounded below and above by 
  $0 < 1/M $ and $ M < \infty$ for some $M < \infty$, respectively. 
\end{assumption}

\subsection{Estimators}
\label{sec:estimators}

Here we introduce the two causal estimators, the full and null estimators, proceeding in an analogous fashion as in the previous section.  They are both based on least-squares estimation (which is maximum likelihood estimation assuming Gaussian errors, which motivates the ``likelihood ratio'' terminology, although we do {\it not} make any such Gaussianity assumption).  For $\theta \in \RR^n$, let
\begin{equation}
  \label{eq:3}
  \wh \phi_n(\theta)
  := \inv{2} \sum_{i=1}^n (\wh \xi_i - \theta_i)^2
\end{equation}
be the least-squares objective function based on the pseudo-observations (where the hat  in $\wh \phi_n$  indicates that we use noisy pseudo-observations as the data points).  %
We define $\wh \theta_n$ to be the argmin  of $\wh \phi_n(\cdot)$ over $\mc{M}_n$
and
$\wh \theta_n^0$ to be the argmin of $\wh \phi_n(\cdot) $ over $\mc{M}^0_n$.

In fact, by the results of \cite{Groeneboom:2015ew}, we can (and do in the following lemma) characterize not just the full estimator but {\it also} the null estimator as a left derivative of a GCM of a certain cusum diagram, meaning that it is a ``generalized Grenander estimator'' in the terminology of \cite{Westling_Gilbert_Carone_2020}.

\begin{lemma} %
  \label{lem:1}
  If $\mle \in \mc{M}^0$ then $\mlevr = \mle$.  If $\mle(a_0) \ne t_0$ then define $\wh \lambda_n$ to be the solution in $\lambda$ of the equation
  \begin{equation}
    \label{eq:4}
    \max_{k \le k_0} \min_{i \ge k_0} \frac{n \lambda + \sum_{j=k}^i \wh \xi_j }{i - k + 1} = t_0.
  \end{equation}
  Then $\wh \theta_n^0$ is the left derivative of the greatest convex minorant of the cusum diagram of the points
  \begin{equation}
    \label{eq:5}
    \{(0,0)\} \cup \lb\lp i, \sum_{j=1}^i \wh \xi_n + n \wh \lambda_n 
    \One{j=k_0} \rp\rb.
  \end{equation}
\end{lemma}
\medskip \noindent If $\wh \theta_{k_0+1} < t_0$ then $\wh \lambda_n > 0$ and if $\wh \theta_{k_0} > t_0$ then $\wh \lambda_n < 0$.  Outside some local neighborhood, the null and full estimators coincide.
The lemma above gives the same characterization as is given  in the non-causal case of
Lemma~\ref{lem:1:non-causal},
and they share a proof (given in Appendix~\ref{sec:app:monoreg}).

A benefit to applying this representation of $\wh \theta_n^0$ as a left derivative of the GCM of a certain cusum diagram from Lemma~\ref{lem:1}  is that it allows us to bring to bear some of the techniques from \cite{Westling_Gilbert_Carone_2020} which apply to such ``generalized Grenander estimators,'' even allowing for the noisy pseudo-outcomes that we use.

\subsection{Consistency}

Under (subsets of) the conditions given in Subsection~\ref{sec:caus-model-assumpt}, we have global consistency of both the full estimator, as was shown by
\cite{Westling_Carone_2020}, and of the null estimator. Here is the consistency theorem for the full estimator.

\begin{theorem}[\cite{Westling_Carone_2020}, Theorem 1]
  If Assumptions
  \ref{assm:EC-I}
  and   \ref{assm:E-II_double-robust-setup}  
  hold
  then $\mle(a) \to_p \theta_0(a)$ for any value $a \in \mc{A}$ such that $F_0(a) \in (0,1)$, $\theta_0$ is continuous at $a$, and $F_0$ is strictly increasing in a neighborhood of $a$.
  If $\theta_0$ is uniformly continuous and $F_0$ is strictly increasing on $\mc{A}$  then $\sup_{a \in \mc{A}_0} | \mle(a) - \theta_0(a)| \to_p 0$ for any bounded strict subinterval $\mc{A}_0 \subsetneq \mc{A}$.
\end{theorem}

\smallskip \noindent The null estimator is also locally and uniformly consistent, which we show in the next theorem.
Although it leads to a slight loss of parallelism in the results between the full and null estimators, for the latter we use a slightly more precise assumption on the curvature of the target function at $a_0$   as well as on the nuisance functions
(Assumptions \ref{assumption:estimators} and \ref{assm:model}). These assumptions allow us to see that $\wh \lambda_n $ is $o_p(1)$, and in addition to actually  understand its order of magnitude which later on allows us to derive rates of convergence for the null estimator.

The lemma below shows that the gap between knots is
$O_p(n^{-1/(2 \beta_0 +1)})$ for both estimators and that the Lagrange
multiplier is  $O_p(n^{-(\beta_0 +1)/ (2 \beta_0 + 1)})$.
For a (fixed or random) point $\alpha \in \mc A$, let $\tau_+(\alpha)$ be
$\inf \{ t : t \ge \alpha, \ur{\theta}(t-) \ne \ur{\theta}(t+) \}$
(notationally ignoring dependence on $n$) where
$f(t \pm)$ denotes the right or left limits of a function $f$, respectively.
Let $\tau^0_+(\alpha)$ be defined similarly but with $\vr{\theta}$ in place of
$\ur{\theta}$.  And let $\tau_-(\alpha)$ and $\tau^0_-(\alpha)$ be defined analogously but for $t \le \alpha$.
Recall we let $t_n := n^{-1 / (2\beta_0+1)}$.
\begin{lemma}
  \label{lem:lambda_knot_size}
  Let Assumptions~\ref{assm:setup}, \ref{assm:model},
  \ref{assm:model2}, %
  \ref{assm:EC-I}, %
  \ref{assm:E-II_double-robust-setup}, and
  \ref{assumption:estimators}, 
  hold,
  and assume $H_0: \theta_0(a_0) = t_0$ is true.
  Let $a \in \mc{A}$ and assume $F_0(a) \in (0,1)$ and is strictly increasing at $a$. %
  We have for any $M > 0$ that
  $\tau_+(a_0 + M t_n ) - a_0 = O_p(t_n)$,
  $a_0 - \tau_-(a_0 - M t_n) = O_p(t_n)$.
  The same statement holds with $\tau_{\pm}$ replaced by $\tau_{\pm}^0$. We also have $\wh \lambda_n = O_p( n^{-(\beta_0+1)/(2\beta_0+1)})$,
  all  as
  $n \to \infty$. 
\end{lemma}

\medskip \noindent
\begin{all-in-one-file}
  Proofs are given in
  Appendix~\ref{app:sec:proofs-caus-estim-lemm}.
\end{all-in-one-file}
\begin{main-file}
  Proofs are given in
  Appendix~\ref{app:sec:proofs-caus-estim-lemm}
  in 
  \cite{DRDRlocalmono-supp}. 
\end{main-file}
Now, using the previous lemma, we show in the next theorem the consistency of the null estimator. 
We present the main parts of the proof here, since it is relatively short and shows the novel way we combine the results  of
\cite{Westling_Carone_2020} and the CDF representation given in Lemma~\ref{lem:1}
\modCRD{(together with Lemma~\ref{lem:lambda_knot_size}). For this consistency result, all we need from
  Lemma~\ref{lem:lambda_knot_size} about $\wh \lambda_n$ is that it converges to $0$ rather than the precise rate.
In the next section when we study the limit distribution (of the NE) we use
the actual rate of convergence of $\wh \lambda_n$ given in the lemma to show
that (a properly normalized) $\wh \lambda_n$ converges to a tight limit
random variable ($\Lambda$ given in
\eqref{eq:defn-M-m0}); this convergence  characterizes the null limit
distribution.}

\begin{theorem}%
  \label{thm:2}
  Let the assumptions of Lemma~\ref{lem:lambda_knot_size} hold. 
  Then $\mlevr(a) \to_p \theta_0(a)$.
  If, in addition, $\theta_0$ is uniformly continuous and $F_0$ is strictly
  increasing on $\mc{A}$ then
  $\sup_{a \in \mc{A}_0} | \mlevr(a) - \theta_0(a)| \to_p 0$ for any bounded
  strict subinterval $\mc{A}_0 \subsetneq \mc{A}$.
\end{theorem}

\begin{proof}
  The proof relies on Theorem 1 of \cite{Westling_Carone_2020} combined with Lemma~\ref{lem:1}.  Define 
  $\Gamma_n(a)$ and $\Gamma_n^0(a)$  for $a \in \RR$ by
  \begin{align*}
    \Gamma_n(a)  &:=
                   \inv{n} \sum_{i=1}^n \one_{(-\infty, a]}(A_i) \frac{Y_i - \wh \mu_n(A_i, \bs L_i)}{\wh g_n(A_i, \bs L_i)}
                   + \inv{n^2} \sum_{i=1}^n \sum_{j=1}^n \one_{(-\infty, a]}(A_i) \wh \mu_n(A_i, \bs L_j), \\
    \Gamma_n^0(a)  &:=
                     \inv{n} \sum_{i=1}^n  \one_{(-\infty, a]}(A_i) \lp \frac{Y_i - \wh \mu_n(A_i, \bs L_i)}{\wh g_n(A_i, \bs L_i)}
                     + n \wh \lambda_n  \one_{\{i=k_0\}} \rp \\
                 & \qquad + \inv{n^2} \sum_{i=1}^n \sum_{j=1}^n \one_{(-\infty, a]}(A_i) \wh \mu_n(A_i, \bs L_j), 
  \end{align*}
  where $\wh \lambda_n$ solves \eqref{eq:4}.  
  Define $\Gamma_0(\cdot)$ to be the limit of $\Gamma_n$ and $\Gamma_n^0$, namely
  \begin{equation*}
    \Gamma_0(a) :=
    \EE \lp  \one_{(-\infty,a]}(A)  \ls \frac{Y - \mu_\infty(A, \bs L)}{g_\infty(A, \bs L)} \rs +
    \eta_\infty(a, \bs L)
    \rp
  \end{equation*}
  with $ \eta_\infty(a, \bs l) :=   \PP_0 \one_{(-\infty,a]}(A)  \mu_\infty(A, \bs l).$
  By Theorem 1 of \cite{Westling_Carone_2020} we need only to show that $\sup_{a \in \mc{A}} |\Gamma_n^0(a) - \Gamma_0(a)| \to_p 0$, and by the proof of Theorem 1 of \cite{Westling_Gilbert_Carone_2020} we have that $\sup_{a \in \mc{A}} |\Gamma_n(a) - \Gamma_0(a)| \to_p 0$.  Thus, by the definitions of $\Gamma_n$ and $\Gamma_n^0$ (and since $\mc{A}$ is bounded), it suffices to show that $\wh \lambda_n \to_p 0$.  This follows by Lemma~\ref{lem:lambda_knot_size},
  and so the proof is complete.
\end{proof}

\subsection{Estimator limit distributions}

We now study the limit distributions for the two estimators.  Recall the definitions
of $\kappa_0$, $\tkappa_0$, $c_0$, $\wh \theta$, and $\wh \theta^0$
from Subsection~\ref{sec:limit-distr-notat}. 

We can now present the limit distribution results.
\begin{main-file}
  The constant $\gamma_2$ is defined in \eqref{eq:gamma-defns} in the Supplement.  
\end{main-file}
\begin{all-in-one-file}
  The constant $\gamma_2$ is defined in \eqref{eq:gamma-defns} in the Appendix.
\end{all-in-one-file}
When $\beta_0 = 1$, the full estimator limit result in Theorem~\ref{thm:causal-fullestimator-limit} is given by \cite{Westling_Gilbert_Carone_2020}, and for other $\beta_0$ values the result is new.
\begin{theorem}
  \label{thm:causal-fullestimator-limit}
  Let Assumptions~\ref{assm:EC-I}, \ref{assm:E-II_double-robust-setup}, \ref{assumption:estimators}, \ref{assm:setup},
  \ref{assm:cts-diffable},
  \ref{assm:model}, and \ref{assm:model2} hold.  Let $a \in \mc{A}$ and assume $F_0(a) \in (0,1)$ and is strictly increasing at $a$. %
  Then we have that
  $t_n^{-\beta_0} (\mle(a_0 + u t_n) - \theta_0(a_0))$ converges weakly to
  $\cc \wh \theta(\gamma_2  u)$ in $L^\infty[-K,K]$ for any $K>0$.
\end{theorem}

\smallskip\noindent Next we present the limit distribution for $\mlevr$.
\begin{theorem}
  \label{thm:causal-nullestimator-limit}
  Let the conditions of Theorem \ref{thm:causal-fullestimator-limit} hold. 
  Assume also that $H_0: \theta_0(a_0) = t_0$ is true. 
  Then we conclude that
  $t_n^{-\beta_0} (\mlevr(a_0 + u t_n) - \theta_0(a_0))$ converges in distribution to
  $ \cc  \wh \theta^0(\gamma_2 u)$ in $L^\infty[-K,K]$ for any $K>0$.
\end{theorem}

\medskip \noindent
\begin{all-in-one-file}
  The proofs are given in Appendix~\ref{app:sec:results-rates-conv}.
\end{all-in-one-file}
\begin{main-file}
  The proofs are given in Appendix~\ref{app:sec:results-rates-conv} in \cite{DRDRlocalmono-supp}.
\end{main-file}
When $u = 0$ Theorem~\ref{thm:causal-fullestimator-limit}  yields the limit distributions of the full estimator at $a_0$ (the null estimator under the null is trivial to study at $a_0$).
Also, the theorem proofs actually yield a joint limit statement for $\mle$ and $\mlevr$.
The proof relies on Lemma~\ref{lem:lambda_knot_size}; in particular, the rate
of convergence of $\wh \lambda_n$ given there implies that (a properly
normalized) $\wh \lambda_n$ converges to the tight limit random variable
$\Lambda$ (given in \eqref{eq:defn-M-m0}) that characterizes $\wh \theta^0$.

\subsection{Likelihood ratio asymptotics}

We now can study the `log likelihood ratio' statistic and its limit distribution.  The
statistic $S_n$ is defined to be
\begin{equation*}
  S_n := 
  \sum_{i=1}^n (\wh \xi_i - \wh \theta_i^0)^2-
  \sum_{i=1}^n (\wh \xi_i - \wh \theta_i)^2 ,
\end{equation*} which is nonnegative by the definitions of the two estimators.  Under our conditions (those of Theorem~\ref{thm:causal-nullestimator-limit}) which guarantee the negligibility of the remainder terms related to the nuisance parameters, the limit random variable in the limit distribution of $S_n$ is the same as that given in the non-causal case in Theorem~\ref{thm:noncausal-LRS}.  The constant, $\kappa_0$ (from  \eqref{eq:defn:kappa}), depends on the nuisance functions.
\begin{theorem}
  \label{thm:LLR-limit}
  Let Assumptions~\ref{assm:EC-I}, \ref{assm:E-II_double-robust-setup},
  \ref{assumption:estimators}, \ref{assm:setup},
  \ref{assm:cts-diffable}, 
  \ref{assm:model}, and \ref{assm:model2}
  hold.
  Let $a \in \mc{A}$ and assume $F_0(a) \in (0,1)$ and is strictly
  increasing at $a$. %
  Assume  that $H_0: \theta_0(a_0) = t_0$ holds. 
  Then $S_n \to_d \kappa_0(a_0) \DD_{\beta_0}$ as $n \to \infty$.
\end{theorem}

\begin{figure}[bh]
  \centering
  \includegraphics[width=\linewidth]{\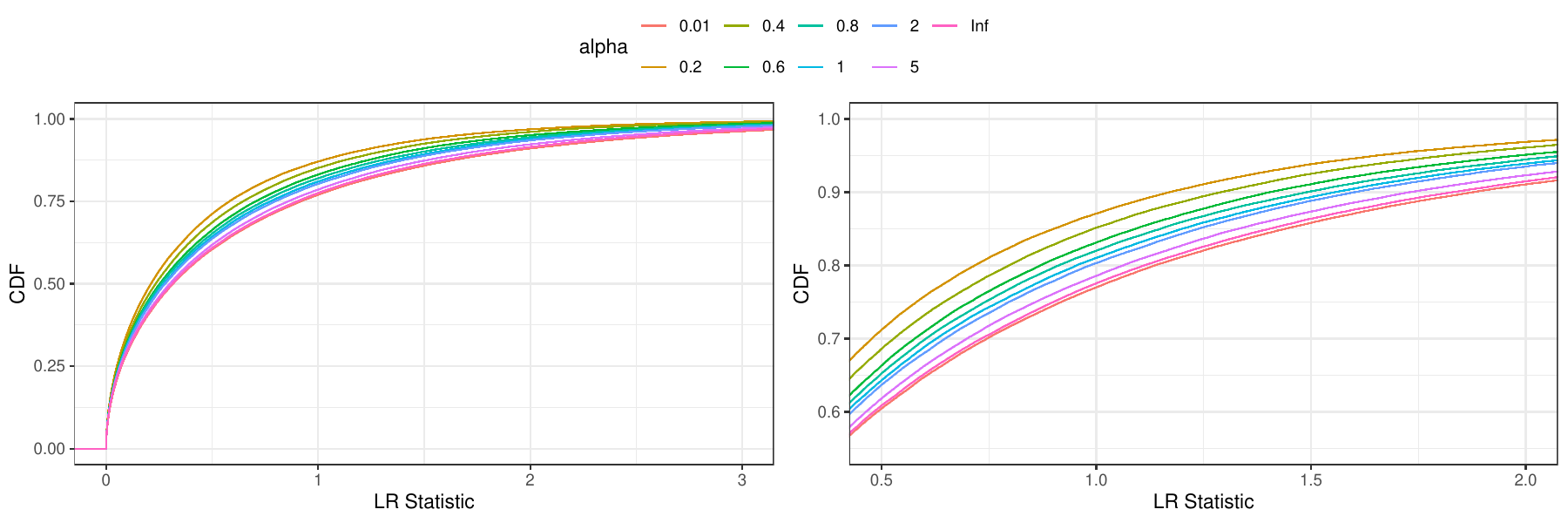}
  \caption{Estimated CDFs (plotted on domains $[0,3]$ [left] and $[.5,2]$ [right]) of $\DD_{\beta}$ for a range of $\beta$ values. \label{fig:LRS-CDFs}}
\end{figure}

The proof is in Subsection~\ref{sec:proof-theor-refthm:l}.
For estimating  %
$\kappa_0(a_0)$  %
\cite{Westling_Gilbert_Carone_2020} propose a doubly robust estimator, $\wh \kappa_0(a_0)$, essentially based on a kernel estimator of the estimated ``residuals". It is doubly robust as long as the original regression estimator is doubly robust. 
For
$\alpha \in (0,1)$ let $q_{\alpha, \beta_0}$ denote the $\alpha$ critical value (quantile) for $\DD_{\beta_0}$, i.e.\ $\PP(\DD_{\beta_0} \le q_{\alpha, \beta_0}) = \alpha$.  Then we can form a  hypothesis test with asymptotic level $1-\alpha$ which
rejects the null whenever $S_n > \wh \kappa(a_0) q_{1-\alpha, \beta_0}$. We can find the critical values of $\DD_{\beta_0}$ by simulation: %
Figure~\ref{fig:LRS-CDFs} presents Monte Carlo'd estimates of the limit
distribution $\DD_{\beta}$ for a range of $\beta$ values, based on $10,000$
Monte Carlos.  Table~\ref{tab:Dbeta-crit-values} presents the corresponding $95$\% critical values.
\modCRD{Table~\ref{tab:Dbeta-crit-values_all} provides critical values at other $\alpha$ levels. }
\remCRD{ Correct here.} 
For each Monte Carlo we simulated a Brownian motion plus
drift, $M_{\beta}$ (defined in \eqref{eq:defn-M-m0}), on domain
$[-5,5]$ %
on an equally spaced grid $\{ x_{i} \}$ with
$10,000$ points ($0.005$ grid width).  The ``derivative'' was computed to yield
data $y_i := (M_{\beta}(x_{i+1})-M_\beta(x_i)) / .005$, which we used to
compute the two estimators and then the likelihood ratio statistic.  (For
computing the full model estimator, this procedure is equivalent to computing
the GCM. And for the null estimator, it is equivalent to computing the two
one-sided GCMs and combining them as described in \cite{Banerjee:2001jy}.)
\begin{table}[bh]
  \centering
  \begin{tabular}{|c|c|c|c|c|c|c|c|c|}
    \hline
    $\beta$ & $0.01$ & $0.2$ & $0.4$ & $0.6$ & $0.8$ & $1$ & $2$ & $5$ \\
    \hline
    $q_{.95, \beta}$ & 1.65 & 1.81 & 1.98 & 2.10 & 2.18 & 2.25 & 2.44 & 2.57 \\ 
    \hline
  \end{tabular}
  \caption[$95$\% critical values for $\DD_\beta$'s]{$0.95$-critical values for $\DD_{\beta}$ for a range of $\beta$ values.\label{tab:Dbeta-crit-values}}  
\end{table}

\subsection{An adaptive procedure}
\label{sec:an-adapt-proc}

The CI's developed so far depend on knowing the flatness parameter $\beta_0$.
It is well known that over general smoothness classes (without any modifications)
adaptation in confidence intervals is impossible \citep{Low_1997}, but for shape-constrained classes, certain types of adaptation, such as to the flatness parameter $\beta_0$, are in fact possible \citep{Cai_Low_Xia_2013}.
\cite{Cai_Low_Xia_2013} develop such adaptive procedures in the white noise model and fixed  equi-spaced design regression, and they present theoretical results but do not implement their procedures or study them in practical settings (nor do they consider random design).

We are able to develop an adaptive procedure here. 
The Monte Carlo results displayed in Figure~\ref{fig:LRS-CDFs} suggest that the distributions of $\DD_\beta$ are stochastically increasing in $\beta$, meaning that $\PP(\DD_{\beta_1} \le d) > \PP(\DD_{\beta_2} \le d)$, $d > 0$, whenever $0 < \beta_1 < \beta_2 < \infty$.
This provides the ability to develop rate-adaptive confidence intervals.
If we know an upper bound $\mathfrak{b}>0$ on $\beta_0$, then we can select the  $1-\alpha$ critical value $q_{\alpha, \mathfrak{b}}$  of $\DD_{\mathfrak{b}}$ and form a test (and then a corresponding CI) that rejects whenever $S_n > \wh \kappa(a_0) q_{\alpha, \mathfrak{b}}$.  
Since for any $\beta_0 \le \mathfrak{b}$, we have  $q_{\alpha, \beta_0} < q_{\alpha, \mathfrak{b}}$, the test is slightly conservative (by a constant factor) when $\beta_0$ is the true flatness parameter. %
\modCRD{(In fact, we conjecture that by considering the case of Brownian motion with no drift, one arises at the ``$\beta = \infty$'' case, which will yield a distribution that is stochastically larger than that of $\DD_\beta$ for all $\beta < \infty$, allowing CI's that adapt over all $\beta \in (0,\infty]$. Since the likelihood ratio statistic on a flat region requires different techniques for its study, %
  we leave theoretical study of that case for separate work. We include that case in the Monte Carlo study presented in   Figure~\ref{fig:LRS-CDFs}.)}
On the other hand,  the confidence interval is the same as if we had just specified a slightly smaller $\alpha$ value and so its  expected length is thus the same order of magnitude. We did not formally study the order of magnitude of the expected length of the confidence intervals here but the likelihood ratio can be expected to yield the optimal order of magnitude which has been shown in other settings \citep{Banerjee:2001jy, Banerjee:2005gw, Banerjee:2005im}.

\section{Simulations}
\label{sec:simulations}

\begin{figure}%
  \centering
  \tcbox{\includegraphics[width=.9\linewidth]{\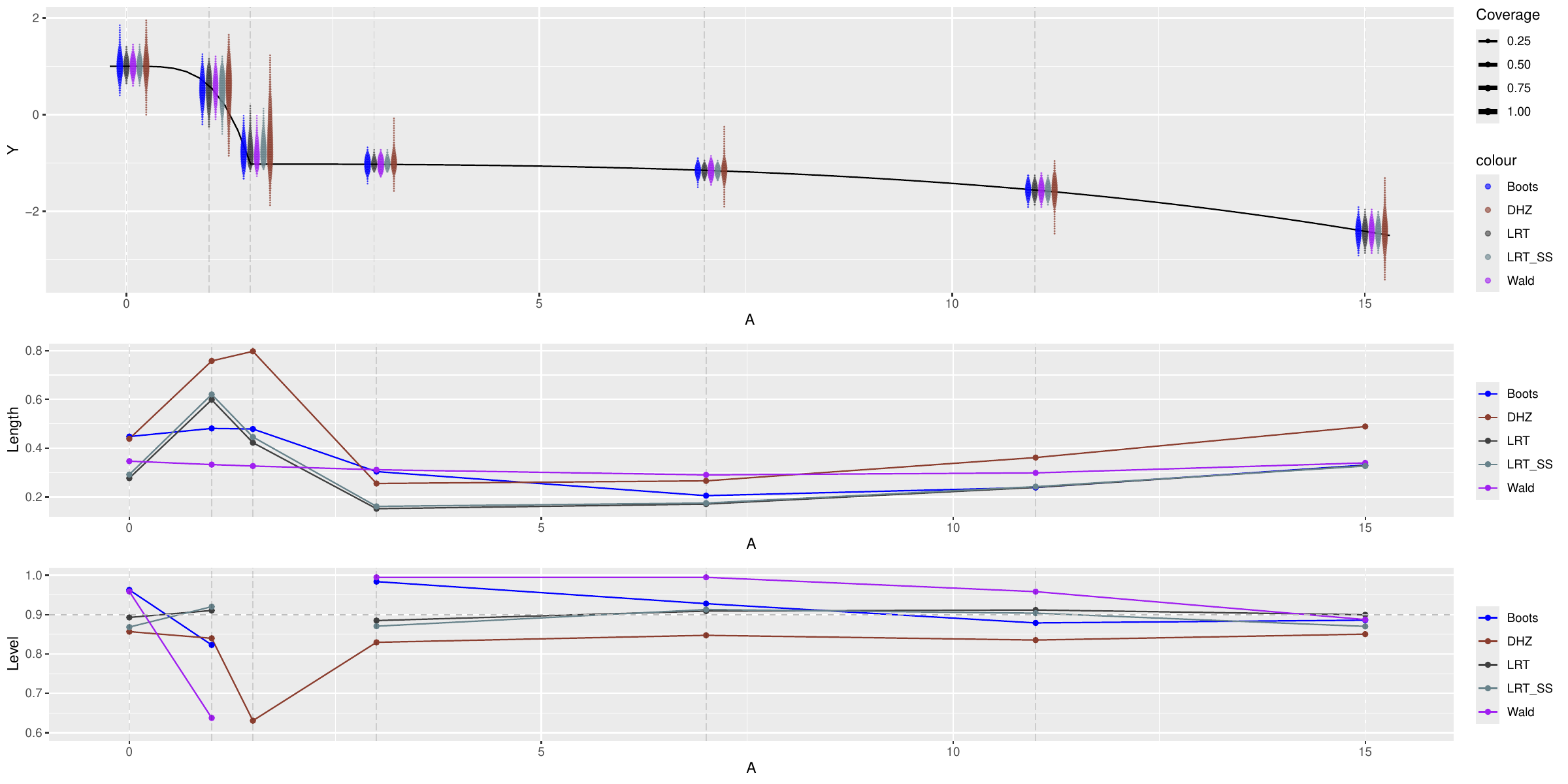}}
  \tcbox{\includegraphics[width=.9\linewidth]{\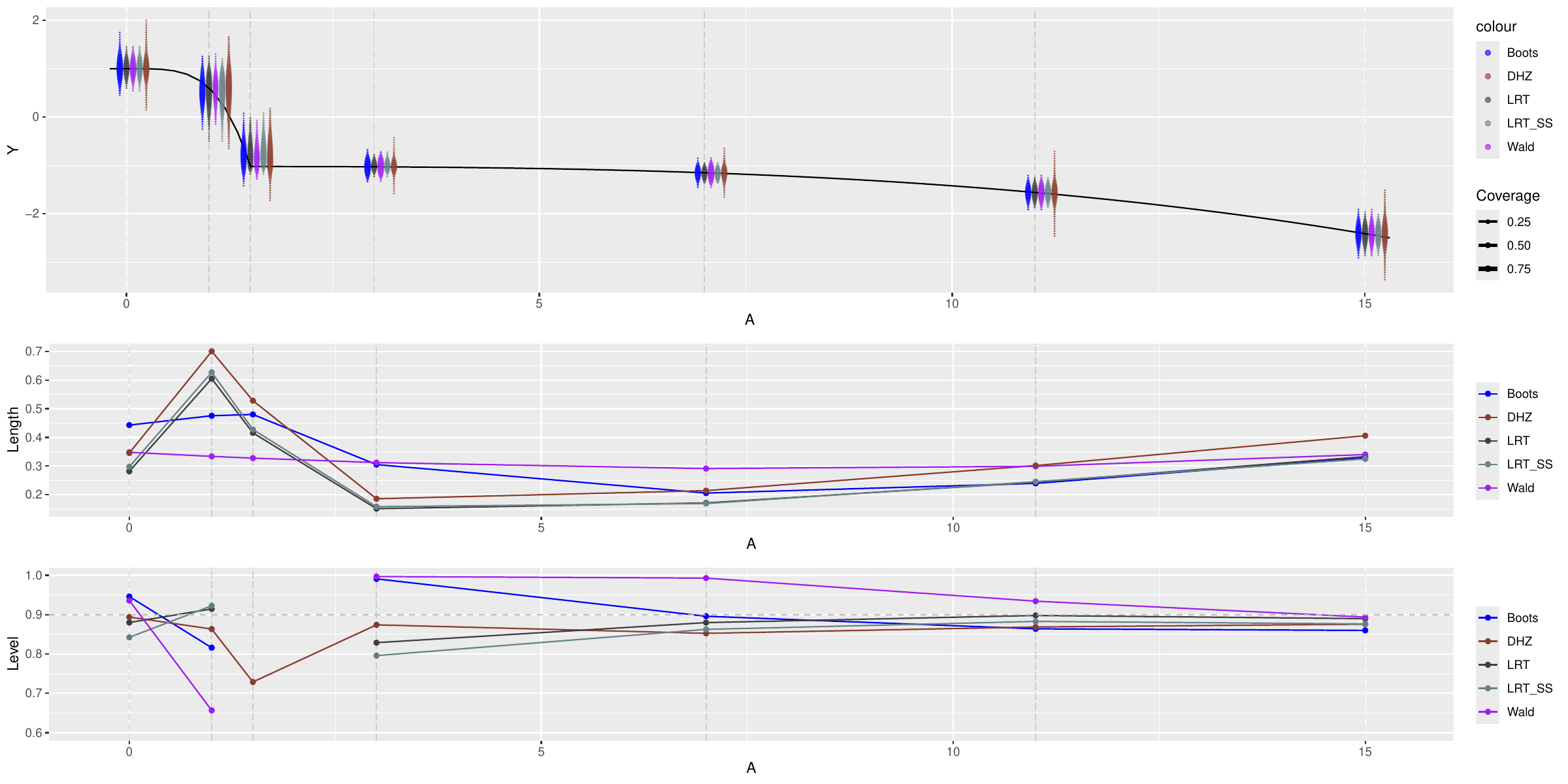}}

  \caption{Simulation study of CI procedures at 7 treatment values. There are 2 sets of 3 plots each. The top plot-triple is at the lower confounding level and the lower plot-triple is at the higher confounding level. In each triple: the top plot visualizes CI coverage (equivalently, test power), the middle plot presents average length, the bottom plot presents estimated confidence level (nominally $90$\%).  A complete description is given in the text. \label{fig:sim-plots_confounding.66}}  
\end{figure}

Here we present some simulation results for our and other procedures.
Our procedure is implemented in the R package \code{DRDRmonoLRT}, available on the author's webpage. 
The data were generated as follows.
We have $d=4$ confounders and
use
normal distributions for $A$ and
$Y|(\bs L, A)$.  Let $ {\bs{L}}=(L_1, L_2, L_3, L_4)^T\sim N(0,\bs{I}_4)$ where $\bs{I}_4$ is the identity matrix.
Then let
$(A|\bs{L}) \sim N(7.5  + \lambda(\bs{L}),  7.5^2),$ %
with $ \lambda(\bs{L}) = s (L_1 + L_2 - L_3 - L_4),$
for a constant $ s \in \RR$.
We simulate the continuous response from a conditional normal distribution as
$ (Y|\bs{L},A) \sim N(\mu(\bs{L},A),0.5^2),$
where
\begin{align*}
  \mu(\bs{L},A)= 1 +  %
  s \cdot (2, 2, -2, -2)\bs{L} +
  0.0025 A\cdot  (1 - L_1 + L_3 - .2A^2) 
  + c(A)
\end{align*}
where $c(a)$ %
is the decreasing (continuous) function that equals
$- 0.4 \sign(a)  a^4$ on $[-1.5,1.5]$ and equals $\pm  0.4  (1.5)^4$ outside of $[-1.5,1.5]$.
(We will focus attention on the curve in the interval $[0,15]$.)
In {\bf Model 1} (lower confounding level) we set $s = 0.1$ and in {\bf Model 2} (higher confounding level) we set $s = 0.2$.
The true dose-response curve is $\theta_0(a) = c(a) - (0.0025 \times .2) a^3$; this is visualized as the solid black curve in the top plots of each plot-triple in Figure~\ref{fig:sim-plots_confounding.66}. 
This curve has a variety of features. It has a point $a=0$ with flatness level $\beta=3$. The next treatment level is a point of significant steepness (large negative first derivative) complicated in finite samples by having nearby points of flatness and nonsmoothness.
The third treatment level point is pathological: the left and right derivatives are different (and so none of the procedures work correctly).  The remaining points all have nonzero derivative which is decreasing (increasing in absolute value) as we move out along  the cubic curve.
In Figure~\ref{fig:sim-plots_confounding.66}, we plot results based on a simulation with $1000$ Monte Carlo replications and a sample size of $1000$ for both Model 1 and Model 2 (with three plots for each model).

We implement our method without and with sample splitting (``LRT", ``LRT\_SS"), we implement the Wald procedure (``Wald"), and we implement the bootstrap assisted procedure of \cite{cattaneo2023bootstrap} %
(``boots'').
\modCRD{We also implement the procedure of \cite{deng2021confidence} with the pseudo-outcomes as the response variables (``DHZ''). That paper implements adaptive confidence intervals in monotone regression, not based on a likelihood ratio.} Sample splitting is implemented with $K=2$ folds and both LRT procedures use the conservative/adaptive $\mathfrak{b} = \beta = 5$.
The two nuisance functions were both estimated parametrically with well specified models.
{(Details of model specification are given in Appendix~\ref{app:sec:simulations}.)}
More extensive simulation results (different sample sizes, nuisance misspecification, and nonparametrically estimated nuisances) are presented in Appendix~\ref{app:sec:simulations}.
\modCRD{In particular, those simulations demonstrate similar performance  when one nuisance is misspecified (and the other is parametrically estimated) as when both are correctly specified, i.e.\ the ``double robustness'' of the procedure. }

Figure~\ref{fig:sim-plots_confounding.66}
has two sets of three plots each.
In each set, the top/first plot visualizes the coverage of $90$\% CI's at $7$ different treatment values (the vertical dashed line (in all the plots) is the treatment value $a$ under consideration, where the values are  0,  1,  1.5,  3,  7, 11, and 15). The black solid line is the true unknown dose-response curve.  For each procedure (at each point $a$), for each $y$ value we present the coverage by shading more thickly according to the power of the test at that point (equivalently, the proportion of time that the CI contains that point).  To present multiple procedures corresponding to a given single treatment level $a$, we slightly shifted the shaded coverage levels for each procedure so they are side-by-side (rather than on top of each other).

The second/middle plot provides the estimated average lengths of each procedure. The third/bottom plot gives the estimated confidence level. %
If a procedure has no dot present its value was off the plot. (Note that there are no dots present for the level at the third treatment value ($a=1.5$): the standard asymptotics do not apply in this pathological situation (with different left and right derivatives) and all methods fail.)

We see very good behavior for the LRT method: it has accurate level and generally the shortest or approximately shortest length (except in one case where the competitor has poor coverage). It does this automatically across the variety of different flatness regimes without any user tuning.  The Wald approach has, essentially by definition, similar widths across the curve, even when shorter or longer widths are called for, and so has incorrect level in some places.
The sample splitting LRT is similar to but slightly less efficient than the non-sample-splitting LRT.  Sample splitting is generally not expected to have significant benefit in the low dimensional regime we use in this simulation study but rather is expected to have significant benefits in higher dimensionality regimes.  We do not provide a high dimensionality simulation study since many such studies of sample splitting procedures do exist across the literature at this point.
{The bootstrap-assisted adaptive method performs better than the generic Wald procedure does, having good performance at some points, but  it does not seem to adapt fully to all the different flatness regimes and has both conservative and anticonservative behavior at other points.}

\modCRD{The DHZ procedure does demonstrate adaptive behavior, like the LRT
  procedures. Interestingly, in some small sample size scenarios (e.g.,
  $n=200$, $S=0.2$, in Appendix~\ref{app:sec:simulations}) DHZ seems to somewhat outperform the LRTs. In most regimes, and especially when sample size is larger, the DHZ procedure on average tends to be longer than the LRTs, and as can be seen from the CI coverage (test power) plots this is caused by heavy tails, meaning that the CI lengths can be quite long with nonnegligible probability.
  This is arguably a detriment to using DHZ in practice. This characteristic
  is related to the fact that the DHZ interval (length) involves division by
  a random `local bandwidth', which may sometimes be small. For some reason
  in the high complexity (SuperLearner) settings, DHZ performed quite
  poorly. Further simulations and discussion about them can be found in Appendix~\ref{app:sec:simulations}; those simulations generally reinforce the story described above. 
}

\section{Data on nursing hours and hospital readmissions}
\label{sec:data-example}

\begin{figure}[t]
  \centering
  \includegraphics[width=.75\linewidth]{\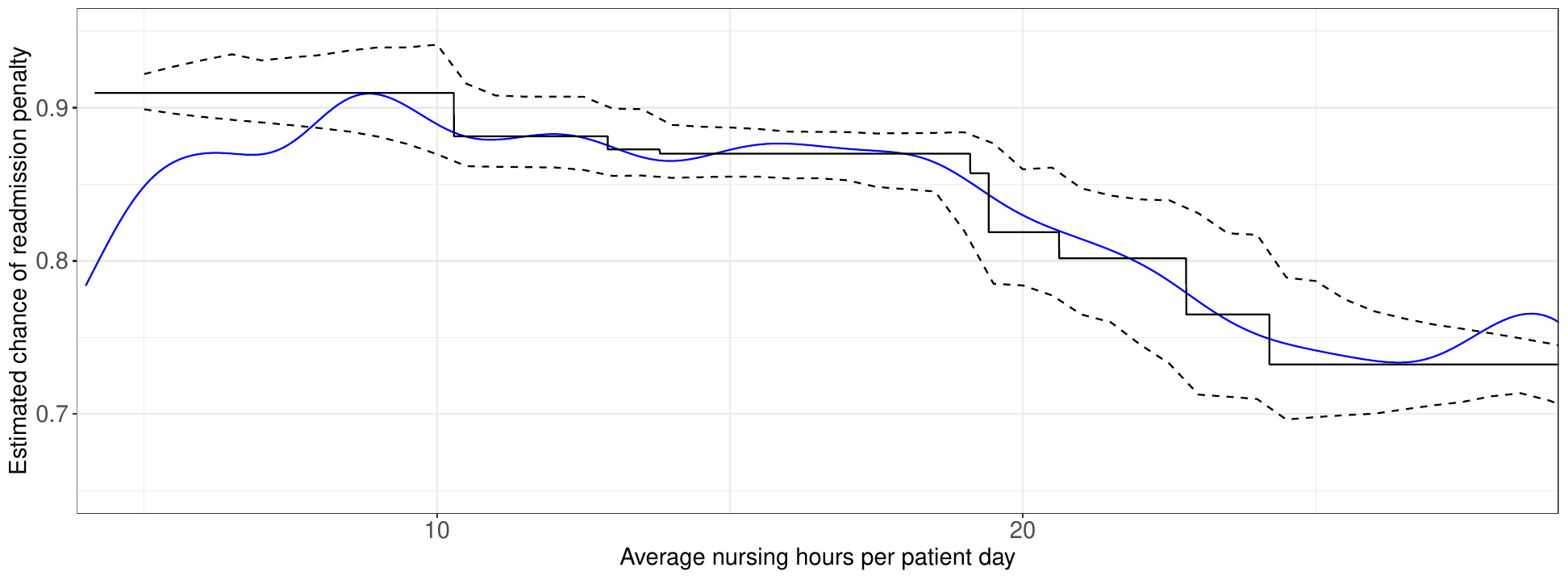}
  \caption{Estimates and CI's for treatment effect of average nursing hours on probability of (readmission) penalty. Solid lines are estimates (blue = \cite{Kennedy:2017cq}, black = \cite{Westling_Gilbert_Carone_2020}) and dotted lines are our 90\% CI's.
    \label{fig:realdata-nursing-main-plot}
  }  
\end{figure}

\begin{figure}[t]
  \centering
  \includegraphics[width=.75\linewidth]{\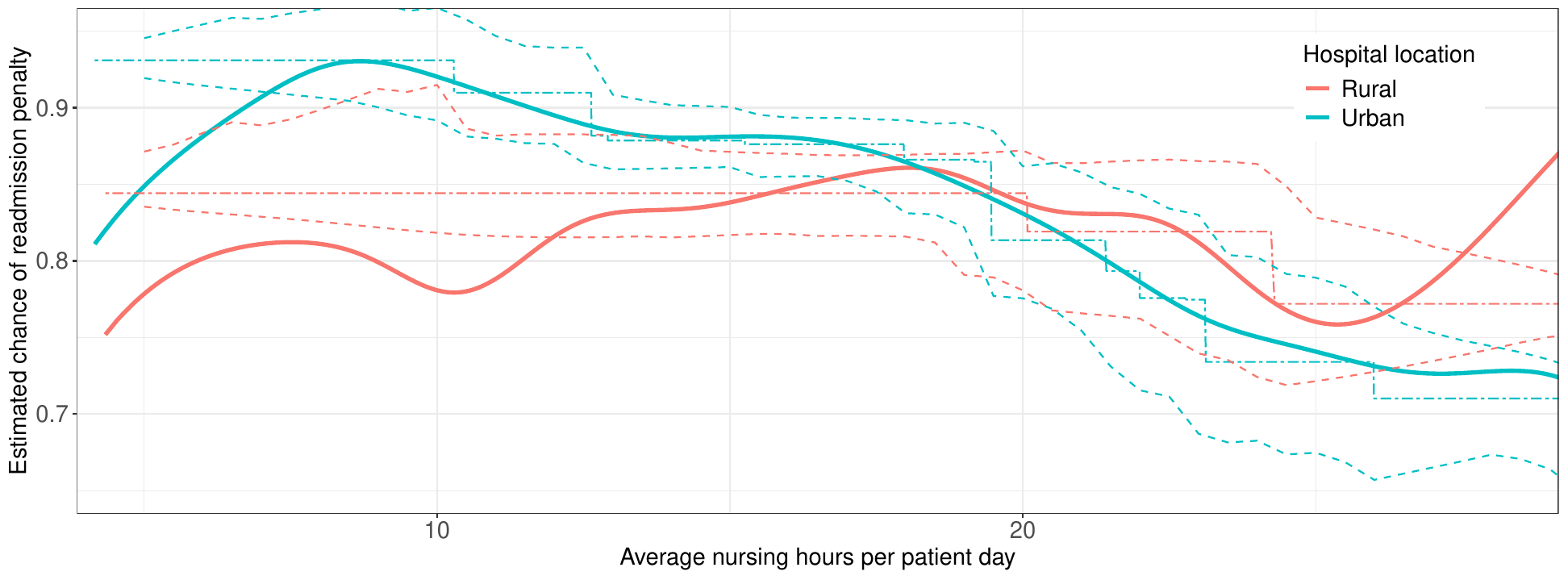}
  \caption{Estimates and CI's for treatment effect of average nursing hours on probability of (readmission) penalty, by hospital location type (urban vs.\ rural). Solid lines are estimates of \cite{Kennedy:2017cq}, dash-dotted lines are monotonic estimates \citep{Westling_Gilbert_Carone_2020}, and dotted lines are our 90\% CI's.
    \label{fig:realdata-nursing-urban-rural-plot}}  
\end{figure}

In this section we present the results of applying our method to a nurse staffing dataset, with plots given in Figure~\ref{fig:realdata-nursing-main-plot} and Figure~\ref{fig:realdata-nursing-urban-rural-plot}.  An important health policy question is whether increasing the number of or hours of nurses in a hospital will improve patient outcomes.  In \cite{McHugh:2013gn} (see also \cite{Kennedy:2017cq, drtest-main}) the authors study this question by looking at data from the American Hospital Association (\url{https://www.aha.org/}) on whether nurse staffing %
affected a hospital's risk of ``excess readmission penalty,'' after adjusting for hospital characteristics as possible confounders.  Under the Affordable Care Act, the Center for Medicare \& Medicaid Services (CMMS; \url{https://www.cms.gov}) penalizes hospitals for whether they have readmissions of patients in excess of a threshold defined by CMMS, with the goal of improving patient care.  Our unit of analysis is a hospital, and the outcome $Y$ is an indicator for whether the hospital was penalized due to excess readmissions by CMMS.
The treatment $A$ measures nurse
staffing hours.
There are nine possible confounder variables $\bs{L}$.
Further details about the variables and their definitions are given in Appendix~\ref{sec:furth-deta-nursing}.
We use Super Learner \citep{van2007super} (with the same implementation as in \cite{Kennedy:2017cq, drtest-main}) to estimate $\pi_0$ and $\mu_0$.
We truncate $\wh\pi_n$ to be $0.01$ if the %
estimate fell below that value.
It is reasonable to assume, or at least of interest for a data analyst to consider, that hospital performance (the probability of readmissions penalty) would not get worse (increase) on average if a hospital were assigned more nurse staffing hours.

In Figure~\ref{fig:realdata-nursing-main-plot} we present estimates and CI's for the treatment effect of nursing hours on readmissions penalty. The solid lines are estimates and the dotted lines are our 90\% CI's.
The black lines are based on the assumption of non-increasingness
and the blue (smooth solid) line is the estimate of \cite{Kennedy:2017cq}.  The CI's do not use sample splitting and are based on setting $\beta=5$. Near the edges there are many fewer data points and many of the propensity scores were truncated, so inference is less reliable there.

In Figure~\ref{fig:realdata-nursing-urban-rural-plot}, we present similar output but grouped by hospital location type: rural (569 data points) or urban (2089 data points). Note that we leave out the monotonicity-based estimators (to avoid plot clutter).  In \cite{drtest-main}, the hypothesis test developed in that paper rejected the no-treatment-effect hypothesis for urban hospitals but failed to reject that hypothesis for the rural hospitals.  For the urban hospitals, the (smoothness-based) estimate in Figure~\ref{fig:realdata-nursing-urban-rural-plot} mostly (away from the edges) falls within the CI's, and the overall trend of the estimate and CI's is similar to the overall (downward) trend of the combined data presented in Figure~\ref{fig:realdata-nursing-main-plot}.  But for the rural hospitals, the picture is somewhat different. The CI's are somewhat wide and it is relatively clear from seeing the CI's why the global test of \cite{drtest-main} did not reject the null, which is not so obvious just from looking at the estimate.  Also, the monotonicity-based CI's
diverge somewhat from the smoothness-based estimate. If we do believe that we have adequately captured the confounders and that monotonicity is a reasonable assumption, then this illustrates the benefit of using the monotonicity assumption.

%

%
%
%
%

\bigskip
\bigskip
\noindent {\LARGE \bf Appendices}

\appendix

\section{Sample splitting}
\label{sec:sample-splitting}

The causal estimators developed in Section~\ref{sec:caus-isot-regr}
are based on nuisance functions that use the entire sample.
This requires entropy conditions that limit the complexity of the class of nuisance functions.  A technique for allowing higher complexity nuisance functions is so-called sample splitting (or cross fitting) in which the nuisances are trained on a separate part of the sample than is used for estimating/testing the target parameter, and since there is no dependence there are no complexity restrictions. This is effective in high dimensional or high complexity data generating regimes \citep{belloni2018uniformly, Chernozhukov_Chetverikov_Demirer_Duflo_Hansen_Newey_Robins_2018}.

\cite{Westling_Gilbert_Carone_2020} study a cross fitting estimator (see also \cite{Banerjee:2016uz} who consider a monotonic sample splitting problem without nuisances).  In cross fitting we split the sample into $K$ approximately equally sized folds (subsets of indices), $V_1, \ldots, V_K$ (leaving off the $n$ subscript). Let $V_{-k} := \cup_{j \ne k} V_j$ be all indices outside the $k$th fold.  For $k \in \{1, \ldots, K\}$, we estimate nuisance functions based on the data points in $V_{-k}$ (outside the $k$th fold) and then plug those estimates in to \eqref{eq:pseudo-outcome-hat} to form pseudo-outcomes based on the samples in $V_k$ (within the $k$th fold).  We then can form $K$ test statistics $S_{n,k}$ and then average them together to yield $\overline S_{n} := K^{-1} \sum_{k=1}^K S_{n,k}$.  Assuming $K$ is fixed and $n$ is large, $\overline{S}_n$ is approximately distributed as $\kappa_0(a_0) K^{-1}\sum_{k=1}^K \DD_{\beta_0, k}$ where $\DD_{\beta_0,k}$, $k=1,\ldots, K$ are $K$ independent variables distributed as $\DD_{\beta_0}$.  As with $K=1$, for any fixed $K>1$ we can simulate the distribution of $ \sum_{k=1}^K \DD_{\beta_0, k}$. %
This approximation holds under Assumptions
\ref{assm:E-II_double-robust-setup}, \ref{assumption:estimators},
\ref{assm:setup}, \ref{assm:cts-diffable},
\ref{assm:model}, and \ref{assm:model2};
we do not need Assumption~\ref{assm:EC-I} to hold.

One can estimate $\kappa_0(a_0)$ similarly. For doubly robust estimation of $\kappa_0(a)$ we start by defining the ``residual''
$\eta_\infty(y,a,\bs l) := ((y - \mu_\infty(a, \bs l))/g_\infty(a, \bs l) + \theta_\infty(a) - \theta_0(a) )^2$ (where recall $\theta_\infty(a):= \PP_0 \mu_\infty(a, \bs L)$). To estimate via sample splitting, for the $k$th fold we have estimates $\wh \mu_{n,k}$ and $\wh g_{n,k}$ which are based on $V_{-k}$. We
then form $\theta_{\mu, n, k}(a) := \PP_{n,k} \wh \mu_{n,k}(a, \bs L)$
where $\PP_{n,k}$ is the empirical distribution based on the samples in $V_k$. We let $\wh \theta_{n,k}(a)$
be  the doubly robust isotonic estimator based on $\wh \mu_{n,k}, \wh g_{n,k}$ and the samples in $V_k$.  Plugging these in we let
$\wh \eta_{n,k, i} :=  ((Y_i - \wh \mu_{n,k}(A_i, \bs L_i))/ \wh g_{n,k}(A_i, \bs L_i) + \wh \theta_{\mu, n, k}(A_i) - \wh \theta_{n,k}(A_i) )^2$ for $i \in V_k$. Then for each $k$ we can compute based on some local smoothing method (see e.g.\ the discussion in Subsection~4.3 of  \cite{Westling_Gilbert_Carone_2020}) an estimate $\wh \kappa_{n,k}(a_0)$ which can be aggregated into $\overline{\kappa}_n := K^{-1} \sum_{k=1}^K \wh \kappa_{n,k}(a_0)$.  This yields a (doubly robust) estimate of $\kappa_0$ which again does not rely on entropy conditions.

\section{Monotone  regression}
\label{sec:non-causal-isotonic-regression}
In this section we present some results about the classical (non-causal) monotone regression problem.  They are new and of interest in their own right, and are also necessary for the development of analogous causal results developed in the sections following this one.  
\begin{all-in-one-file}
  Proofs are given in Appendix~\ref{sec:app:monoreg}. %
\end{all-in-one-file}
\begin{main-file}
  Proofs are given in
  Appendix~\ref{sec:app:monoreg} in 
  \cite{DRDRlocalmono-supp}. 
\end{main-file}
We consider the univariate regression problem based on $(\tilde A_i, \tilde Y_i) \in \RR^2$,  $i=1,\ldots, n$.
We let $\mc M$ be the set of nondecreasing functions $\{ \theta(\cdot) : \theta(x) \le \theta(y), \text{ if } x \le y \}$.
We let $\{ A_{(i)}\}$  be the (sorted) order statistics of the $A_i$'s, and then let $Y_{(i)}$ denote the observation corresponding to $A_{(i)}$ (so $Y_{(1)}$ is the outcome corresponding to the smallest $A_i$).
Define $k_0$ to be the index such that $a_0 \in [\tilde A_{(k_0)}, \tilde A_{(k_0+1)})$. 
Then we let $\mc{M}^0 := \{ \theta(\cdot) : \theta(\tilde A_{(k_0)}) = t_0, \theta \in \mc{M}\}$.
Here, rather than forcing $\theta_0(a_0) = t_0$ we require $\theta_0(\tilde A_{(k_0)}) = t_0$ where $A_{(k_0)}$ is the nearest treatment less than or equal to $a_0$.  This difference will not be relevant  in our asymptotic results. %
We assume
\begin{equation}
  \label{eq:reg-defn}
  \tilde Y_i = r_0(\tilde A_i) + \epsilon_i
\end{equation}
with $r_0 \in \mc{M}$, $\tilde A_1, \ldots, \tilde A_n$ independent and identically distributed,
$\epsilon_1,\ldots, \epsilon_n$ independent with $\EE( \epsilon_i | \tilde A_i) = 0$ and $\tilde \sigma_0^2(a)  \le \sigma_{\text{max}}^2 < \infty$
where
$\tilde \sigma_0^2(a) := \Var(\epsilon_i |\tilde  A_i=a)$.  For $r \in \RR^n$ define
\begin{equation}
  \label{eq:phi-defn-noncausalreg}
  \phi_n(r) := \inv{2} \sum_{i=1}^n (\tilde Y_i - r_i)^2.
\end{equation}
We sometimes overload notation and consider the argument to $\phi_n(\cdot)$ to be a function $r(x)$ which is evaluated at the data points yielding $r_i = r(\tilde A_i)$.
Assume that $\tilde A_i$ have cumulative distribution function (CDF) $\tilde F_0$  on a set $\mc{A} \subset \RR$.
When it exists, we denote the derivative of $\tilde{F}_0(\cdot)$ by  $\tilde f_0(\cdot)$ .

As is well known, the full estimator $\mlereg$ is uniquely defined at the data points $\mlereg(A_i)$ and can be characterized as the {\it left derivative} of the {\it greatest convex minorant} (GCM) of the so-called ``cusum'' (cumulative sum) diagram which consists of the set of points
$(0,0) \cup \{ (i, \sum_{j=1}^i \tilde Y\ord{i}) \colon i \in \{1,\ldots,n\} \}$ \citep[p. 20]{Groeneboom:2014hk}.
In fact, by the results of \cite{Groeneboom:2015ew}, we can characterize not just the full estimator but {\it also} the null estimator as a left derivative of a GCM of a certain cusum diagram, meaning that it is a ``generalized Grenander estimator'' in the terminology of \cite{Westling_Gilbert_Carone_2020}.
This is very helpful, as it allows us to use
some of the results of \cite{Westling_Gilbert_Carone_2020} that apply to generalized Grenander estimators in our proofs,
which is a novel approach to studying shape constrained likelihood ratio statistics.

\modCRD{Another way to put it is that one perspective of the null estimator is that it is a ``null'' projection operator applied to the same data. The other (Lagrange multiplier perspective of \cite{Groeneboom:2015ew}) is  that we can apply the standard (``full'') projection operator to a modified (by the Lagrange multiplier) dataset. This latter approach allows us to re-use proofs more directly. }

Below is a characterization of the null estimator.
This is similar to Lemma 2.2 of \cite{Groeneboom:2015ew}, which is analogous but about the interval censoring problem.
Define $\wh r_n$ 
and $\wh r_n^0 $ by
\begin{equation}
  \label{eq:defn-estimators-noncausal}
  \wh r_n := \argmin_{r \in \mc{M}} \phi_n(r)
  \quad   \text{ and } \quad
  \wh r_n^0 := \argmin_{r \in \mc{M}^0} \phi_n(r).
\end{equation}
Also, we let $\wt \lambda_n$  be the solution in $\lambda $ of the equation
\begin{equation}
  \label{eq:4:non-causal}
  \max_{k \le k_0} \min_{i \ge k_0} \frac{n \lambda + \sum_{j=k}^i \tilde Y_i }{i - k + 1} = t_0.
\end{equation}
The following characterizes $\wh r_n^0$.
\begin{lemma} \label{lem:1:non-causal} Assume the regression model \eqref{eq:reg-defn} with $r_0 \in \mc{M}$, and $\wh r_n$ and $\wh r_n^0$ defined by
  \eqref{eq:defn-estimators-noncausal}.
  If $\wh r_n \in \mc M^0$ then $\wh r_n^0 = \wh r_n$.  Otherwise,
  with $\wt \lambda_n$ defined in \eqref{eq:4:non-causal}, 
  $\wh r_n^0$ is the left derivative of the greatest convex minorant of the cusum diagram of the points
  \begin{equation}
    \label{eq:5:non-causal}
    \{(0,0)\} \cup \lb\lp i, \sum_{j=1}^i \tilde Y_i + n \wt \lambda_n 
    \One{j=k_0} \rp\rb_{i=1}^n.
  \end{equation}
\end{lemma}
\medskip \noindent The proof is given in Appendix~\ref{sec:app:monoreg}. %
A minor remark is that when $a_0$ is not a data point, we enforce the
constraint at the nearest data point below $a_0$.  There are very slightly
different other options, including enforcing the equality at exactly $a_0$,
but they are negligible for our theoretical results and require some
complication in the notation so we proceed in this fashion.  Next, we analyze
the order of magnitude of $\wt \lambda_n$. %
To do so, we rely on the
(Monotonicity) model assumptions described in the main paper, Assumptions
\ref{assm:model} and \ref{assm:model2}.

\begin{lemma}
  \label{lem:2}
  Assume the regression model \eqref{eq:reg-defn} with $r_0 \in \mc{M}$, and $\wh r_n$ and $\wh r_n^0$ as defined  by
  \eqref{eq:defn-estimators-noncausal}.
  Assume
  the null hypothesis $r_0(a_0) = t_0$ holds.
  Assume $r_0(a)$ satisfies Assumption~\ref{assm:model} at $a_0$
  with $\beta_0 > 0$.
  Assume that the CDF of $\tilde A_i$ is positive and differentiable at
  $a_0$ and $\tilde A_i$ satisfies
  Assumption~\ref{assm:model2}.  Define $\wt \lambda_n$ as in the previous lemma.
  Then we can conclude that
  $\wt \lambda_n = O_p(n^{-(\beta_0 +1) / (2\beta_0+1)})$.
\end{lemma}
Next we present the  asymptotic statement for the estimators as  local processes around $a_0$.
Let
\begin{equation}
  \label{eq:defn:tn}
  t_n := n^{-1 / (2\beta_0 + 1)}.
\end{equation}
Recall that we let $\tilde \sigma_0^2(a) = \Var(\epsilon | A = a)$.
\cite{Banerjee:2001jy,Banerjee:2000wh,Groeneboom_Jongbloed_2015} studied the likelihood ratio in the current status problem and found, when $\beta_0 = 1$, the limit distribution of their corresponding null hypothesis estimator. The limit distribution depends on a standard Brownian motion $W$ on $\RR$ with $W(0)=0$.  Then define  $X(t) := W(t) + |t|^{\beta_0}$. We need to ``isotonize'' $X$;
\begin{all-in-one-file}
  recall from Section~\ref{sec:notat-probl-setup}
\end{all-in-one-file}
\begin{supplement-file}
  recall from Section~\ref{sec:notat-probl-setup} in the main document
\end{supplement-file}
that we let
\begin{equation}
  \label{eq:1}
  \wh \theta(\cdot)  \equiv \wh \theta_{\beta_0}(\cdot) 
  :=
  \mono(X),
\end{equation}
where $\mono(X)$ denotes the
left derivative of the greatest convex minorant of $X$.
\cite{Banerjee:2001jy} also define a null hypothesis version of the isotonization operator (see their Theorem 2.3), which we will denote by $\mono^0$, which is an isotonization that satisfies $\mono^0(\cdot)(0) = 0$. 
Using this, we  define
$ \wh \theta^0(\cdot) \equiv \wh \theta^0_{\beta_0}(\cdot) := 
  \mono^0(X).$
Now we can and do state limit theorems for $\wh r_n$ and $\wh r_n^0$. The result for the former   is from \cite{Wright_1981}.
\begin{mylongform}
  \begin{longform}
    Limit statement constants coming from Wright 1981.  I guess need be sure they match up right.
    Old version: $    \lp \frac{4 \rho_0(a_0) \tilde \sigma^2_0(a_0)}{\tilde f_0(a_0)} \rp^{1/(2 \beta_0 + 1)}$. Wright has constant of 1 in front of monomial term. 
  \end{longform}
\end{mylongform}
\begin{theorem}[\cite{Wright_1981}]
  Assume the regression model \eqref{eq:reg-defn} with $r_0 \in \mc{M}$, and $\wh r_n$ as defined above.
  Assume $r_0 \in \mc{M}$.
  Assume that the CDF of $\tilde A_i$ is positive and differentiable at
  $a_0$ with density $\tilde{f}_0(a_0) > 0$ and $\tilde{A}_i$ satisfies
  Assumption~\ref{assm:model2}.  Assume $r_0(\cdot)$ satisfies
  Assumption~\ref{assm:model} at $a_0$ with $\beta_0 > 0$.  Then
  \begin{equation*}
    t_n^{-\beta_0} (\wh r_n(a_0 +   t_n \cdot ) - r_0(a_0)) \to_d
    \lp \frac{ \rho_0(a_0) \tilde \sigma^2_0(a_0)}{(\beta_0+1)\tilde f_0(a_0)} \rp^{1/(2 \beta_0 + 1)} \wh{\theta}_{\beta_0}( \cdot )   
  \end{equation*}
  in $L^\infty[-c,c]$, for any $c>0$, with $t_n$ from \eqref{eq:defn:tn}.
\end{theorem}

\begin{theorem}
  \label{thm:noncausal-null-hyp}
  Assume the regression model \eqref{eq:reg-defn} with $r_0 \in \mc{M}$, and  $\wh r_n^0$ as defined above.
  Assume $r_0 \in \mc{M}$ and assume
  the null hypothesis $r_0(a_0) = t_0$ holds.
  Assume that the CDF of $\tilde A_i$ is positive and differentiable at
  $a_0$ with density $\tilde{f}_0(a_0) > 0$ and $\tilde{A}_i$ satisfies Assumption~\ref{assm:model2}.  Assume $r_0(a)$ satisfies Assumption~\ref{assm:model} at $a_0$
  with $\beta_0 > 0$.
  Then
  \begin{equation*}
    t_n^{-\beta_0} (\wh r^0_n(a_0 +   t_n \cdot ) - r_0(a_0)) \to_d
    \lp \frac{\rho_0(a_0) \tilde \sigma^2_0(a_0)}{(\beta_0+1)\tilde f_0(a_0)} \rp^{1/(2 \beta_0 + 1)} \wh{\theta}_{\beta_0}^0( \cdot )
  \end{equation*}
  in $L^\infty[-c,c]$, for any $c>0$, with $t_n$ from \eqref{eq:defn:tn}.
\end{theorem}

\bigskip \noindent The two convergences in the two theorems are actually a joint convergence based on the same Brownian motion. When $\beta_0=1$, Theorem~\ref{thm:noncausal-null-hyp} gives a similar limit statement as is given  in \cite{Banerjee:2001jy} with $\rho_0 = r_0'$. 
\begin{mylongform}
  \begin{longform}
    \remCRD{ Note: $\beta_0$- ``H\"older'' is not quite enough as I sometimes
      very absent mindedly forget when i setup terminology/def'ns quickly;
      need $\beta_0$ flatness ie Holder but also with lower derivatives set
      to 0}

    \remCRD{The usual definition of beta continuity can be I guess modified
      slightly if we are considering a fixed point $x_0$.  Then, can just
      consider monomials or piecewise-at-$x_0$ monomials.  It is then not
      true that ``if $\beta > 1$ the function is constant.'' }

  \end{longform}
\end{mylongform}

Finally, we can study the LRS in the (non-causal) regression setting and give
its limit distribution.  The limit distribution is pivotal except for the parameter $\tilde \sigma_0^2(a_0)$ which is generally easy to estimate (e.g., \cite{Rice_1984} for the homoscedastic case or \cite{Muller_Stadtmuller_1987} for the heteroscedastic case),
\modCRD{and in particular does not require estimating $\theta_0'(a_0)$ (or rather, $\rho_0(a_0)$), which is known to create difficulties for inference, as was discussed in the Introduction.}
Define
\begin{equation}
  \label{eq:non-causal-LRS}
  \tilde S_n :=
  \sum_{i=1}^n ( \tilde Y_i - \wh r_n^0)^2 -
  ( \tilde Y_i - \wh r_n)^2 > 0.
\end{equation}
We also let
\begin{equation}
  \label{eq:defn-DD_beta}
  \DD_{\beta} :=
  \int_{\RR} \lp \wh{\theta}_{\beta}(s)^2 -  \wh{\theta}_{\beta}^0(s)^2 \rp \, ds.
\end{equation}
\begin{theorem}
  \label{thm:noncausal-LRS}
  Let the assumptions from the two previous theorems hold. Then
  $\tilde S_n \to_d \tilde \sigma^2_0(a_0) \DD_{\beta_0}
  \, \text{ as } \,
  n \to \infty.$
\end{theorem}

\medskip
\noindent
This statistic can be used to test or form confidence intervals for $r_0(a_0)$ (after estimating $\tilde \sigma_0^2(a_0)$).
A downside to using the previous theorem for CI's is that it requires knowledge of $\beta_0$.  However, we are able to circumvent this and provide confidence intervals that {\it adapt} to an unknown $\beta_0$.  We discuss this
\begin{all-in-one-file}
  in Subsection~\ref{sec:an-adapt-proc}  
\end{all-in-one-file}
\begin{supplement-file}
  in Subsection~\ref{sec:an-adapt-proc}  in the main document
\end{supplement-file}
(the discussion there applies to both the non-causal and causal estimators).

\subsection{Monotone regression proofs}
\label{sec:appendix}
\label{sec:app:monoreg}

Here we present proofs for the new results in the classical (non-causal) regression setting.

\begin{proof}[Proof of Lemma~\ref{lem:1:non-causal}]
  In the case where $\wh r_n \in \mc{M}^0$, there is nothing to prove. We
  consider the case $\wh r_n \notin \mc{M}^0$.
  \begin{mylongform}
    \begin{longform}
      ``and focus on the case where $\wh r_n(a_0) > t_0$.  (The case where $\wh r_n(a_0) < t_0$ is analogous.)''

      GJ 2015 consider these  cases separately.  But I don't see why. 
    \end{longform}
  \end{mylongform}
  The objective function is $l(r) = (1/2) \sum_{i=1}^n (\tilde Y_i - r_i)^2 $ and we define a modified version with a Lagrange multiplier, $$\phi_\lambda(r ) := l(r) + n \lambda( r_{k_0} - t_0)$$ for $\lambda \in \RR$.  Optimizing $l$ over the null-constrained class is equivalent to optimizing $\phi_\lambda$ over the full model $r \in \mc{M}_n$.  Note that the convex cone $\mc{M}_n$ has generators $g_1 = (0,\ldots, 0,1)$, $g_2 = (0, \ldots, 1,1), \ldots,$ and $g_n = (1,\ldots,1)$, meaning that all elements of $\mc{M}_n$ can be represented as linear combinations of these generators with nonnegative coefficients.  Let $\nabla \phi_{\lambda}$ denote the gradient vector of $\phi_\lambda$.  Then a vector $\wh r_n^0$ is the optimum of $\phi_{\lambda}(r)$ if and only if
  \begin{equation}
    \label{eq:11}
    \la \nabla \phi_{\lambda}(\wh r_n^0), g_i\ra
    = \sum_{j=i}^n (\wh r^0_{n,j} - \tilde Y_j + n  \lambda \One{j=k_0} ) \le 0 \quad \text{ for } i=1,\ldots, n,
  \end{equation}
  with equality rather than inequality whenever $i=1$ or $i$ is a bend point of $\wh r_n^0$. 

  This entails that $\wh r_n^0 $ is the vector of left derivatives of the
  greatest convex minorant of the cusum diagram given by
  \eqref{eq:5:non-causal}, where $\wt \lambda_n$ solves \eqref{eq:4:non-causal}.
  This is because any $\wh r_n^0$ satisfying the inequalities and equalities
  given by \eqref{eq:11} is a left derivative of a corresponding greatest
  convex minorant of the cusum diagram (by, say, Lemma 2.1 and the following
  Remark of \cite{Groeneboom:2014hk}).  The left derivative of the GCM of the
  cusum diagram at a point is given by the max-min characterization, which at
  the point $k_0$ is the left hand side of \eqref{eq:4:non-causal}.  Now
  $\wt \lambda_n$ is such that $\wh r_n^0 \in \mc{M}_n^0$, so therefore
  \eqref{eq:4:non-causal} is satisfied by the max-min characterization.
\end{proof}

\bigskip

\begin{proof}[Proof of Lemma~\ref{lem:2}]
  Recall that $a_0 \in [A_{(k_0)}, A_{(k_0+1)})$.  Define
  \begin{equation*}
    \phi(\lambda) :=
    \max_{k \le k_0} \min_{i \ge k_0} \frac{\sum_{j=k}^i \tilde Y_j + n \lambda }{i - k +1}
  \end{equation*}
  This means, by the max-min characterization of the full MLE $\wh r_n $, that we have
  \begin{equation*}
    \phi(0) = \max_{k \le k_0} \min_{i \ge k_0} \frac{\sum_{j=k}^i \tilde Y_j  }{i - k +1}
    = \wh r_n(A_{(k_0)}).
  \end{equation*}
  So we let $k_1 \le k_0$ and $i_1 \ge k_0$ be the indices that satisfy
  \begin{equation*}
    \wh r_n(A\ord{k_0}) =
    \frac{\sum_{j=k_1}^{i_1} \tilde Y_j}{i_1 - k_1 + 1}
    = \max_{k \le k_0} \min_{i \ge k_0} \frac{ \sum_{j=k}^i \tilde Y_i}{ i-k +1}.
  \end{equation*}
  Assume first that $t_0 \ge \wh r_n(A\ord{k_0})$ and for any $\lambda > 0$ let $i_\lambda \ge k_0$ be the index such that
  \begin{equation}
    \label{eq:6}
    \frac{    \sum_{j=k_1}^{i_\lambda}  \tilde Y_j + n \lambda  }{i_{\lambda} -k_1 + 1}
    =\min_{i \ge k_0}      \frac{    \sum_{j=k_1}^{i} \tilde Y_j  + n \lambda  }{i  -k_1 + 1}
    =: \phi_{k_1}(\lambda)
  \end{equation}
  with $\phi_{k_1}(\cdot)$ defined by the previous display.  Then since $\phi_{k_1}(\lambda)$ is continuous, increasing in $\lambda$ (see the proof of Lemma 2.3 of \cite{Groeneboom:2015ew}), and approaches $\infty$ as $\lambda$ gets large,  by the Intermediate Value Theorem there must exist a (random) $\lambda_1 \equiv \lambda_{1,n} > 0$ such that
  \begin{equation*}
    \frac{\sum_{j=k_1}^{i_\lambda} \tilde Y_i + n \lambda_1 }{ i_\lambda - k_1 + 1}
    =  \phi_{k_1}(\lambda_1)
    = t_0.
  \end{equation*}
  Since the null holds,
  letting    $\tilde \PP_n(c,y)$ be the empirical measure of $\{ (\tilde A_j, \tilde Y_j) \}$, we have
  \begin{equation}
    \label{eq:7}
    \lambda_1 %
    =
    \int_{c \in [\tilde A_{(k_1)}, \tilde  A_{(i_\lambda)}]} ( r_0(a_0) - y) d \tilde \PP_n(c, y).
  \end{equation}
  Now, for any $\epsilon > 0$ we can choose an $M >0$ such
  that for $n$ large enough 
  \begin{mylongform}
    \begin{longform}
      In Groeneboom Wellner page 97 they have the sup inside the probability statement.  In Groeneboom Jongbloed 2015 it is outside.  I guess inside is needed. 
    \end{longform}
  \end{mylongform}
  \begin{equation}
    \label{eq:8}
    \PP \lp
    \sup_{b_0 \ge a > a_0 + M n^{-1  / (2\beta_0 + 1)}}   
    \int_{u \in [A_{(k_1)}, a]} (r_0(a_0) - y ) d \tilde \PP_n(u, y)
    < 0 \rp > 1 - \epsilon. 
  \end{equation}
  \begin{mylongform}
    \begin{longform}

      Define
      \begin{equation*}
        \tau^-(b) := \sup \{ c \le b : \wh r_n(b-) \ne \wh r_n(b+) \}
      \end{equation*}
      and similarly
      $\tau^+(b) := \inf \{ c \ge b : \wh r_n(b-) \ne \wh r_n(b+) \}.$
      We will show that $a_- - \tau^-(a_-)  = O_p(n^{-1/ (2\beta_0 + 1)})$  as $n \to \infty$.

      By the well-known characterization of $\wh r_n$, 
      (see, e.g., Lemma 2.1 of \cite{Groeneboom:2014hk}) for any $a > \tau^-$ we have
      
      $\sum_{\tau^- \le \tilde A_i < a} (\wh r_n(\tilde A_i) - Y_i) \le 0$.

      \remCRD{ no,   $\sum_{\tau^- \le \tilde  A_i < a} (\wh r_n(\tilde  A_i) - Y_i) \ge 0$.}

      Now if $r_0(a_0) \le \wh r_n(a_0 - M)$ then for any $a > \tau^-$, by monotonicity,
      \begin{equation*}
        0 \ge \sum_{\tau^- \le \tilde  A_i < a} (\wh r_n(\tilde  A_i) - \tilde Y_i)
        \ge \sum_{\tau^- \le \tilde  A_i < a} ( r_0(a_0) - \tilde Y_i),
      \end{equation*}
      which equals
      \begin{equation*}
        \sum_{\tau^- \le \tilde  A_i < a} ( r_0(a_0) - r_0(\tilde A_i) - \epsilon_i ).
      \end{equation*}
      \modCRD{Now, by using continuity and strict monotonicity of $r_0$ in a left-neighborhood of $a_0$,  a Kim-Pollard type argument \citep{Kim:1990ue} applies to show that  this last term    $\sum_{\tau^- \le \tilde  A_i < a} ( r_0(a_0) - r_0(\tilde A_i) + \epsilon_i )$ is strictly positive with high probability if $a_0 - \tau^- \ge \delta > 0$ for fixed $\delta$. }
      Thus we conclude that $a_0 - \tau^-(a_0) = O_p(n^{-1  / (2\beta_0 + 1)})$.
      Using an analogous argument,  we  can also show that $\tau^+(a_+) - a_+ = O_p(n^{-1  / (2\beta_0 + 1)})$.

    \end{longform}
  \end{mylongform}
  This is shown as follows.  
  We let
  $f_b(a,e) := \One{ a_0 \le a \le a_0 + b} e$ so that
  \begin{equation*}
    n^{-1} \sum_{a_0 \le \tilde  A_i < a_0+b}  \epsilon_i  = \int f_b(a,e) d \tilde \PP_n(a, e).
  \end{equation*}
  Let $\mc{F}_R := \{ f_b : 0 \le a_0 + b \le R \}$.  Then  $\mc{F}_R$ is a VC class,
  since  Example 2.5.4 of \cite{vanderVaart:1996tf}
  shows that indicator functions of intervals are VC,
  and multiplication by a single function preserves the VC property (Lemma 2.6.18, \cite{vanderVaart:1996tf}).  The envelope of $\mc{F}_R$ is of order $R$ since we assume $F_0$ is differentiable at $a_0$, so $\PP(\tilde A \in [ a_0, a_0+R]) = F'(a_0)R + o(R)$, and since $E( \epsilon_i^2 | \tilde A_i) \le \sigma_{max}^2 < \infty$.    
  Then by Lemma A.1 of \cite{Balabdaoui:2007jj} (with $d=1$ and $s = \beta_0$ for any $\beta_0 > 0$), for any  $\epsilon > 0$ there exists an $M_n = O_p(1)$ such that
  $|(\tilde \PP_n - P_0)f_b| = |\tilde \PP_n f_b| \le \epsilon |b- a_0|^{\beta_0 + 1} + n^{-(\beta_0 + 1) / (2 \beta_0 + 1)} M_n.$

  And, on the other hand,
  since $|r_0(a_0) - r_0(a)| = L|a_0 - a|^{\beta_0} + o((a_0 - a)^{\beta_0})$
  by Assumption~\ref{assm:model} (with $\beta_0 > 0$),
  we have that
  \begin{equation*}
    \int_{[a_0, a_0 + b]} (r_0(a_0) - r_0(a) ) d \tilde \PP_n(a,s) \le
    - \max(\epsilon L (a_0 - b)^{1 + \beta_0}, M n^{-(\beta_0 + 1)/(2 \beta_0 + 1)}),
  \end{equation*}
  for all $b \in [M n^{-1/( 2 \beta_0+1)}, b_0]$, for some $b_0 > 0$ fixed, and some $\epsilon > 0$, with high probability.
  This follows similarly as above by considering a class of functions $g_b(a,s) = \One{b \le a \le a_0} (r_0(a_0) - r_0(a))$ for $b \in [a_0, b_0]$ for some $b_0 \ge a_0$.  It is again VC (again by Example 2.5.4 and Lemma 2.6.18 of \cite{vanderVaart:1996tf}) and has constant envelope $\max(|r_0(a_0)|, |r_0(\tilde b)|)$ so is a Donsker class (Theorem 2.5.2  of \cite{vanderVaart:1996tf}) which means that $ \int_{[a_0, a_0+b]} (r_0(a_0) - r_0(a) ) d \tilde \PP_n(a,s)$ equals
  \begin{equation*}
    \int_{[a_0, a_0+b]} (r_0(a_0) - r_0(a) ) d( \tilde \PP_n - P_0)(a,s)
    +      \int_{[a_0, a_0+b]} (r_0(a_0) - r_0(a) ) d P_0 (a,s)
  \end{equation*}
  where the first term is negligible and the second term is  bounded above by
  $-\max(\epsilon L (a_0 - b)^{1 + \beta_0}, M n^{-(\beta_0 + 1)/(2
    \beta_0 + 1)})$ with high
  probability, for all $b \in [-M n^{-1/( 2 \beta_0+1)}, b_0]$.  This
  shows that \eqref{eq:8} holds if $A_{(k_1)}$ is replaced by $a_0$ in the
  integral expression.  The full statement of \eqref{eq:8} then follows by an
  extension of the above argument, using that
  $a_0 - A_{(k_1)} = O_p(n^{-1/(2\beta_0 +1)})$.  So \eqref{eq:8} has been
  shown.

  Now since $\lambda_1 > 0$ by assumption and by
  \eqref{eq:7} we have 
  \begin{equation}
    \label{eq:9}
    0 <     \int_{t \in [\tilde A_{(k_1)}, \tilde  A\ord{i_\lambda}]} ( r_0(a_0) - y) d \tilde \PP_n(t, y).
  \end{equation}
  This allows us to conclude that
  $|\tilde A\ord{i_\lambda} - a_0|  = O_p(n^{-1 / (2\beta_0 + 1)})$ (by comparing \eqref{eq:8} and \eqref{eq:9}).
  Continuing, the right side of
  \eqref{eq:9} equals
  $     \int ( r_0(a_0) - r_0(a) + r_0(a) - y) d \tilde \PP_n(a,y) $ which equals 
  \begin{equation}
    \label{eq:10}
    \begin{split}
      \int (r_0(a_0) - r_0(a) ) d(\tilde \PP_n - P_0 + P_0)(a,y)
      + \int (r_0(a) - y)
      d \tilde \PP_n (a,y),
    \end{split}
  \end{equation}
  where the integrals are over
  $a \in [\tilde A\ord{k_1}, \tilde A\ord{i_\lambda}]$; using the same arguments as
  above, we conclude that the first term of \eqref{eq:10} is
  $O_p(n^{-(\beta_0+1) / (2 \beta_0 + 1)})$ and the second term is also
  $O_p(n^{-(\beta_0+1) / (2 \beta_0 + 1)})$
  since $ \tilde  A\ord{i_\lambda} - \tilde A\ord{k_1} = O_p(n^{-1 / (2\beta_0 + 1)})$.
  \begin{mylongform}
    \begin{longform}
      Last term does not need to be centered since it is already mean 0.

      Note for simply concluding the order of magnitude is
      $O_p(n^{-2 \beta_0 / (2\beta_0 + 1)})$, simple empirical process arguments are fine without needing a kim-pollard arg.  Of course, we've already done  the kim-pollard argument. 
    \end{longform}
  \end{mylongform}

  Thus, we conclude that $\lambda_1 = O_p(n^{- ( \beta_0 + 1) / (2\beta_0 + 1)})$.  This allows us to conclude that  (when $\wh r_n(A\ord{k_0}) \le t_0$) $\wh \lambda_n$ is also $O_p(n^{- ( \beta_0 +1) / (2\beta_0 + 1)})$, since
  \begin{equation*}
    \phi(\lambda)
    = \max_{k \le k_0} \min_{i \ge k_0} \frac{\sum_{j=k}^i \tilde Y_i + n \lambda t_0 (1-t_0)}{ i  - k + 1}
    \ge  \min_{i \ge k_0} \frac{\sum_{j=k_1}^i \tilde Y_i + n \lambda t_0 (1-t_0)}{ i  - k_1 + 1}
    = t_0,
  \end{equation*}
  and by the monotonicity and continuity of $\phi(\cdot)$, we can see that $0 \le \wh \lambda_n \le \lambda_1 = O_p(n^{- (\beta_0 +1) / (2\beta_0 + 1)})$.  An analogous argument holds for the
  case when $\wh r_n(A\ord{k_0}) > t_0$ and $\wh \lambda_n < 0$. This completes the proof. 
\end{proof}

\section{Causal estimator: lemmas,  remainder term analysis, and proofs}
\label{app:sec:proofs-caus-estim-lemm}

\subsection{Results for rates of convergence and limit distributions}
\label{app:sec:results-rates-conv}

\begin{mylongform}
  \begin{longform}
    \remCRD{\bf Strategy is just to re-do the entire proof, not to do a direct comparison between causal and non causal, which seems to complicate things. }      
  \end{longform}
\end{mylongform}
  \begin{proof}[Proof of Lemma \ref{lem:lambda_knot_size}]
  We need to consider
  \begin{equation}
    \label{eq:12}
    \int \one_{[A\ord{k_1}, A\ord{i_\lambda}]}(a) (\theta_0(a_0) - \wh \xi(\bs w)) \, d\PP_n(\bs w)
  \end{equation}
  where $\bs w = (l,a,y)$.
  As in the proof of Lemma~\ref{lem:2}, this leads us to consider 
  $\int \one_{I_{n,M}}(a) (\theta_0(a_0) - \wh \xi(\bs w)) \, d\PP_n(\bs w)   $
  where we again write
  $\theta_0(a_0) - \wh \xi(\bs w) = \theta_0(a_0) - \theta_0(a) + \theta_0(a)
  - \wh \xi(\bs w)$ and consider
  \begin{equation}
    \label{eq:K-P_2terms}
    \int \one_{I_{n,M}}(a) (\theta_0(a_0) - \theta_0(a)) \, d\PP_n(\bs w)    
    \text{ and } 
    \int \one_{I_{n,M}}(a) (\theta_0(a) - \wh \xi(\bs w)) \, d\PP_n(\bs w).
  \end{equation}
  The second term is the one we consider now.  (The first term can be managed just as it was in
  the proof of Lemma~\ref{lem:2}.)
  Decompose the second term
  as 
  \begin{equation}
    \label{eq:100}
    \int \one_{I_{n,M}}(a) (\theta_0(a) - \wh \xi(\bs w)) \, d\PP_n(\bs w) =
    E(M) 
    + R_V(M) + R_S(M),      
  \end{equation}
  where we have 
  $R_V(M) := \PP_n ( \wh \xi(\bs W; \wh \eta) -  \xi(\bs W ; \wh \eta) \one_{I_{n,M}}(A))$     (the V-process remainder term),
  and  $R_S(M) := \PP_0(  \xi(\bs W; \wh \eta) -  \xi(\bs W; \eta_\infty) )\one_{I_{n,M}}(A))$ (the second order remainder term), and
  $E(M) := (\PP_n-\PP_0)(\xi(\bs W; \wh \eta) \one_{I_{n,M}}(A))$ (the main empirical process term).
  {\bf The term $R_V(M).$} In Lemma~\ref{lem:RvM} we show that   $R_V(M)$ is  $O_p(n^{-1/2})$ (uniformly in $M$ in fact).

  {\bf The term $R_S(M)$.} %
  We will show
  \begin{equation}
    \label{eq:RsM-order}
    R_S(M) = o_p(n^{-(\beta_0+1)/(2\beta_0+1)}),
  \end{equation}
  for any fixed $M > 0$. 
  We begin by analyzing a conditional version,
  $\PP_0( (\wt \xi(\bs W; \wh \eta) - \wt \xi(\bs W; \eta_\infty)  |  A = b)$, which equals
  \begin{align*}
    & \PP_0 \lp [\mu_0(\bs L , b) - \wh \mu_n(\bs L, b)] \frac{g_0(b | \bs L)}{\wh g_n(b | \bs L)} \rp
      + \PP_0(\wh \mu(\bs L, b) - \mu_0(\bs L, b)) \\
    & =  \PP_0 \lp (\mu_0(\bs L, b) - \wh \mu(\bs L, b) ) 
      \ls \frac{g_0(\bs L, b)}{\wh g_n(\bs L, b)}  
      -  1 \rs \rp \\
    & =  \PP_0 \lp (\mu_0(\bs L, b) - \wh \mu(\bs L, b) ) 
      \ls \frac{g_0(\bs L, b) - \wh g_n(\bs L, b)}{\wh g_n(\bs L, b)}  
      \rs \rp
  \end{align*}
  whose absolute value is bounded above by 
  \begin{equation}
    \label{eq:second-order-term-RSM}
    \| \mu_0(\bs L, b) - \wh \mu(\bs L, b) \|_2
    \| g_0(\bs L, b) - \wh g_n(\bs L, b)\|_2
  \end{equation}
  since $\wh g_n$ is bounded below by Assumption~\ref{assm:EC-I}.  Thus, $|R_S(M)|$ is bounded above by $\int_{I_{n,M}} \| \mu_0(\bs L, b) - \wh \mu(\bs L, b) \|_2 \| g_0(\bs L, b) - \wh g_n(\bs L, b)\|_2 \, f_0(b) db$.  By Assumption~\ref{assm:model2}  on $f_0$ and Assumption~\ref{assumption:estimators}, this is of order %
  $t_n r_{n,M} s_{n,M} = o_p(n^{-(\beta_0+1)/(2\beta_0+1)})$. %

  \begin{mylongform}
    \begin{longform}
      Initially tried to consider the `asymptotic equicontinuity' remainder
      term, but in fact I think it becomes much simpler (albeit probably
      equivalent on a fundamental level) to apply kim-p asymptotics to the
      `whole' term, rather than to consider the remainder (which I had named $R_E(M)$).
    \end{longform}
  \end{mylongform}
  {\bf The term $E(M).$} We can apply Kim-Pollard asymptotics to $E(M)$.
  We define a class of  functions $\mc{F}_\xi$ to contain the semi-oracle pseudo-outcomes, $\xi(\bs W; \wh \eta)$.  With a slight overloading of notation, let $Y(w)$ be the function  $Y(l,a,y) = y$.
  Then we define
  \begin{equation}
    \label{eq:mcFxi-defn}
    \mc{F}_\xi :=
    \{(Y - \mu) h + \PP_0 \mu(\bs L, \cdot) \colon
    \mu \in \mc{F}_\mu, \, h \in \mcFpi
    \}.
  \end{equation}
  By Lemma~\ref{lem:pseudo-outcome-entropy},
  $J_1(1, \mc{F}_\xi, L_2) < \infty$ and the class admits an envelope $F_\xi$
  with $\EE( F_\xi^2(\bs W) |A=a) \le K$ for some $K>0$ and all $a \in \mc{A}$.   
  Thus, now let $\mc{F}_{a_0,R} $ be the class
  $\{ \bs w \mapsto \zeta(\bs w) \One{I_{n,M}}(a) \colon
  \zeta \in \mc{F}_\xi, \, M \le R \}$.  For any $R < \infty$, this class has finite uniform entropy integral:
  by Example 2.5.4 of \cite{vanderVaart:1996tf}, the class $\{ \One{I_{n,M}}(a) : 0 \le M \le R \}$ is a VC class (see \cite{vanderVaart:1996tf} for the definition of a VC class) which entails that it has bounded uniform entropy integral,  and then Lemma~\ref{lem:andrews-mult} implies that
  $\mc{F}_{a_0,R}$ has bounded uniform entropy integral.
  An envelope $F_{a_0,R}$ is then given by $F_{a_0,R}(\bs w) := F_\xi(\bs w) \One{I_{n,R}}(a)$ which satisfies
  $\PP_0 F_{a_0,R}(\bs W)^2 \le \EE (\One{I_{n,R}}(A) \EE( F_{\xi}^2(\bs W) | A))  \le  K R$.
  This is by taking expectation conditional on $A$, using the inequality $|ab| \le a^2 + b^2$, using that $\EE(Y^2 |A=a)$ is uniformly bounded over $a \in \mc{A}$ (Assumption~\ref{assm:setup}),  using that $\mc{F}_\mu$ and $\mcFpi$ are uniformly bounded above (Assumption~\ref{assm:EC-I}),
  and using that $A$ has a density bounded away from infinity and zero on $\mc{A}$ (Assumption~\ref{assm:model2}). 

  Thus we can apply Lemma~\ref{emptermlemma},
  with $l=\beta_0$ and $t=1$, to conclude that for any $\epsilon > 0 $, 
  \begin{equation*}
    |(\PP_n - \PP_0) \zeta \one_{I_{n,M}} |
    \le \epsilon M^{1+\beta_0} + n^{- (\beta_0+1)/(2\beta_0 +1)} A_n
  \end{equation*}
  for all $M \le R_0$, some $R_0$, and where $A_n = O_p(1)$ and does not depend on $M$.

  Now, the first term in 
  \eqref{eq:K-P_2terms} can be analyzed exactly as in
  the proof of Lemma~\ref{lem:2}.   
  Thus the same arguments made to complete the proof of Lemma~\ref{lem:2} apply now,
  and this completes the proof.   
\end{proof}

\medskip
\noindent The following lemma shares some similarities with Lemma~C.1 of \cite{drtest-main}. 
Recall that $R_V(M) := \PP_n ( \xi(\bs W; \wh \eta) -  \xi(\bs W ; \wh \eta) \one_{I_{n,M}}(A))$.
\medskip

\begin{mylongform}
  \begin{longform}
    Use assumption N (nuisance assumption rates) here. 
  \end{longform}
\end{mylongform}
\begin{lemma}
  \label{lem:RvM}
  Under the conditions of Theorem~\ref{thm:2}  we can conclude that
  $R_V(M) = O_p(n^{-1/2})$ uniformly in $M > 0$.
\end{lemma}
\begin{proof}
  We analyze $\{ R_V(M): M > 0 \}$ by considering it as a V-process.  We can write $R_V(M)$ as
  \begin{equation}
    \label{eq:RvM-sum}
    n^{-2} \sum_{i=1}^n \sum_{j=1}^n  \one_{I_{n,M}}(A_i) \lp \wh \mu_n(\bs L_j, A_i) - \int \wh \mu_n(\bs l, A_i)  d\PP_0(\bs l) \rp.    
  \end{equation}
  Recall the definitions of $J_m$ (and $N(\cdot,\cdot,\cdot)$) given in Subsection~\ref{subsec:empirical-proc-results}.
  We consider the class (of `V-process functions')
  $      \mc{F}_\mu^V $ defined to be the class
  of functions on $\mc{W}^2$ of the form
  $$(\bs l_1, a_1, y_1, \bs l_2, a_2, y_2) \mapsto (\mu(\bs l_1, a_2) - \PP_0 \mu(\bs L, a_2))  \one_{I_{n,M}}(a_2) $$
  for $\mu \in \mc{F}_\mu$ and all $0 \le M.$
  Recall that $I_{n,M} := [a_0 + M t_n]$.
  We will check that $J_2(1, \mc{F}^V_\mu, L_2) < \infty$. 

  By Assumption~\ref{assm:EC-I}, $\mc{F}_\mu$ is uniformly bounded and satisfies $J_2(1, \mc{F}_\mu, L_2) < \infty$, so, by the proof of Lemma 20 of \cite{Nolan:jw}, %
  the class $\{ \PP_0 \mu(\bs L, \cdot) : \mu \in \mc{F}_\mu \}$ has uniform entropy bounded above by that of $\mc{F}_\mu$.  By Lemma~\ref{lem:function-domain-extension-covering}, both of these classes when extended to the domain $\mc{W}^2$ (e.g., the class %
  of functions $(\bs l_1, a_1, y_1, \bs l_2, a_2, y_2) \mapsto \mu(\bs l_1, a_2)$ on $\mc{W}^2$ for $\mu \in \mc{F}_\mu$) have the same uniform covering numbers.  By Example 2.5.4 of \cite{vanderVaart:1996tf}, the set of indicator functions $\mc{I} := \{ \one_{I_{n,M}}(a_2) : M > 0 \}$ (with domain $\mc{W}^2$) has $J_2(1, \mc{I}, L_2) < \infty$.  Combining these classes by addition and multiplication yields the class $\mc{F}_\mu^V$, and then by Lemma~\ref{lem:andrews-add} and Lemma~\ref{lem:andrews-mult} we have that $J_2(1, \mc{F}_\mu^V, L_2)$ remains bounded. %

  We need to consider the symmetrized version, which can always be done, by noting that the sum
  \eqref{eq:RvM-sum} can be written in the form
  \begin{equation}
    \label{eq:RvM-sum-symmetrized}
    n^{-2} \sum_{i=1}^n h(\bs W_i, \bs W_i)
    + n^{-2} \sum_{1 \le i < j \le n} h(\bs W_i, \bs W_j) + h(\bs W_j, \bs W_i)
  \end{equation}
  (for $h \in \mc{F}_\mu^V$), and so we can consider the symmetric class
  $\mc{ F}_\mu^{V,s}$ of functions $h(w_1, w_2) + h(w_2,w_1)$ for
  $h \in \mc{F}_\mu^V$.  Then for all $\epsilon>0$,
  $N(\epsilon \sqrt{2}, \mc{F}_\mu^{V,s}, L_2) = N(\epsilon,
  \mc{F}_\mu^{V}, L_2) $ so the same entropy bounds as above apply.
  Then, $\mc{F}^{V,s}_\mu$ is uniformly bounded by Assumption~\ref{assm:EC-I},
  so we can apply Proposition~\ref{prop:1}.  This shows the first term on the right side of \eqref{eq:U-process-moment-bound} is finite (and is $O(n^{-1/2})$ in fact).

  Finally, for the  second term on the right side of \eqref{eq:U-process-moment-bound}
  we consider the class of functions $\PP_0 \mc{F}_\mu^{V,s} := \{ \PP_0 f(\bs W_1, \cdot) : f \in \mc{F}_\mu^{V,s} \}$.  By the existence of the envelope  $F$ for $\mc{F}_\mu^{V,s}$, we have an envelope $\PP_0 F(\bs W_1, \cdot)$ for $\PP_0 \mc{F}_\mu^{V,s}$.
  And again by   
  Lemma 20 of
  \cite{Nolan:jw} applied to the uniformly bounded $\PP_0 \mc{F}_\mu^{V,s}$,  we can conclude that $J_1(1, \PP_0 \mc{F}_\mu^{V,s}, L_2) < J_2(1, \PP_0 \mc{F}_\mu^{V,s}, L_2) < \infty$. 
  
  This bounds the off-diagonal terms in the sum \eqref{eq:RvM-sum}.  The diagonal sum (i.e., the
  first summand in \eqref{eq:RvM-sum-symmetrized}) is of smaller order, by an empirical process argument using the above entropies and Theorem~2.14.1 of \cite{vanderVaart:1996tf} (and the uniform boundedness of $\mu \in \mc{F}_\mu$ by Assumption~\ref{assm:EC-I}).
  So the proof is complete.
\end{proof}

\begin{proof}[Proof of Theorems   \ref{thm:causal-fullestimator-limit} and  \ref{thm:causal-nullestimator-limit}]
  We will use Theorem~3 of \cite{Westling_Carone_2020} to study both estimators.
  (Note that the convergence in the conclusion of the theorem is in the space $L^\infty[-K,K]$, any $K>0$; see the proof.)
  We let $\Phi_n(a) := n^{-1} \sum_{i=1}^n \one_{(-\infty, a]}(A_i)$ and $\Phi_0(a) := \PP_0( A \le a)$, for any $a \in \mc{A}$.  We will argue along subsequences.  Recall that $t_n := n^{-1 / (2\beta_0 + 1)}.$ Since
  $ t_n^{-(\beta_0+1)} \wh \lambda_n = O_p(1)$,
  along every subsequence there is a subsubsequence such that $t_n^{-(\beta_0+1)}  \wh \lambda_n$ converges in distribution to some limit random variable, $\Lambda_{\PP_0}$.
  \begin{all-in-one-file}
    Recall the definitions of $\Gamma_n, \Gamma_n^0$, and $\Gamma_0$ from the
    proof of Theorem~\ref{thm:2}.
  \end{all-in-one-file}
  \begin{supplement-file}
    Recall the definitions of $\Gamma_n, \Gamma_n^0$, and $\Gamma_0$ from the
    proof of Theorem~\ref{thm:2} in the main document.
  \end{supplement-file}
  Let $\Gamma_{n,0} := \Gamma_n - \Gamma_0$ and let $\Gamma_{n,0}^0 := \Gamma_n^0 - \Gamma_0$. 
  Define 
  \begin{equation*}
    \label{eq:Wna-defn}
    W_{n,a}(u) :=
    t_n ^{-(\beta_0+1)}\lp \Gamma_{n,0}(a + u t_n) - \Gamma_{n,0}(a)
    - \theta_0(a) ( \Phi_{n,0}(a + u t_n) - \Phi_{n,0}(a))
    \rp 
  \end{equation*}
  and define $W_{n,a}^0(\cdot)$ similarly
  except with $\Gamma_{n,0}(a+u t_n)$ replaced (twice) by $\Gamma_{n,0}^0(a + ut_n) - t_n^{-(\beta_0+1)}\wh \lambda_n \one_{[A_{k_0},a_0)}(a+ut_n)$.  (The indicator function term is to account for  the discrepancy between $a_0$ and  the data point $A_{k_0}$ at which we enforce the constraint, and this term is negligible since $n(A_{k_0}-a_0) = O_p(1)$.)

  We will let 
  $\phi_{\infty,b} := \phi_{\mu_\infty,g_{\infty}, b}$ and
  \begin{equation}
    \label{eq:phi-Gamma-AL}
    \begin{split}
      \MoveEqLeft \phi_{\mu,g, b}(\bs l, a, y) := \one_{(-\infty,b]}(a) \lp \frac{ y -
        \mu(a, \bs l)}{g(a, \bs l)} + \int \mu (a, \tilde{\bs l})
      d\PP_0(\tilde{\bs l}) \rp \\
      &     \qquad \qquad \quad + \int_{-\infty}^b \mu(\tilde a, \bs l)    d\PP_0(\tilde a)
      - \int \int_{-\infty}^b \mu(\tilde a, \tilde{\bs l}) d
      \PP_0( \tilde a) d\PP_0 ( \tilde{\bs l})
    \end{split}
  \end{equation}
  and we let $\phi^*_{\infty, b} := \phi_{\infty, b} - \Gamma_0(b)$.
  Then, under our conditions, by Lemma~1 of (the supplementary material of) \cite{Westling_Gilbert_Carone_2020},  $\Gamma_{n,0}(a_0 + b t_n) $ is asymptotically linear and is equal to
  $\PP_n \phi^*_{\infty,a_0 + b t_n} + R_{n,a_0+b t_n}$, $b \in \RR$. The latter term $R_{n, a_0+b t_n}$ is a remainder term that we will show to be negligible.

  In more detail, we will apply Theorem 3 of \cite{Westling_Carone_2020} to yield the desired limit distribution statements. We need to verify the conditions (A1)--(A5) of that theorem, which we refer to as WCA1--WCA5.  WCA4 is just from a classical Donsker theorem on a univariate empirical cumulative distribution function.  Condition WCA5 for $\Gamma_{n,0}$ and $\Gamma_{n,0}^0$ is established in our Theorem~\ref{thm:2}.

  Conditions WCA1--WCA3 are about the process $W_{n,a_0}$ or $W_{n,a_0}^0$. From the definitions of $W_{n,a_0}$, $W_{n,a_0}^0$ %
  the two processes can be decomposed (analogously to $\Gamma_{n,0}$, $\Gamma_{n,0}^0$)  into asymptotically linear terms and remainder terms.   Let $I_{a_0,u}(a) := \one_{(-\infty, a_0+u]}(a)   -  \one_{(-\infty,a_0]}(a) $ for $u \in \RR$.
  Then the asymptotically linear part of  $W_{n,a_0}(b)$ is $t_n^{-(\beta_0+1)} \PP_n (\phi_{\infty, a_0 + b t_n} - \theta_0(a_0) \gamma_{a_0+b t_n}^*)$ 
where   $\gamma_s^*(\bs w) := \one_{(-\infty, s]}(a) - F_0(s)$.
  The localized version of $ (\phi_{\infty, a_0 + b t_n} - \theta_0(a_0) \gamma_{a_0+b t_n}^*)$ is 
  the function
  \begin{equation*}
    \begin{split}
      \MoveEqLeft    f_{u}(\bs w) :=
      I_{a_0,u}(a) \lp \frac{y - \mu_\infty(a,\bs l)}{g_\infty(a, \bs l)} + \theta_\infty(a) - \theta_0(a)\rp 
      + \int I_{a_0,u}(v) \mu_\infty(v,\bs l) dF_0 (v) \\
      & \qquad - ( \Gamma_\infty(a_0 + u) - \Gamma_\infty(a_0) \\
      & \qquad - (\Gamma_0(a_0+u) - \Gamma_0(a_0))
      - \theta_0(a_0)( F_0(a_0 + u) - F_0(a_0))
      ,
    \end{split}
  \end{equation*}
  where we let
  $\theta_\infty(b) := \int \mu_\infty(b, \bs w) d\PP_0(\bs w)$ and
  $\Gamma_\infty(b) := \int_{-\infty}^b \theta_\infty(z) dF_0(z)$.
  Then $W_{n, a_0}(b)$ equals
  $t_n^{-(\beta_0+1)} \PP_n (f_{a_0, b t_n})  + R_{n, a_0 + b t_n}$
  {and $W_{n,a_0}^0(u) = W_{n,a_0}(u) +  t_n^{-(\beta_0 + 1)}  \wh \lambda_n \one_{[0,\infty)}(u)$.}
  We verify the conditions  WCA1--WCA3 separately for the main term  $\PP_n f_{a_0, b}$ and for the remainder term $R_{n, a_0 + b t_n}$. 

  WCA1 and WCA2 are about
  $W_{n,a_0}$ (and $W_{n,a_0}^0$) and we need to show the negligibility of $R_{n, a_0 + b t_n}$ in its contributions.
  This is shown by
  \cite{Westling_Gilbert_Carone_2020,Westling_Gilbert_Carone_2020_supp} under
  our current assumptions. (In particular, they do not rely on their
  assumption that $\mu_0, \mu_\infty, g_0, g_\infty$ are continuously
  differentiable, which we do not assume here; see the analysis of the terms
  $K_{n,j}$, $j=1,2,3,$ in the proof of Theorem 2.)  Similarly, their proof
  (pages 8--10 of the supplement \cite{Westling_Gilbert_Carone_2020_supp})
  also shows that $R_{n, a_0 + b t_n}$ satisfies assumption WCA3 (for $c_n$
  in WCA3 given by $t_n^{-1}$).

   {\bf Condition WCA3 holds for the main term.}  Condition WCA3 is about $\EE \sup_{|u| \le t_n \delta} |W_{n,a_0}(u)|$, for $0 < \delta,$ and here we focus on the asymptotically linear term, so we need to consider $\EE \sup_{|u| \le t_n \delta} |t_n^{-(\beta_0+1)} \PP_n f_{u t_n}|$, for $0 < \delta.$ Note that $t_n^{-(\beta_0+1)} \PP_n f_{u t_n} = t_n^{-1/2} \GG_n f_{u t_n}$ with $\GG_n = \sqrt{n}(\PP_n-\PP_0)$.  Let $\mc{G}_R := \{f_u : |u| \le R\}$.  Using Assumption~\ref{enum:item:1} and Assumption~\ref{assm:cts-diffable}(i), \cite[Proof of their theorem 2]{Westling_Gilbert_Carone_2020_supp} show that $\mc{G}_R$ has envelope $G_R$ and that $\sup_{Q} \log N( \epsilon \| G_R\|_{Q,2} , \mc{G}_R, L_2(Q)) \lesssim \log(1/\epsilon)$, and that $\PP_0 G_R^2 \lesssim R$ for $R$ small enough. This then implies that $t_n^{-1/2} \EE \sup_{|u| \le t_n \delta} |\GG_n f_{u t_n}| \lesssim \delta^{1/2} t_n^{-1/2}$ by Theorem~2.14.1 of \cite{vanderVaart:1996tf}, so that WCA3 is satisfied (with $f_n(u) = u^{1/2}$ and $\beta$ taken to be any value in $(1, 1+\beta_0)$).

\begin{mylongform}
  \begin{longform}
    They prove B4 and B5, which are used (see eg the text after the definitions of B4 and B5 in WC) to show the negligibility of remainder terms for $W_n$ and that remainders satisfy A3. 
  \end{longform}
\end{mylongform}

Define the `residual' $\delta_\infty(\bs W) := \frac{Y - \mu_\infty(A, \bs L)}{g_\infty(A, \bs L)} + \theta_\infty(A) - \theta_0(a_0)$. 
Now from the proof of Theorem 2 of \cite{Westling_Gilbert_Carone_2020}, we have that
the  linear term $t_n^{-(\beta_0+1)} \PP_n f_{ b t_n} $
(and so $W_{n,a_0}(b)$ itself)
converges weakly in $L^\infty[-M,M]$ to the process
  $\sqrt{\tkappa_0(a_0)} W(\cdot)$
  where $W$ is standard Brownian motion  on $\RR$ started at $0$ and  
  \begin{equation}
    \label{eq:wt-kappa-defn}
    \tkappa_0(a_0)
    := \EE_0 \lp \EE_0 \ls  \delta_\infty(\bs W)^2
    \Bigg\vert A=a_0, \bs L  \rs g_0(a_0, \bs L) \rp f_0(a_0).
  \end{equation}
  We can also then see that along the subsubsequence $W_{n,x}^0(\cdot)$ converges in distribution in the space $L^\infty[-K,K]$, for any $K > 0,$ from the definition of $\Gamma_{n,0}^0$ (and the convergence of $t_n^{-(\beta_0+1)} \wh \lambda_n$).
  Thus we define the process $M_{\PP_0}$ (notationally suppressing dependence on $a_0$) which appears in the conclusion of Theorem 3 of \cite{Westling_Carone_2020} as
  \begin{equation*}
    M_{\PP_0}(v)
    := \sqrt{\tkappa_0(a_0)}W(v) + \frac{\rho_0(a_0) f_0(a_0)}{\beta_0 +1} |v|^{\beta_0+1}.
  \end{equation*}
  (Recall the definition of $\rho_0$ in Assumption~\ref{assm:model}.)
  \begin{mylongform}
    \begin{longform}
      \remCRD{ The easiest way to finish it all up is to argue via characterization.  But I may be in a bit of a bind, not sure how easy it is to connect these dots.  Want to have a unique limit process.  Then can argue just from {\it Brownian rescaling} rather than having to argue about rescaling lambda.

        Actually, maybe not in a bind.  {\it On the other hand, I guess the limit of lambda may be easy to study.  It looks like a limit of fhat, so rescales in the same way.}}
    \end{longform}
  \end{mylongform}
  Now the process $t_n^{-(\beta_0+1)}\Gamma_{n,0}^0(a_0 + b t_n)$ is equal to $t_n^{-(\beta_0+1)}\Gamma_{n,0}(a_0 + b t_n)$ except for the Lagrange multiplier summand; thus, along the subsubsequence (along which $n^{-(\beta_0+1)/(2\beta_0+1)} \wh \lambda_n $ converges), the process $M_{\PP_0}^0(b)$ to which $t_n^{-(\beta_0+1)}\Gamma_{n,0}^0(a_0 + b t_n)$ converges is 
  \begin{equation*}
    M_{\PP_0}^0(v) :=
    M_{\PP_0}(v) + \Lambda_{\PP_0} \one_{[0,\infty)}(v).
  \end{equation*}
  Formally, we replace $M_{\PP_0}^0$ by its lower semi-continuous version, to accommodate the conditions of
  Theorem 3 of \cite{Westling_Carone_2020}.  This is allowable because it  does not change the $\GCM$ of $M_{a_0}^0$ (it only changes the value of $    M_{\PP_0}^0(v)$ possibly at the one point $v=0$).
  Define 
  \begin{equation*}
    \wh \theta_{\PP_0}(b) := \mono (M_{\PP_0})(b) 
    \quad   \text{ and } \quad
    \wh \theta^0_{\PP_0}(b) := \mono(M_{\PP_0}^0)(b) .
  \end{equation*}  %
  Since $\{ W_{n,a_0}(u) : |u| \le K\}$ converges weakly in $L^\infty[-K,K]$ to the limit process $\{\sqrt{\tkappa_0(a_0)} W(u) : |u| \le K\}$, WCA1 and WCA2 for the main terms   are satisfied.
  Thus  we have met the five conditions WCA1--WCA5 of Theorem 3 of \cite{Westling_Carone_2020}.
  Therefore  by that theorem
  we have the joint convergence
  \begin{equation}
    \label{eq:estimator-limits}
    \begin{split}
      n^{\beta_0 / (2\beta_0+1)}
      \begin{pmatrix}
        \wh \theta_n(a_0 + b t_n) - \theta_0(a_0)        \\
        \wh \theta^0_n(a_0 + b t_n) - \theta_0(a_0       )
      \end{pmatrix}
      \to_d 
      \densA_0(a_0)^{-1} 
      \begin{pmatrix}
        \wh \theta_{\PP_0}(b) \\
        \wh    \theta_{\PP_0}^0(b)
      \end{pmatrix}
    \end{split}
  \end{equation}
  in $L^\infty[-M,M]$, for any $M>0$. (The proof of Theorem 3 of \cite{Westling_Carone_2020} yields not just marginal but joint  convergence.)
  Now, by the representation \eqref{eq:4} and the proof of Lemma~\ref{lem:2}, we can show that when
  $  t_n^{-(\beta_0+1)} \wh \lambda_n $ has a limit distribution along a subsequence, which we denote $\Lambda_{\PP_0}$, 
  then
  \begin{equation}
    \label{eq:Lagrangemult-constants}
    \gamma_1^{-1}
    t_n^{-(\beta_0+1)}
    \wh \lambda_n 
    \to_d \Lambda \equiv \Lambda_{\beta_0},
    \text{ or equivalently }
    \gamma_1 \Lambda =_d  \Lambda_{\PP_0} 
  \end{equation}
  where $\gamma_1$ is defined below in \eqref{eq:gamma-defns} and where
  $\Lambda_{\beta_0}$ is universal (is independent of  $\PP_0$, except through $\beta_0$).
  We postpone showing
  \eqref{eq:Lagrangemult-constants} for the moment and proceed with it as given.

  We can relate the process $M_{\PP_0}$ to a universal process $M$
  and similarly we can relate $M_{\PP_0}^0 $ to a universal process $M^0$ 
  by \eqref{eq:lambda-representation} (and the argument after).
  Recall the definitions
  $M(t) \equiv M_{\beta_0}(t) := W(t) + |t|^{\beta_0+1}$
  and 
  $M^0(t)  \equiv M_{\beta_0}^0(t) := M(t) + \Lambda \one_{(0,\infty)}(t),$
  where $\Lambda$ is as described in \eqref{eq:Lagrangemult-constants}.
  By Lemma~\ref{lem:brownian-scaling} and \eqref{eq:Lagrangemult-constants}, we have
  \begin{equation}
    \label{eq:limit-process-rescaling}
    \{ M_{\PP_0}(t) \} \stackrel{d}{=}  %
    \{ \gamma_1  M_{\beta_0}(\gamma_2 t) \}
    \quad \text{and} \quad
    \{ M_{\PP_0}^0(t) \}
    \stackrel{d}{=}  %
    \{ \gamma_1  M_{\beta_0}^0(\gamma_2 t) \}
  \end{equation}
  where
  \begin{equation}
    \label{eq:gamma-defns}
    \gamma_1 := \lp \frac{(\beta_0+1) \tkappa_0(a_0)^{\beta_0+1} }{\rho_0(a_0) \densA_0(a_0)} \rp^{1/ (2\beta_0+1)},
    \; 
    \gamma_2 := \lp \frac{ \rho_0(a_0) \densA_0(a_0)}{(\beta_0+1) \sqrt{\tkappa_0(a_0)}} \rp^{2  / (2\beta_0+1)}.
  \end{equation}
  Finally, define (suppressing dependence of $\wh \theta,$ $\wh \theta^0$ on $\beta_0$)
  \begin{equation*}
    \wh \theta(b) := \mono(M_{\beta_0})(b) 
    \quad   \text{ and } \quad
    \wh \theta^0(b) := \mono(M_{\beta_0}^0)(b) .
  \end{equation*}
  It now follows by the equivariance of the greatest convex minorant (and the chain rule of differentiation) that
  \begin{equation}
    \label{eq:limit-estimator-rescaling}
    \wh \theta_{\PP_0}(t) \stackrel{d}{=} \gamma_1 \gamma_2 \wh \theta(\gamma_2 t)
    \quad \text{and} \quad
    \wh \theta_{\PP_0}^0(t) \stackrel{d}{=} \gamma_1 \gamma_2 \wh \theta^0(\gamma_2 t).
  \end{equation}
  Note that
  \begin{equation}
    \label{eq:cc-gamma-identity}
    \densA_0(a_0)^{-1} \gamma_1 \gamma_2 = \densA_0^{-1}(a_0) (\densA_0(a_0) \rho_0(a_0) \tkappa_0(a_0)^{\beta_0} / (\beta_0+1)  )^{1 / (2\beta_0+1)} = \cc.
  \end{equation}
  Thus we have shown by \eqref{eq:estimator-limits} and  \eqref{eq:limit-estimator-rescaling} that
  \begin{equation}
    \label{eq:limitthms-finalconclusion}
    t_n^{-\beta_0}(\mle(a_0+u t_n) - \theta_0(a_0), \mlevr(a_0+ut_n)-\theta_0(a_0))
    \to_d \cc ( \wh \theta(\gamma_2 u), \wh \theta^0(\gamma_2 u)),
  \end{equation}
  in $L^\infty[-K,K]^2$, as desired.

  It now remains to complete the proof of \eqref{eq:Lagrangemult-constants}.
  Note that there is no circularity in completing this argument after establishing \eqref{eq:limitthms-finalconclusion} because we will only use results about $\mle$ (not about $\mlevr$). 
  By  \eqref{eq:4} we can
  write
  \begin{equation}
    \label{eq:lambda-representation}
    \wh \lambda_n
    =    %
    t_0 (\Phi_n(\eta_{+,n}) - \Phi_n(\eta_{-,n}-)) - (\Gamma_n(\eta_{+,n})  - \Gamma_n(\eta_{-,n}-))    
  \end{equation}
  for knot points $\eta_{\pm,n}$.
  If we define $M_n$ by $M_n(u) := t_n^{-(\beta_0+1)} ( \Gamma_n(a_0 + u t_n)
  - t_0 \Phi_n(a_0 + u t_n))$ then $t_n^{-(\beta_0+1)} \wh \lambda_n$ equals
  $-( M_n( (\eta_{+,n}-a_0) t_n^{-1} ) - M_n ((\eta_{-,n}-a_0)t_n^{-1} - ) )$. By the arguments above (i.e., Theorem 3 of  \cite{Westling_Carone_2020}) this converges to
  $M_{\PP_0}(\eta_-) - M_{\PP_0}(\eta_+)$ where
  $(\eta_{\pm,n}-a_0)t_n^{-1} \to_d \eta_{\pm}$ along a subsubsequence by tightness (Lemma~\ref{lem:lambda_knot_size}).
  By   \eqref{eq:limit-process-rescaling},
  $(M_{\PP_0}(\eta_-) - M_{\PP_0}( \eta_+)  ) =_d \gamma_1 (M_{\beta_0}(\gamma_2 \eta_- ) - M_{\beta_0}(\gamma_2  \eta_+)).$ This shows \eqref{eq:Lagrangemult-constants} once we note that
  $(M_{\beta_0}(\gamma_2 \eta_- ) - M_{\beta_0}(\gamma_2  \eta_+))$ is universal (does not depend on $\gamma_2$). This  is true because
  of the scaling \eqref{eq:limit-process-rescaling} which shows that $\gamma_2 \eta_+$ is a knot of $M_{\beta_0}$, meaning it is a functional of $M_{\beta_0}$, meaning it is independent of $\PP_0$ (except through $\beta_0$). This completes the proof.
\end{proof}
\begin{mylongform}
  \begin{longform}
    In shorthand
    \begin{align*}
      f^{-1} \gamma_1 \gamma_2
      =     f^{-1} \lp \frac{(\beta+1) \tilde \kappa^{\beta+1}}{\rho f } \rp^{1/(2\beta+1)} \lp  \frac{\rho f }{ (\beta+1) \sqrt {\tilde \kappa}} \rp^{2/(2\beta+1)}
    \end{align*}
  \end{longform}
\end{mylongform}

\subsection{Proof of Theorem~\ref{thm:LLR-limit}}
\label{sec:proof-theor-refthm:l}

\begin{proof}[Proof of Theorem~\ref{thm:LLR-limit}]

Recentering $\wh \theta_i$ and $\wh \xi_i$ at $t_0$ and
expanding the squares,
  we can write $S_n$ as 
  \begin{equation*}
    \sum_{i=1}^n \lp(\wh \theta_i^0 - t_0)^2 - (\wh \theta_i - t_0)^2 - (2 (\wh \theta_i^0 - t_0) (\wh \xi_i - t_0) - 2 (\wh \theta_i - t_0) (\wh \xi_i - t_0))\rp    
  \end{equation*}
  from which we can see
  \begin{equation}
    \label{eq:Sn-expression}
    S_n = \sum_{i=1}^n (\wh \theta_i - t_0)^2 - (\wh \theta_i^0 - t_0)^2.
  \end{equation}
  This used the fact that from the characterizing equations
    (as in \eqref{eq:11}) or the max-min representation, we have
    \begin{equation}
      \label{eq:basic-identity}
      \sum \wh \xi_i - \check \theta_i = 0 
    \end{equation}
    where $\check \theta$ is either one of the two estimators, and the sum is
    taken over an interval of constancy for that estimator, except this
    expression does not hold when the estimator is $\wh \theta_n^0$ and the
    interval is the one on which $\wh \theta_n^0$ equals $t_0$.  Thus (since
    $\wh \theta_i - t_0$ is constant on the interval of summation) we have
    $\sum (\check \theta_i - t_0) (\wh \xi_i - t_0 - (\check \theta_i - t_0))
    = 0$; this follows trivially for $\wh \theta_n^0$ on the interval where
    it equals $t_0$ and otherwise it follows by \eqref{eq:basic-identity}.
    Thus \eqref{eq:Sn-expression} holds.

  Now,
  based on a standard argument (e.g., see the proof of Theorem~2.1 of \cite{Groeneboom:2015ew}) and Lemma~\ref{lem:lambda_knot_size}, the two estimators $\wh \theta$ and $\wh \theta_n$ can be shown to be identical except for on an $O(t_n)$ neighborhood of $a_0$.  Thus, letting $D_n := [\tau_{n,-}, \tau_{n,+}]$ be the largest interval such that the two estimators are identical on $\RR \setminus D_n$, we can write
  \begin{align}
    \label{eq:Sn-main-expr}
    0 \le
    S_n =
    n  \int_{D_n} \lp (\wh \theta_n(v) - \theta_0(a_0))^2 - (\wh\theta_n^0(v) - \theta_0(a_0))^2 \rp \, d\Phi_n(v).
  \end{align}

  Now for any subsequence there exists a subsubsequence such that $t_n (\tau_{n,\pm} - a_0)$ converge weakly to limit variables, denoted $\tau_{\pm}.$ These variables are characterized as being the endpoints of the largest interval on which $\wh \theta_{\PP_0} \equiv \wh \theta$ and $\wh \theta_{\PP_0}^0 \equiv \wh \theta^0$ are not equal, which are uniquely defined.  Since the limit distributions are the same along every subsubsequence, they are the limits as $n \to \infty$.

  Now we return to \eqref{eq:Sn-main-expr}.
  After a change of variables
  $ t_n (v-a_0) = u$,
  this can be seen
  by Theorems~\ref{thm:causal-fullestimator-limit}
  and \ref{thm:causal-nullestimator-limit}
  to  converge weakly to
  \begin{align}
    \MoveEqLeft \cc^2 \int_{\tau_-}^{\tau_+} ( \wh \theta(\gamma_2 u)^2 - \wh \theta^0(\gamma_2 u)^2 ) \densA_0(a_0) du \nonumber \\
    & =
      \densA_0(a_0)^{-1}  \gamma_1^2 \gamma_2 \int_{\tau_- / \gamma_2}^{\tau_+ / \gamma_2} ( \wh \theta(w)^2 - \wh \theta^0(w)^2 )  dw
      \nonumber
  \end{align}
  by a change of variables $w = \gamma_2 u$,
  where 
\begin{main-file}
  the constants $\gamma_i$, $i=1,2$, are defined in \eqref{eq:gamma-defns} in the Supplement,
\end{main-file}
\begin{all-in-one-file}
  the constants $\gamma_i$, $i=1,2$, are defined in \eqref{eq:gamma-defns} in Appendix~\ref{app:sec:proofs-caus-estim-lemm}
\end{all-in-one-file}
and  from \eqref{eq:cc-gamma-identity} in Appendix \ref{app:sec:proofs-caus-estim-lemm}, $\cc = \densA_0(a_0)^{-1} \gamma_1 \gamma_2$.  The final integral on the right hand side above is the universal limit variable $\DD_{\beta_0}$.  By \eqref{eq:gamma-defns} compute $ \gamma_1^2 \gamma_2 = \tkappa_0(a_0).$ Thus, the previous display equals (recall that $\kappa_0(a_0) := \tkappa_0(a_0) / \densA_0(a_0)$)
  \begin{equation*}
    \densA_0(a_0)^{-1} \tkappa_0(a_0) \DD_{\beta_0}
    = \kappa_0(a_0) \DD_{\beta_0}.
  \end{equation*}
  This completes the proof.
  \begin{mylongform}
    \begin{longform}
      Note that $\kappa_0(a)$ is equal to the conditional variance,
      $\Var( \xi(\bs W; \eta_\infty ) | A = a)$.  (Add more derivations here)
    \end{longform}
  \end{mylongform}
  \begin{mylongform}
    \begin{longform}
      We have, in shorthand,
      \begin{align*}
        f \gamma_1^2 \gamma_1
        = f \lp \frac{(\beta_0+1) \tkappa_0^{\beta_0+1}}{\gamma f} \rp^{2 / (2\beta_0+1)}
        \lp \frac{\gamma f}{(\beta_0+1) \sqrt{ \tkappa_0}} \rp^{2 / (2\beta_0+1)}
      \end{align*}
    \end{longform}
  \end{mylongform}  
\end{proof}

\section{Empirical process and entropy results}
\label{subsec:empirical-proc-results}

In this section we present various empirical process and Brownian motion results on which we rely.
First, we introduce basic definitions.  Let
$$J_m(\delta, \mc{F}, L_2) := \int_0^\delta \sup_{Q} ( 1 + \log N(\epsilon
\|F \|_{Q,2}, \mc{F}, \|\cdot \|_{2,Q}))^{m/2} \, d\epsilon$$ for $m=1,2$,
where the sup is over all probability measures $Q$, $\| \cdot \|_{Q,2}$ is
the $L_2(Q)$ semimetric under distribution $Q$, and $N(\epsilon,\mc{F}, d)$
is the so-called covering number, i.e.\ the minimal number of $d$-balls (for
some (semi-)metric $d$) of size $\epsilon$ needed to cover $\mc{F}$.

The following two lemmas are from (slight modifications of the result given in) Theorem 3 of \cite{andrews1994empirical}.

\begin{lemma}
  \label{lem:andrews-add}
  For two classes of measurable functions $\mc{G},$ $\mc{H}$, with envelopes
  $G$ and $H$, respectively, for any $\epsilon > 0$ and probability measure $Q$, we have
  \begin{align*}
    \MoveEqLeft
    N(\epsilon \| G + H \|_{Q,2}, \mc{G} + \mc{H},   L_2(Q))
    \\&    \le  %
    N(2^{-1} \epsilon \|G \|_{Q,2}, \mc{G}, L_2(Q))
    N( 2^{-1} \epsilon \|H \|_{Q,2},  \mc{H},L_2(Q)).
  \end{align*}
\end{lemma}

\begin{lemma}
  \label{lem:andrews-mult}
  For two classes of measurable functions $\mc{G},$ $\mc{H}$, with envelopes
  $G$ and $H$, respectively, and for any $\epsilon > 0$, we have
  \begin{align*}
    \MoveEqLeft    \sup_Q N(     \epsilon \| (G \vee 1) (H \vee 1) \|_{Q,2}, \mc{G} \mc{H},
    L_2(Q))
    \\& \le \sup_Q N(2^{-1} \epsilon \|G \|_{Q,2}, \mc{G}, L_2(Q))
    \sup_Q N(2^{-1} \epsilon \|H \|_{Q,2}, \mc{H}, L_2(Q)).
  \end{align*}
\end{lemma}

The following lemma is proved in the proof of Lemma~C.1 of \cite{drtest-main}.
\begin{lemma}
  \label{lem:function-domain-extension-covering}
  Let $\mc F$ be a class of measurable functions on a measure space $\mc X$
  with finite covering number $N( \mc F, \| \cdot \|_{2,Q}, \tau)$ and
  envelope $F$.  Then the class $\mc{F}^\circ$ of
  functions $f^\circ(x,z) := f(x)$ defined on the extended space
  $\mc X \times \tilde{\mc X} $, for a measurable space $\tilde{\mc X}$, has
  $N( \mc F, \| \cdot \|_{2,Q}, \tau) = N( \mc{F}^\circ , \| \cdot
  \|_{2,Q^\circ}, \tau)$ for any $Q^\circ$ on $\mc X \times \tilde{\mc X}$
  that extends $Q$ in the sense that $Q$ is the marginal of $Q^\circ$ on
  $\mc X$.  In particular,
  $\sup_Q N( \mc F, \| \cdot \|_{2,Q}, \tau) = \sup_{Q^\circ} N( \mc{F}^\circ
  , \| \cdot \|_{2,Q^\circ}, \tau)$.
\end{lemma}

The following lemma is from  Lemma A.1 in \cite{Balabdaoui:2007jj}, %
which is itself based on  \cite{Kim:1990ue}. The version here was given in \cite{ham2024doubly}. 
\begin{lemma}\label{emptermlemma}
  Let $\mc F$ be a collection of functions defined on $[s_0-\delta,s_0+\delta]^2\times \mathbb{R}^m$ with small $\delta>0$ and arbitrary positive integer $m$. Suppose that for a fixed $s_1\in [s_0-\delta,s_0+\delta]$ and $R>0$, such that $s_0-\delta \le s_1 \le s_2 \le s_1+R\le s_0+\delta$, the collection 
  \begin{align*}
    \mathcal{F}_{s_0,R}=\{ f_{s_1,s_2}(\mathbf{x})=f(s_1,s_2,\mathbf{x})\in \mathcal{F} \colon
    s_0-\delta \le s_1 \le s_2 \le s_1+R\le s_0+\delta\}
  \end{align*}
  admits an envelope $F_{s_0,R}$, such that
  \begin{align*}
    \mathbb{E} F^2_{s_0,R}(\mathbf{X})\le K_0R^{2t-1}, ~~R\le R_0
  \end{align*}
  for some $t\ge 1/2$ and $K_0>0$, depending only on $s_0$ and $\delta$. Moreover, suppose that 
  \begin{align*}
    \sup_{Q}\int_0^1\sqrt{{\rm log}N(\eta\|F_{s_1,R}\|_{Q,2},\mathcal{F}_{s_0,R},L_2(Q))}d\eta <\infty.
  \end{align*}
  Then, for each $\epsilon>0$, there exist random variables $M_n$ of order $O_P(1)$ which does not depend on $s_1,s_2$ and $R_0>0$, such that
  \begin{align*}
    |(\PP_n-\PP)f_{s_1,s_2}|\le \epsilon|s_2-s_1|^{l+t}+n^{-(l+t)/(2l+1)}M_n~~~~{\rm for}~|s_2-s_1| \le R_0
  \end{align*}
  for $f\in \mathcal{F}_{s_0,R}$ and $l>0$.
\end{lemma}

The below lemma shows that the `semi-oracle pseudo-outcomes' (involving nuisance parameters but using $d\PP_0$ rather than $d\PP_n$) satisfy uniform entropy conditions and have a square integrable envelope, when similar conditions are assumed on the nuisance estimator classes, so that empirical process results apply. 
\begin{lemma}
  \label{lem:pseudo-outcome-entropy}
  Under the assumptions of Theorem~\ref{thm:2}, the class $\mc{F}_\xi$ defined in \eqref{eq:mcFxi-defn} satisfies $J_1(1, \mc{F}_\xi, L_2) < \infty$ and has an envelope $F_\xi$ with with $\EE( F_\xi^2(\bs W) |A=a) \le K$ for some $K>0$ and all $a \in \mc{A}$.
\end{lemma}
\begin{proof}
  By Assumption \ref{assm:EC-I}, $J(1, \mc{F}, L_2) < \infty$ for $\mc{F}$ equal to  $\mc{F}_\mu$ or $\mcFpi$, and both of these classes are uniformly bounded.
  By Lemma 20 of \cite{Nolan:jw} (which applies to uniformly bounded classes) we can conclude that $\PP_0 \mc{F}_\mu := \{ \PP_0 \mu(\bs L, \cdot) : \mu \in \mc{F}_\mu \}$ also has finite uniform entropy integral.  Now, by Lemmas \ref{lem:andrews-add} and \ref{lem:andrews-mult}, it follows that $\mc{F}_\xi$ has finite uniform entropy integral as desired.

  Now, using that $\sup_{a \in \mc{A}} E(Y^2 |A =a ) <\infty$, and that $\mc{F}_\mu$,
  $\mcFpi,$ and $\PP_0 \mc{F}_\mu$ are all uniformly bounded by
  Assumptions~\ref{assm:EC-I},
  we can conclude by Cauchy-Schwarz that $\mc{F}_\xi$ has an envelope satisfying the needed conditional second moment condition.
\end{proof}

For a function $f$ on $\mc{W} \times \mc{W}$ with measure $\PP \times \PP$, that is symmetric in its arguments, we let $\PP f$ be the function $w \mapsto \PP f(W,w)$.
And given $n$ variables $W_1, \ldots, W_n \in \mc{W}$, let
\begin{equation}
  \label{eq:Un-defn}
  U_n(f) := n^{-3/2} \sum_{1 \le i < j \le n} f(W_i, W_j).
\end{equation}

\begin{proposition}[Proposition K.1 of \cite{drtest-supp}]
  \label{prop:1}
  Assume that $\mc{F}$ is a class of measurable functions on a measure space $\mc{W} \times \mc{W}$ with (measurable) envelope $F$, and measure $\PP$ on $\mc{W}$.  Assume that $f \in \mc{F}$ satisfies $f(w_1, w_2) = f(w_2 , w_1)$ and $\PP f(W_1, W_2) = 0$.  Assume that $W_1, \ldots, W_n$ are i.i.d.\ and that $U_n$ is defined by \eqref{eq:Un-defn}.  Let $F_1(w)$ be an envelope for $\PP \mc {F}$.  Then for a universal constant $C>0$,
  \begin{equation}
    \label{eq:U-process-moment-bound}
    \PP \| U_n \|_{\mc F} \le
    C J_2(1, \mc {F}, L_2) \sqrt{\PP F(W_1, W_2)^2} n^{-1/2}
    +  C J_1(1, \PP \mc{F}, L_2) \sqrt{\PP F_1(W)^2}.
  \end{equation}
\end{proposition}

The following lemma is via basic properties of Brownian scaling.  It is included for completeness.
\begin{lemma}
  \label{lem:brownian-scaling}
  Let $W(t)$ be a two-sided standard Brownian motion with $W(0)=0$.  For $a,b, \beta > 0$,
  let $Z_{a,b, \beta}(t) := a W(t) + b |t|^{\beta+1}$ for $ t \in \RR$.
  Then
  \begin{equation}
    \label{eq:brownian-scaling}
    \{Z_{a,b,\beta}(t)\}_{t \in \RR} =_d \{ a (a/b)^{1/(2\beta+1)} Z_{1,1,\beta}(
    (b/a)^{2/(2\beta+1)}
    s ) \}_{s \in \RR}. 
  \end{equation}
\end{lemma}
\begin{proof}
  The proof is just via Brownian scaling (that is, by
  $\{ \sigma W(\cdot) \}_{\RR} =_d \{  W(\sigma^2 \cdot) \}_{\RR}$, for $\sigma > 0$).
  Let $\gamma = 2 / (2\beta+1)$.  Then by the change of variables $t = (a/b)^\gamma s$,
  we have $a W(t)  + b |t|^{\beta+1} = a W( (a/b)^\gamma s ) + b (a/b)^{\gamma(\beta+1)} |s|^{\beta+1}$,
  which is equal in distribution to
  $a (a/b)^{1/ 2\beta+1} W(s) + |s|^{\beta+1} a^{(2\beta+2)/(2\beta+1)}/ b^{1/2\beta+1}$ as desired.
\end{proof}

\section{Further details on nursing hours and readmissions data}
\label{sec:furth-deta-nursing}

Here we provide some further details about the definitions and calculation of the variables in the data analysis for the nursing hours and hospital readmissions data. 
We calculate $A$ as the ratio of registered nurse hours to inpatient days (which is slightly different from \cite{Kennedy:2017cq} and \cite{McHugh:2013gn}, because we don't have access to the hospitals' financial data so cannot calculate their ``adjusted inpatient days'').
Another reason our data  is slightly different than that of those two earlier papers %
is that we use updated data from the year 2018.

We measure covariates $\bs{L}$ as possible confounders. These are the following nine variables:
the number of beds, the teaching intensity, an indicator for not-for-profit status, an indicator for whether the location is urban or rural, the proportion of patients on Medicaid, the average patient socioeconomic status, a measure of market competition (see \cite{drtest-supp} for details on how these last two variables are calculated), an indicator for whether the hospital has a skilled nursing facility (because our measure of nurse staffing hours $A$ will unfortunately include hours worked in such a skilled nursing facility), and whether open heart or organ transplant surgery is performed (which serves  as a measurement of whether the hospital is high technology).  We did not include  patient race proportions and operating margin variables %
(present in \cite{Kennedy:2017cq} and \cite{McHugh:2013gn}) because we don't have access to those features. The data we use here are discussed in more detail in
\cite{drtest-main},
along with a discussion of possible missing confounders. For more detail about the background of the policy problem see \cite{McHugh:2013gn}.

\section{Simulations}
\label{app:sec:simulations}

Here we present further simulation results beyond those in the main document. Also, in
Table~\ref{tab:Dbeta-crit-values_all}, we tabulate further quantiles for the limit distributions $\DD_{\beta}$ for various $\beta$ values. 
\begin{table}%
  \centering
  \begin{tabular}{|c|c|c|c|c|c|c|c|c|}
    \hline
    $\beta$ & $0.01$ & $0.2$ & $0.4$ & $0.6$ & $0.8$ & $1$ & $2$ & $5$ \\
    \hline
    $q_{.99, \beta}$ & 2.85 & 3.06 & 3.34 & 3.56 & 3.75  & 3.89 &4.19  & 4.45 \\ 
    $q_{.975, \beta}$ & 2.17 & 2.33 & 2.55 & 2.73 & 2.86 & 2.92  & 3.16 & 3.35 \\ 
    $q_{.95, \beta}$ & 1.65 & 1.81 & 1.98 & 2.10 & 2.18 & 2.25 & 2.44 & 2.57 \\
    $q_{.90, \beta}$ & 1.17 & 1.29 & 1.40 & 1.49 & 1.55 & 1.60 & 1.73 & 1.83 \\
    $q_{.85, \beta}$ & 0.90 & 0.99 & 1.09 & 1.15 & 1.20 & 1.24 & 1.33 & 1.40 \\ 
    \hline
  \end{tabular}
  \caption[Critical values for $\DD_\beta$'s]{Critical values for $\DD_{\beta}$ for a range of $\beta$ values.\label{tab:Dbeta-crit-values_all}}  
\end{table}
See Section~\ref{sec:simulations} in the main document for a description of the simulation model setup and of the plots presented here.
\newcommand{\reg}{\backsimeq} %
We present simulation results with sample size $n \in \{200, 500, 1000, 2000\}$ and $S \in \{0.1, 0.2\}$.  We consider parametric and machine learning / nonparametric fits.  We fit (i) with both parametric models well specified, (ii) with only $\mu_0$ well specified, (iii) with only $\pi_0$ well specified, and (iv) with Super Learner \citep{van2007super}.  For Super Learner, we use the same implementations as in \cite{Kennedy:2017cq, drtest-main} to estimate $\pi_0$ and $\mu_0$. We truncate $\wh\pi$ to be 0.01 if any of the estimating procedures fell below that value.  We use Monte Carlo replication sizes of $1000$ for the parametric fits and $500$ when using SuperLearner (due to computational constraints).  We do not consider $n=200$ when using SuperLearner, which does not perform well with small sample sizes.
Thus, there are $(4\times 2 \times 3) + (3 \times 2) = 30$ simulation settings. The results are shown in Figures~\ref{fig:sim-plots_confounding1-200-param-well}--\ref{fig:sim-plots_confounding2-2000-SL}.

Our parametric models for $\mu_0$ and $\pi_0$ are all based on linear regression models with $Y$ or $A$ as response.  Then $\pi_0$ is specified as the corresponding true normal density with the modeled mean and the known true variance (specified in Section~\ref{sec:simulations}).   When we use a
well specified model it is as follows.  The well-specified regression model for $A$ is $A \reg L_1 + L_2 + L_3 + L_4$, where $\reg$ is used to denote a linear regression model with the variables on the right side included as covariates and a constant term included.  The well-specified regression model for $Y$ is $Y \reg A*\bs{L} + A^3 + A^4\one_{\{-1.5 \le A \le 1.5\}} + \one_{\{A > 1.5\}} + \one_{\{A<-1.5\}} $ where $A* \bs{L}$ is shorthand for all linear terms of the variables in $(A, \bs L)$ and all interaction terms $AL_i$.  For the misspecified models, we model $Y \reg L_1$ and $A \reg L_1$.  Below we present plots from the simulation studies. Each plot of three figures is similar to the two such plots in Figure~\ref{fig:sim-plots_confounding.66}, described in Section~\ref{sec:simulations}.


\modCRD{These supplementary simulation results demonstrate broadly a similar story as those
 (with sample sizes $n=1000$) given in the main paper.  Figures 
\ref{fig:sim-plots_confounding1-1000-param-bothcorrect} and 
\ref{fig:sim-plots_confounding2-1000-param-bothcorrect}
are the same results as in the simulation figure given in the main paper.
Around those two, in
Figures~\ref{fig:sim-plots_confounding1-200-param-well}--\ref{fig:sim-plots_confounding1-2000-param-well} ($S=0.1$)
and Figures~\ref{fig:sim-plots_confounding2-200-param}--
\ref{fig:sim-plots_confounding2-2000-param-bothcorrect} ($S=0.2$)
we can see how behavior improves as $n$ goes from $200$ to $2000$.
Figures~\ref{fig:sim-plots_confounding1-200-param-mucorrect}--\ref{fig:sim-plots_confounding1-2000-param-picorrect} and \ref{fig:sim-plots_confounding2-200-param-mucorrect}--\ref{fig:sim-plots_confounding2-2000-param-picorrect} show the ``double robustness'', namely that misspecifying a nuisance parameter does not affect performance (when the other is estimated correctly at a parametric rate). 
This story is complemented by the results based on nonparametric/machine learning (SuperLearner)
(Figures \ref{fig:sim-plots_confounding1-500-SL}--\ref{fig:sim-plots_confounding1-2000-SL} and \ref{fig:sim-plots_confounding2-500-SL}--\ref{fig:sim-plots_confounding2-2000-SL}).
SuperLearner generally performs slightly worse than, although overall similarly to especially for larger sample sizes, the parametric methods, illustrating that two nonparametric (well specified) learners are (for these sample sizes) still comparable to one (or two) well specified parametric model(s). 
In general, the higher confounding level ($S=0.2$) is unsurprisingly more challenging, particularly when $a = 0$ or $a=3$. Higher sample sizes than we consider are needed for perfectly ideal performance in those settings (but the good asymptotic performance of the procedures is illustrated by considering results with $S=0.1$ or other $a$ values) but these regimes illustrate reasonable performance when asymptopia has not fully kicked in. 

In all cases that we consider, the sample splitting procedure performs generally worse than the non sample splitting procedure. This is true even when we use SuperLearner (except possibly with $S=0.1, n=500$), which may be slightly surprising (this might be considered a ``high complexity setting'' where we may have expected sample splitting to outperform non sample splitting). Nonetheless, we expect this would reverse in even higher complexity or higher dimensional settings particularly with larger sample sizes. 
}

%

\newcommand{\suppplottext}{A complete description is given in the text in Section~\ref{sec:simulations}.}

\begin{figure}[!ht]
  \centering  \includegraphics[width=\linewidth]{\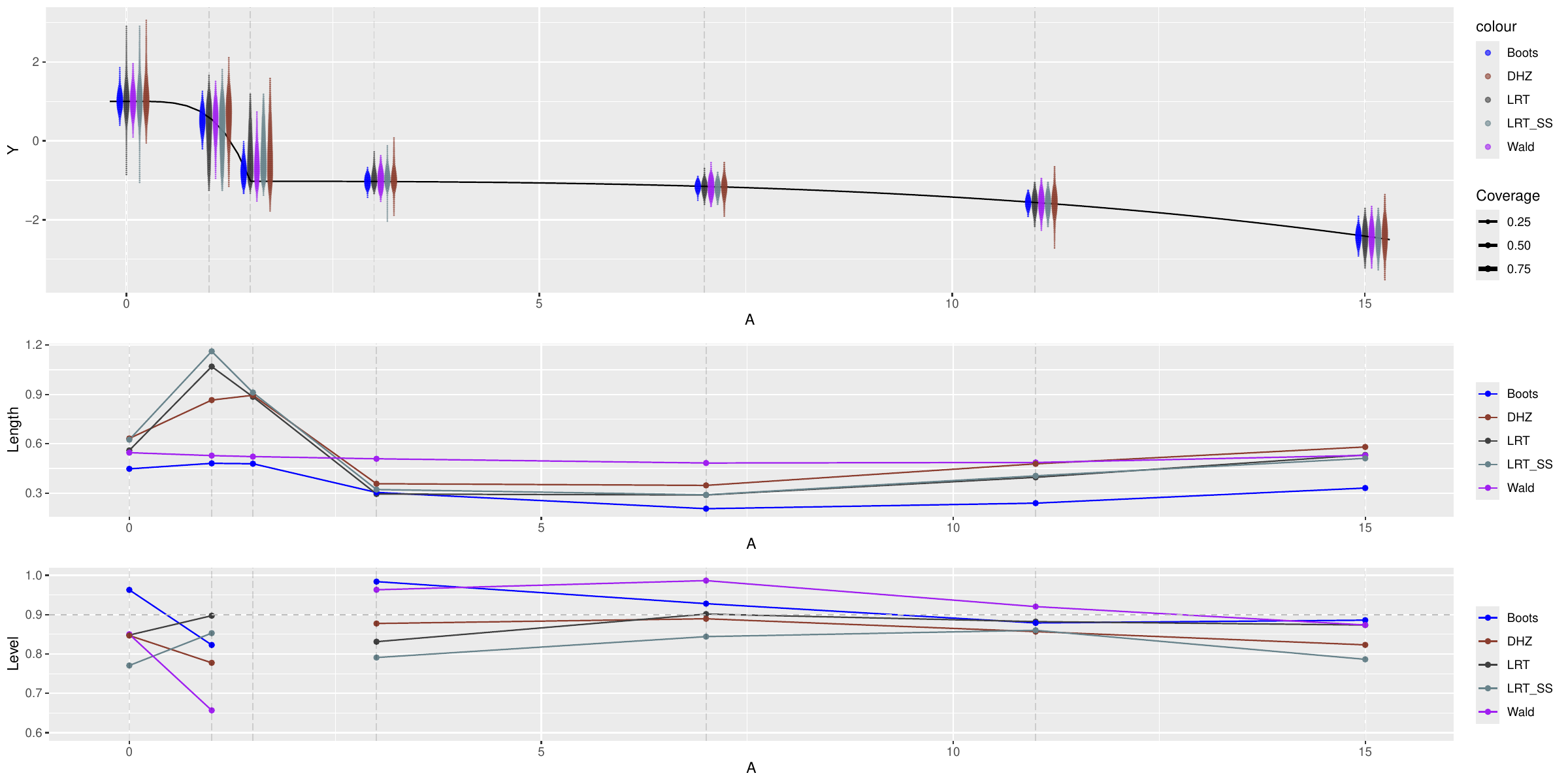}
  \caption{Simulation study ($1000$ Monte Carlos) plots with $n=200$, $S = 0.1$, and $\mu, \pi$ both estimated with well specified (parametric) models. \suppplottext \label{fig:sim-plots_confounding1-200-param-well}}  
\end{figure}

\begin{figure}[ht]
  \centering  \includegraphics[width=\linewidth]{\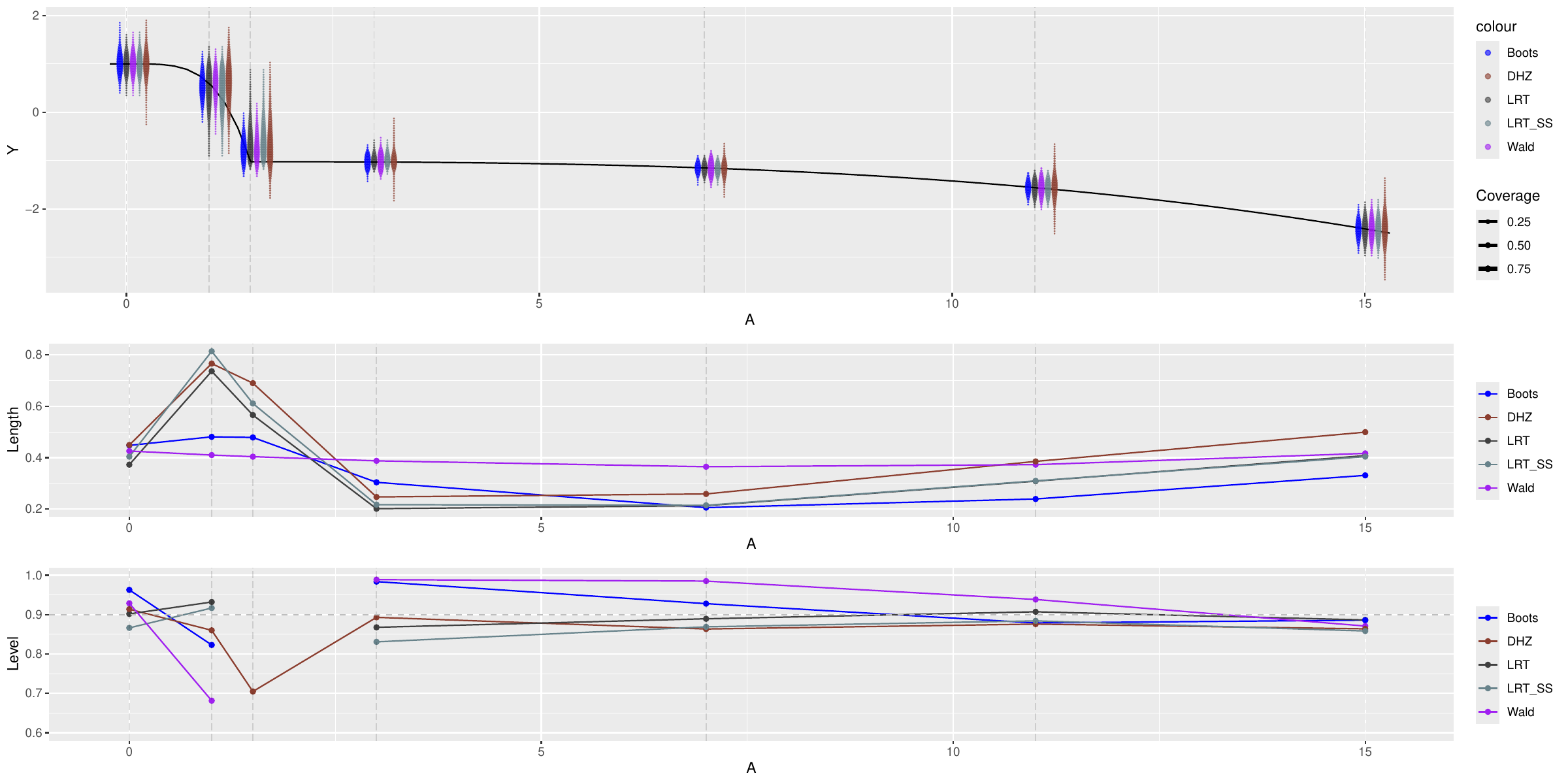}
  \caption{Simulation study ($1000$ Monte Carlos) plots with $n=500$, $S = 0.1$, and $\mu, \pi$ both estimated with well specified (parametric) models. \suppplottext \label{fig:sim-plots_confounding1}}  
\end{figure}

\begin{figure}[ht]
  \centering  \includegraphics[width=\linewidth]{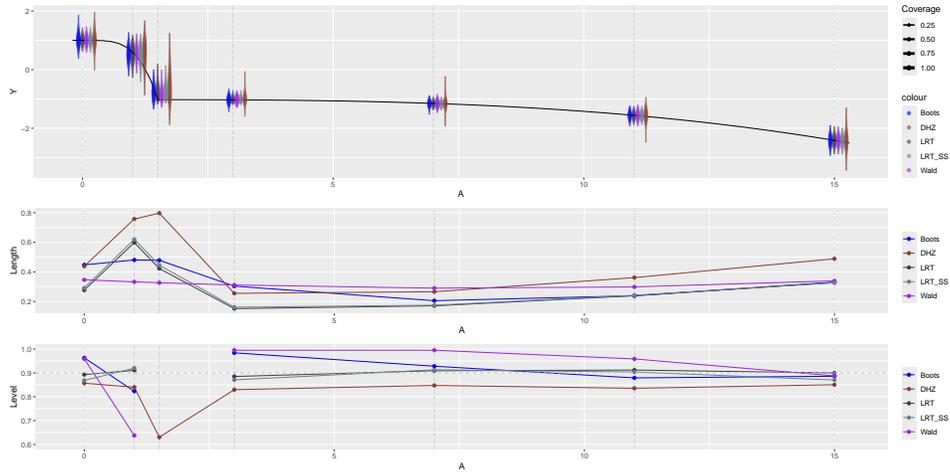}
  \caption{Simulation study ($1000$ Monte Carlos) plots with $n=1000$, $S = 0.1$, and $\mu, \pi$ both estimated with well specified (parametric) models. \suppplottext \label{fig:sim-plots_confounding1-1000-param-bothcorrect}}  
\end{figure}

\begin{figure}[ht]
  \centering  \includegraphics[width=\linewidth]{\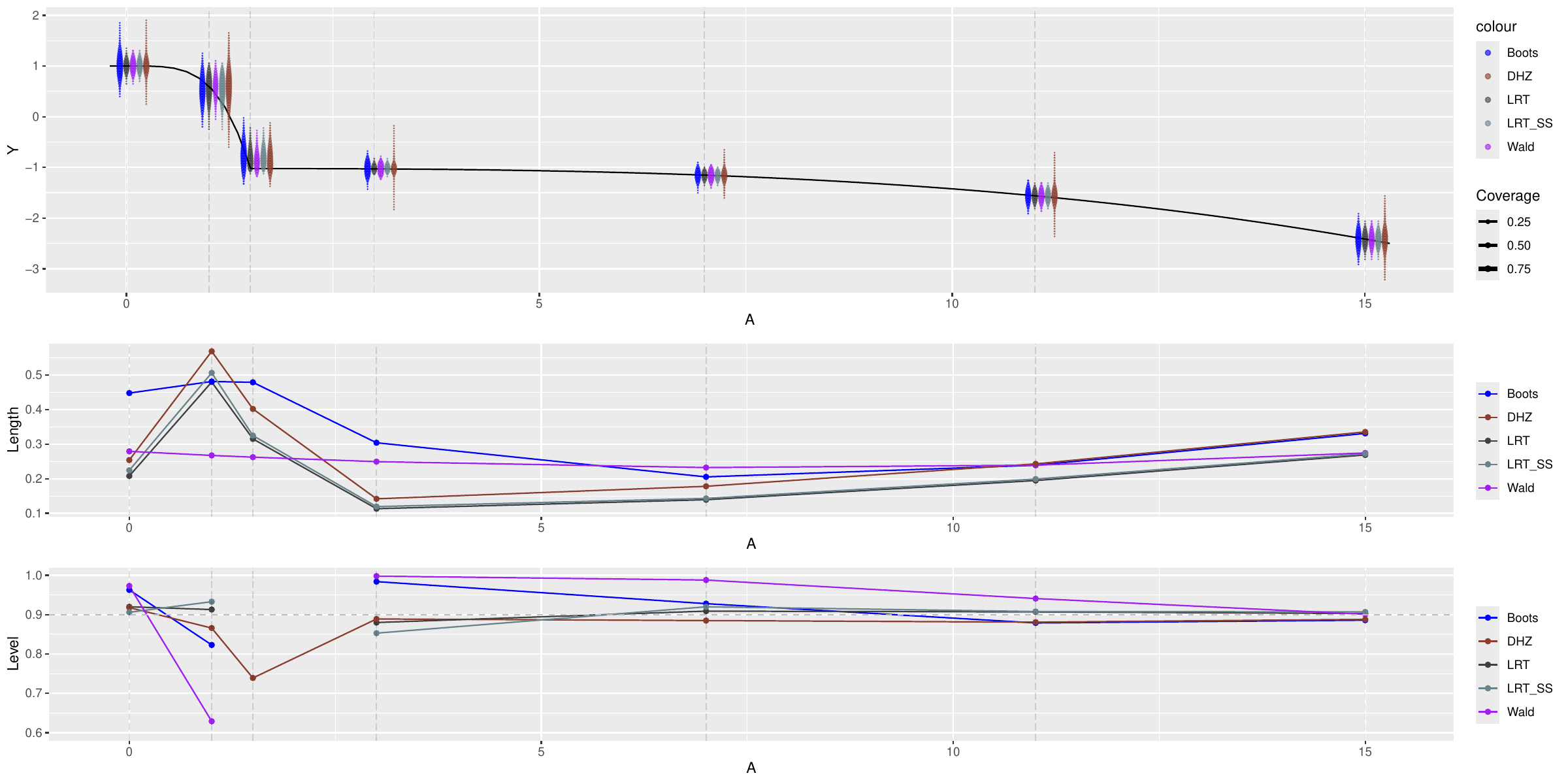} 
  \caption{Simulation study ($1000$ Monte Carlos) plots with $n=2000$, $S = 0.1$, and $\mu, \pi$ both estimated with well specified (parametric) models. \suppplottext \label{fig:sim-plots_confounding1-2000-param-well}}  
\end{figure}

\begin{figure}%
  \centering  \includegraphics[width=\linewidth]{\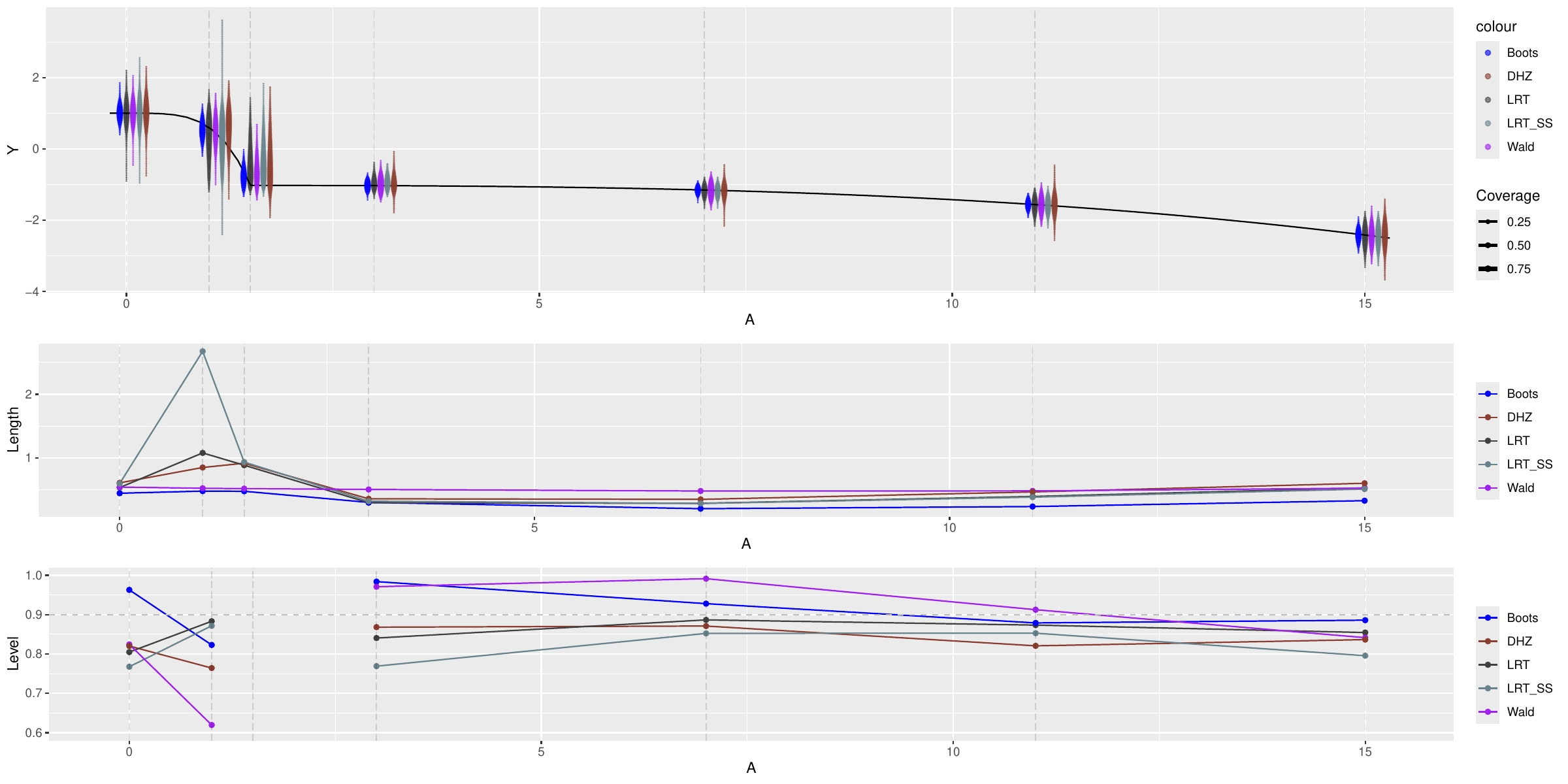}
  \caption{Simulation study ($1000$ Monte Carlos) plots with $n=200$, $S = 0.1$,
    and $(\mu, \pi)$   estimated with (well-, mis-) specified (parametric) models.
    \suppplottext \label{fig:sim-plots_confounding1-200-param-mucorrect}}  
\end{figure}
\begin{figure}%
  \centering  \includegraphics[width=\linewidth]{\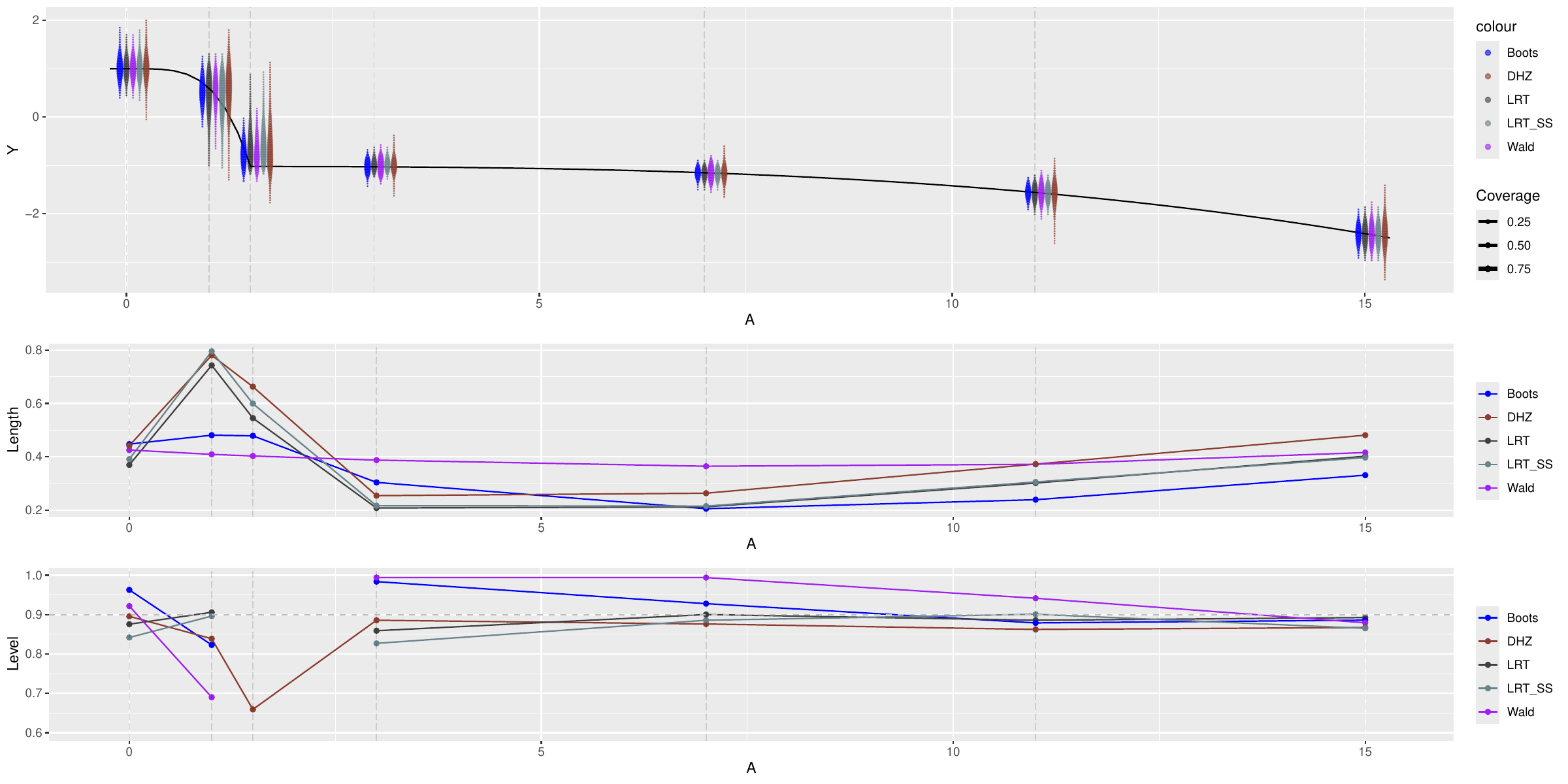}
  \caption{Simulation study ($1000$ Monte Carlos) plots with $n=500$, $S = 0.1$,
    and $(\mu, \pi)$   estimated with (well-, mis-) specified (parametric) models.
    \suppplottext \label{fig:sim-plots_confounding1}}  
\end{figure}

\begin{figure}%
  \centering  %
  \includegraphics[width=\linewidth]{\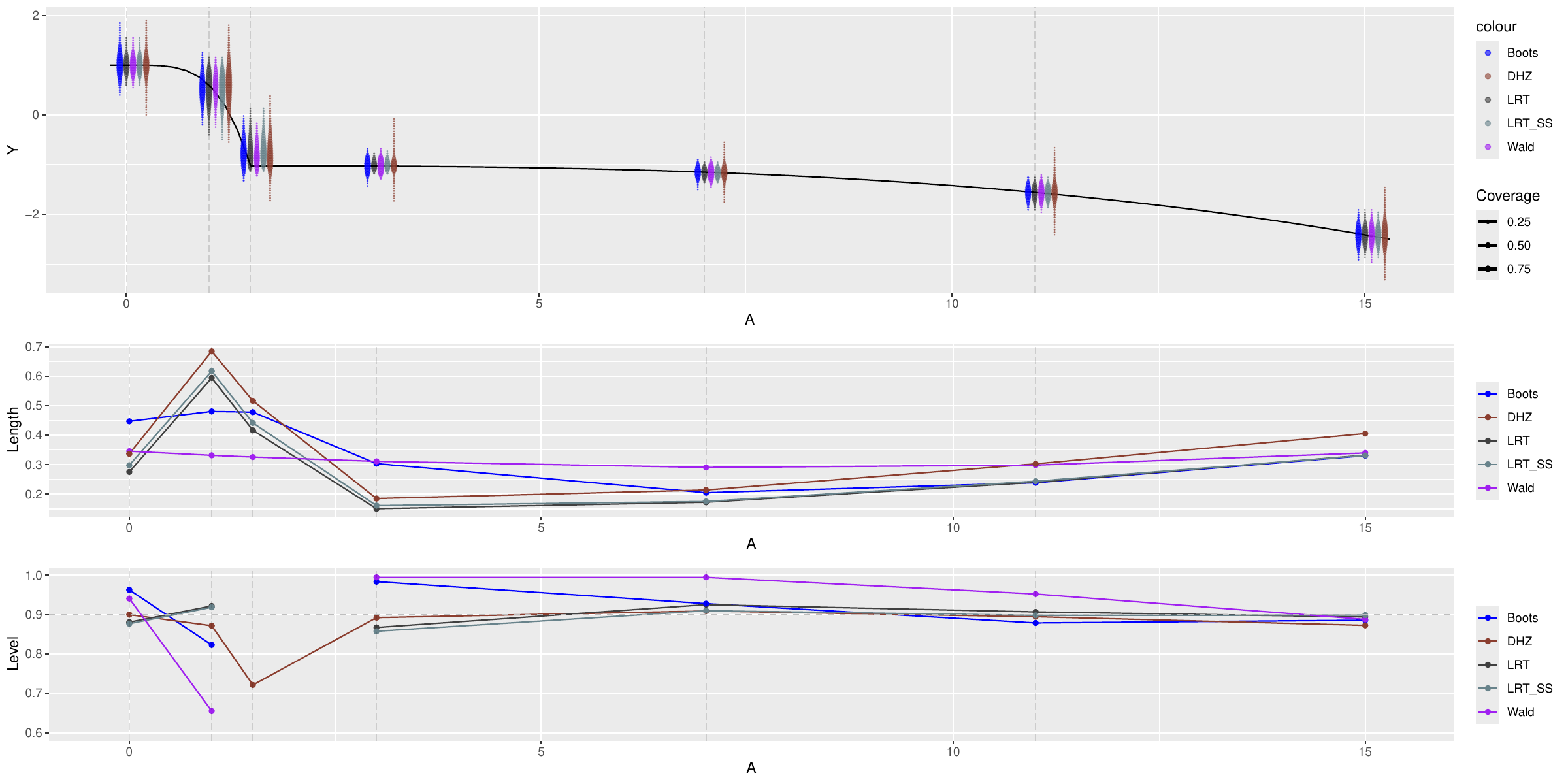}
  \caption{Simulation study ($1000$ Monte Carlos) plots with $n=1000$, $S = 0.1$,
    and $(\mu, \pi)$   estimated with (well-, mis-) specified (parametric) models.
    \suppplottext \label{fig:sim-plots_confounding1}}  
\end{figure}

\begin{figure}%
  \centering  \includegraphics[width=\linewidth]{\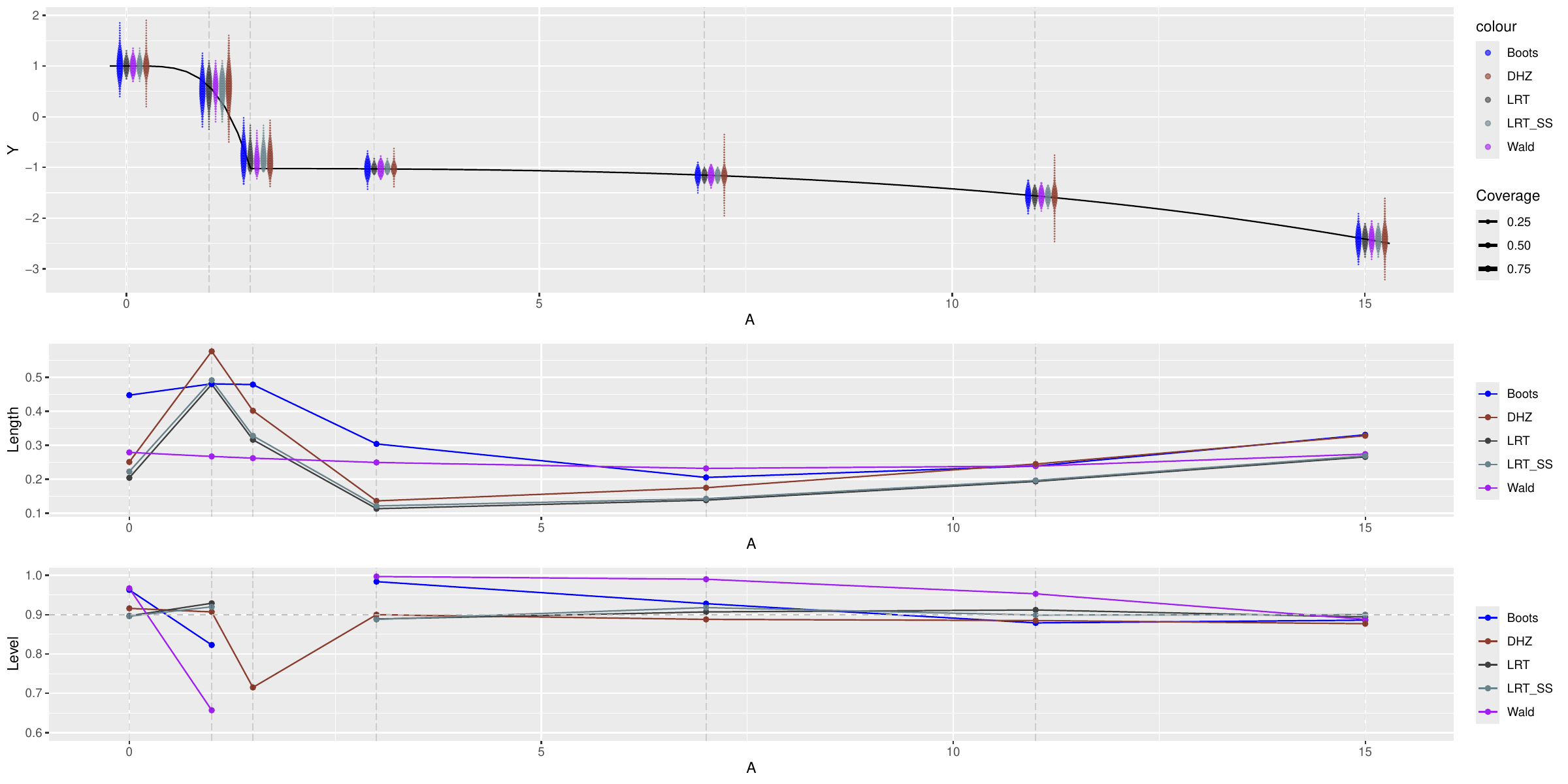} 
  \caption{Simulation study ($1000$ Monte Carlos) plots with $n=2000$, $S = 0.1$,
    and $(\mu, \pi)$   estimated with (well-, mis-) specified (parametric) models.
    \suppplottext \label{fig:sim-plots_confounding1}}  
\end{figure}

\begin{figure}%
  \centering  \includegraphics[width=\linewidth]{\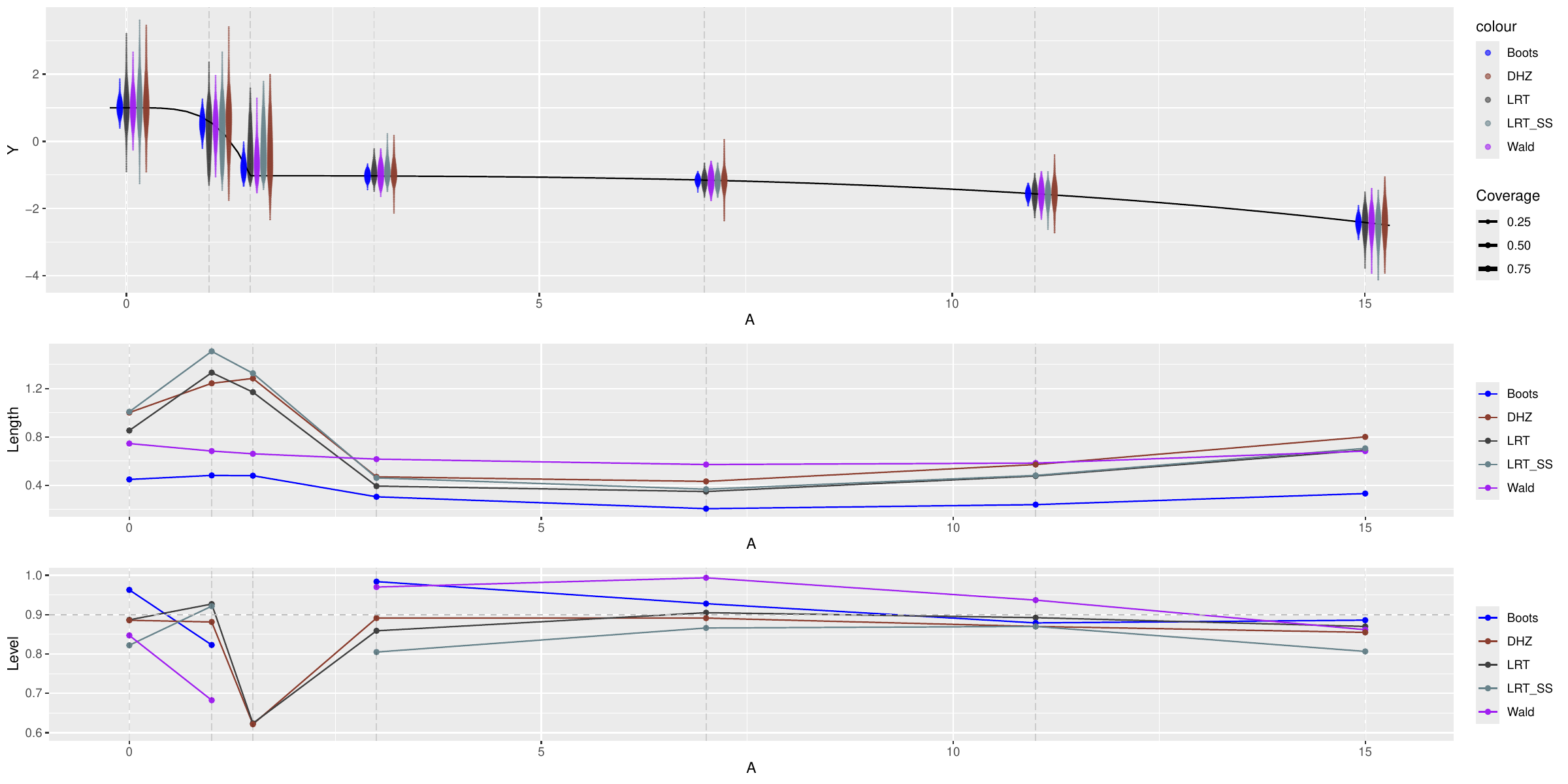}
  \caption{Simulation study ($1000$ Monte Carlos) plots with $n=200$, $S = 0.1$,
    and $(\mu, \pi)$   estimated with (mis-, well-) specified (parametric) models.
    \suppplottext \label{fig:sim-plots_confounding1}}  
\end{figure}
\begin{figure}%
  \centering  \includegraphics[width=\linewidth]{\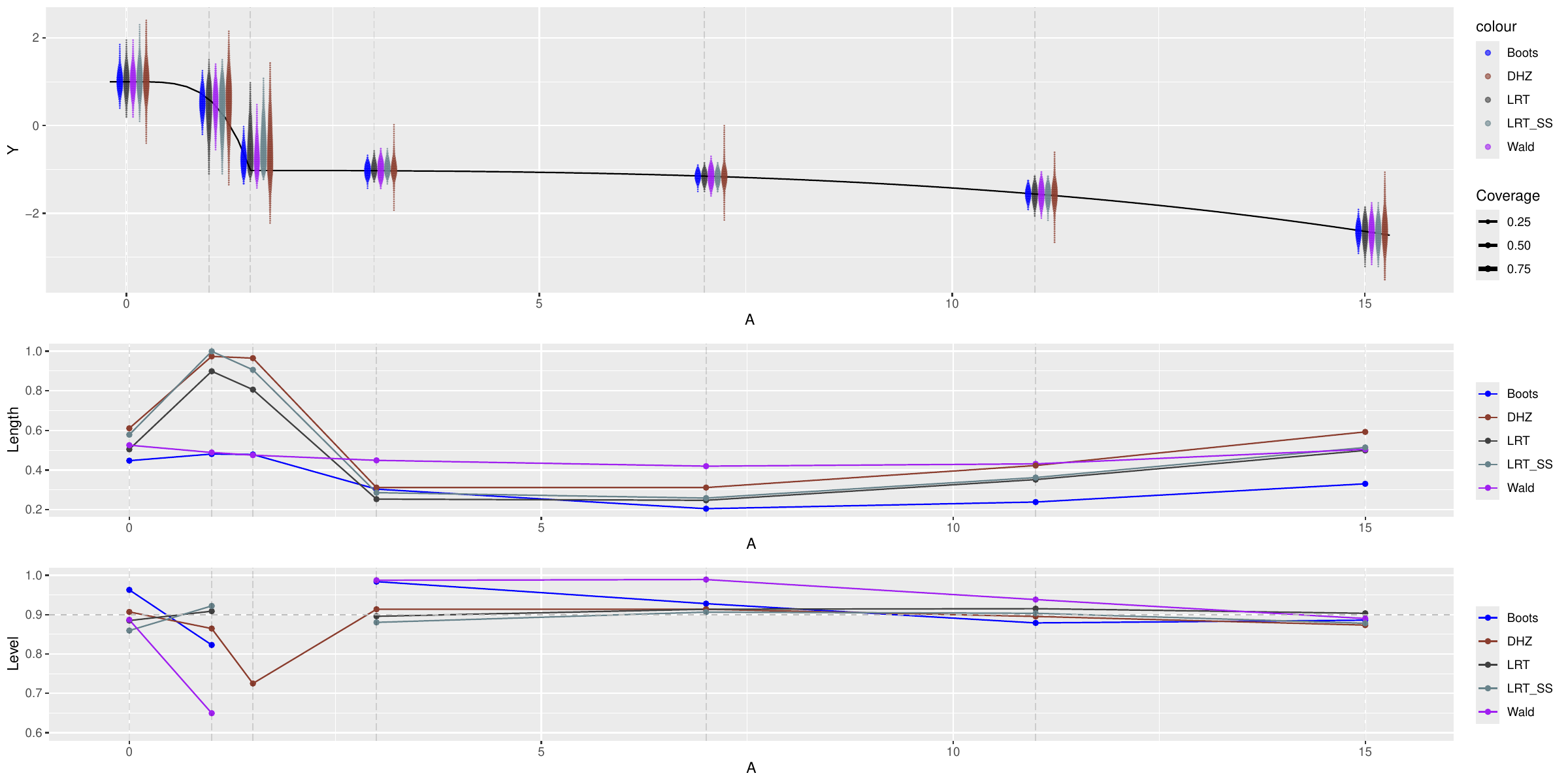}
  \caption{Simulation study ($1000$ Monte Carlos) plots with $n=500$, $S = 0.1$,
    and $(\mu, \pi)$   estimated with (mis-, well-) specified (parametric) models.
    \suppplottext \label{fig:sim-plots_confounding1}}  
\end{figure}

\begin{figure}%
  \centering %
   \includegraphics[width=\linewidth]{\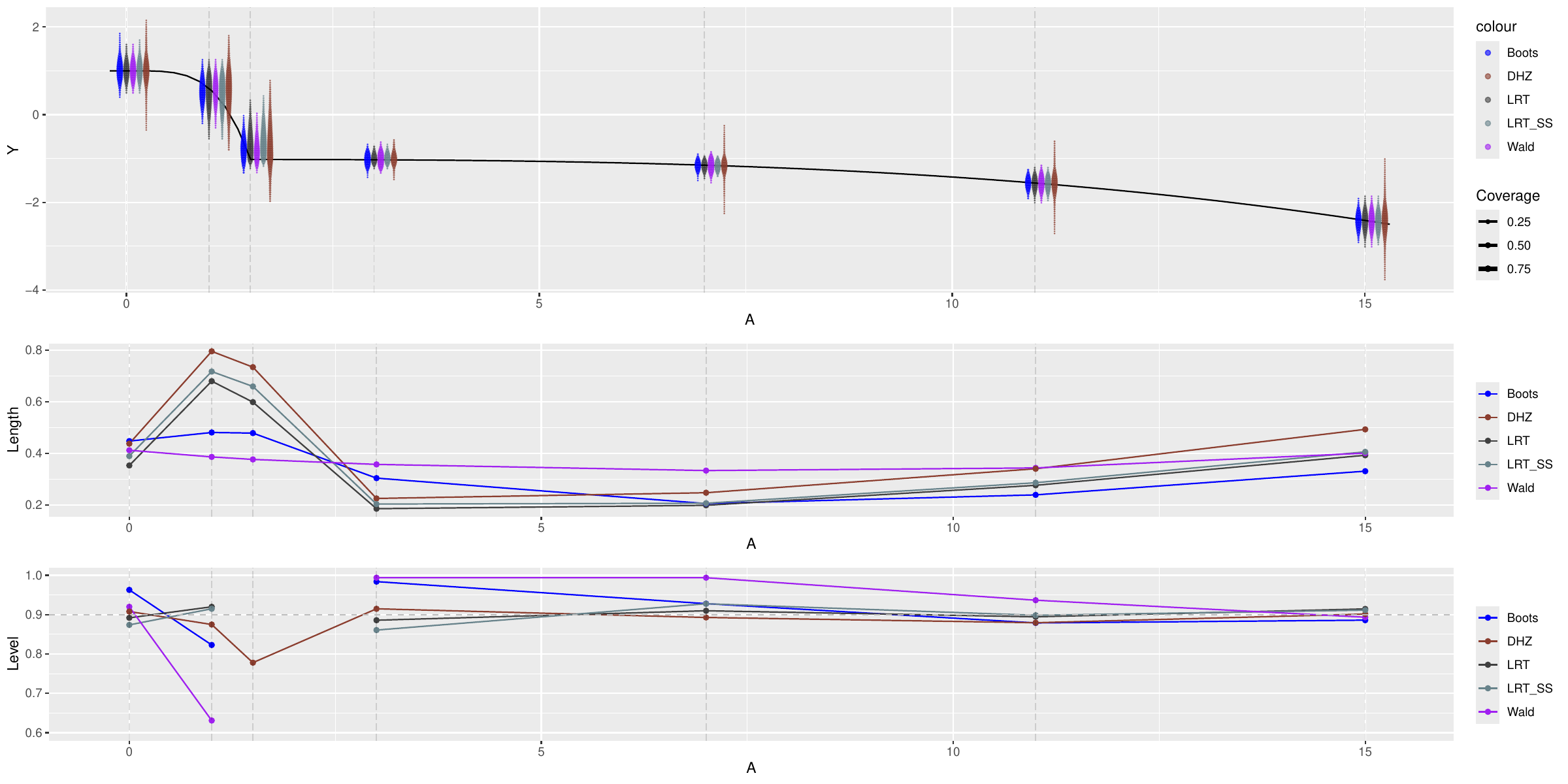}
  \caption{Simulation study ($1000$ Monte Carlos) plots with $n=1000$, $S = 0.1$,
    and $(\mu, \pi)$   estimated with (mis-, well-) specified (parametric) models.
    \suppplottext \label{fig:sim-plots_confounding1}}  
\end{figure}

\begin{figure}%
  \centering  \includegraphics[width=\linewidth]{\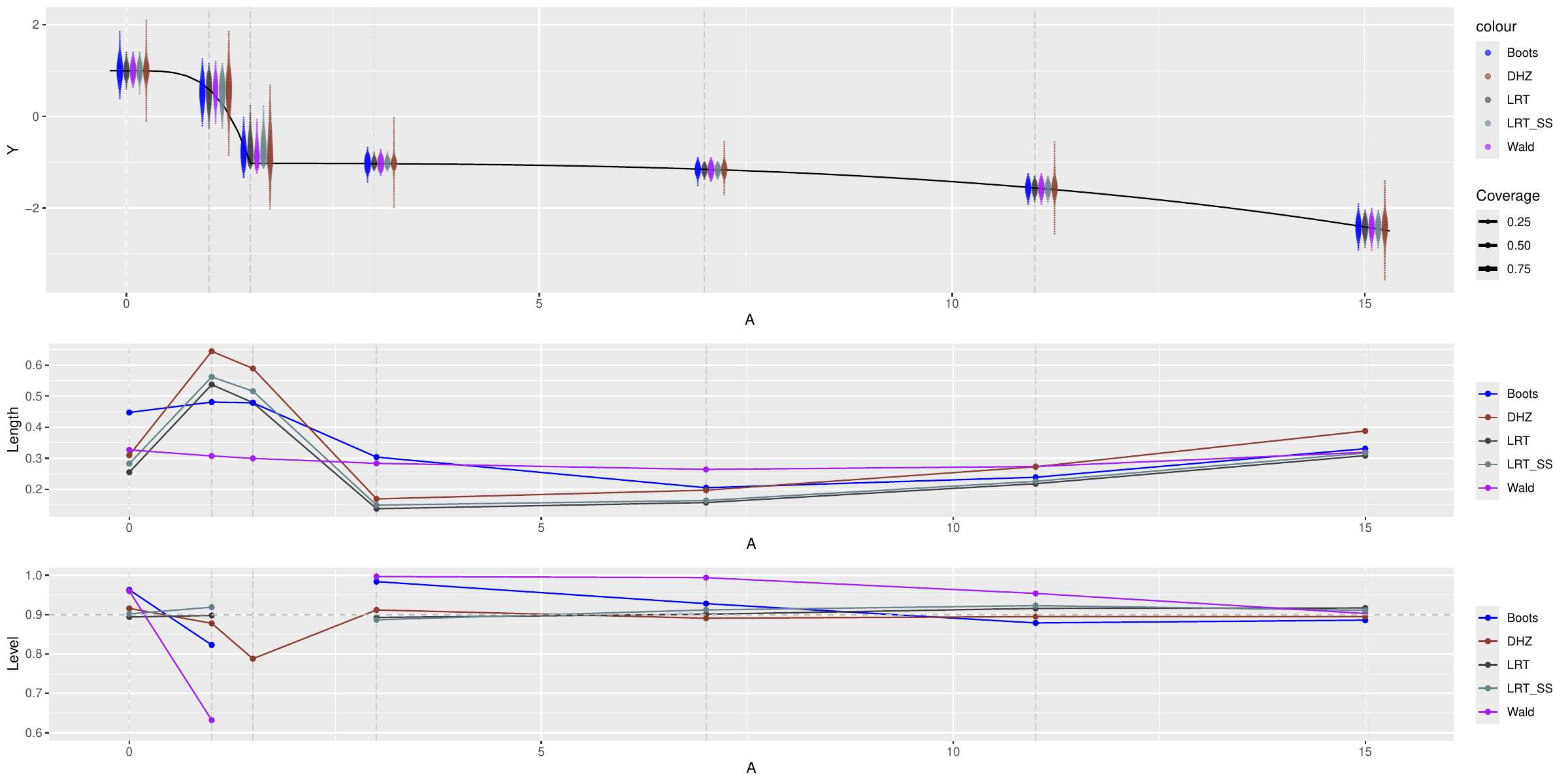} 
  \caption{Simulation study ($1000$ Monte Carlos) plots with $n=2000$, $S = 0.1$,
    and $(\mu, \pi)$   estimated with (mis-, well-) specified (parametric) models.
    \suppplottext \label{fig:sim-plots_confounding1-2000-param-picorrect}}  
\end{figure}

\begin{figure}%
  \centering  \includegraphics[width=\linewidth]{\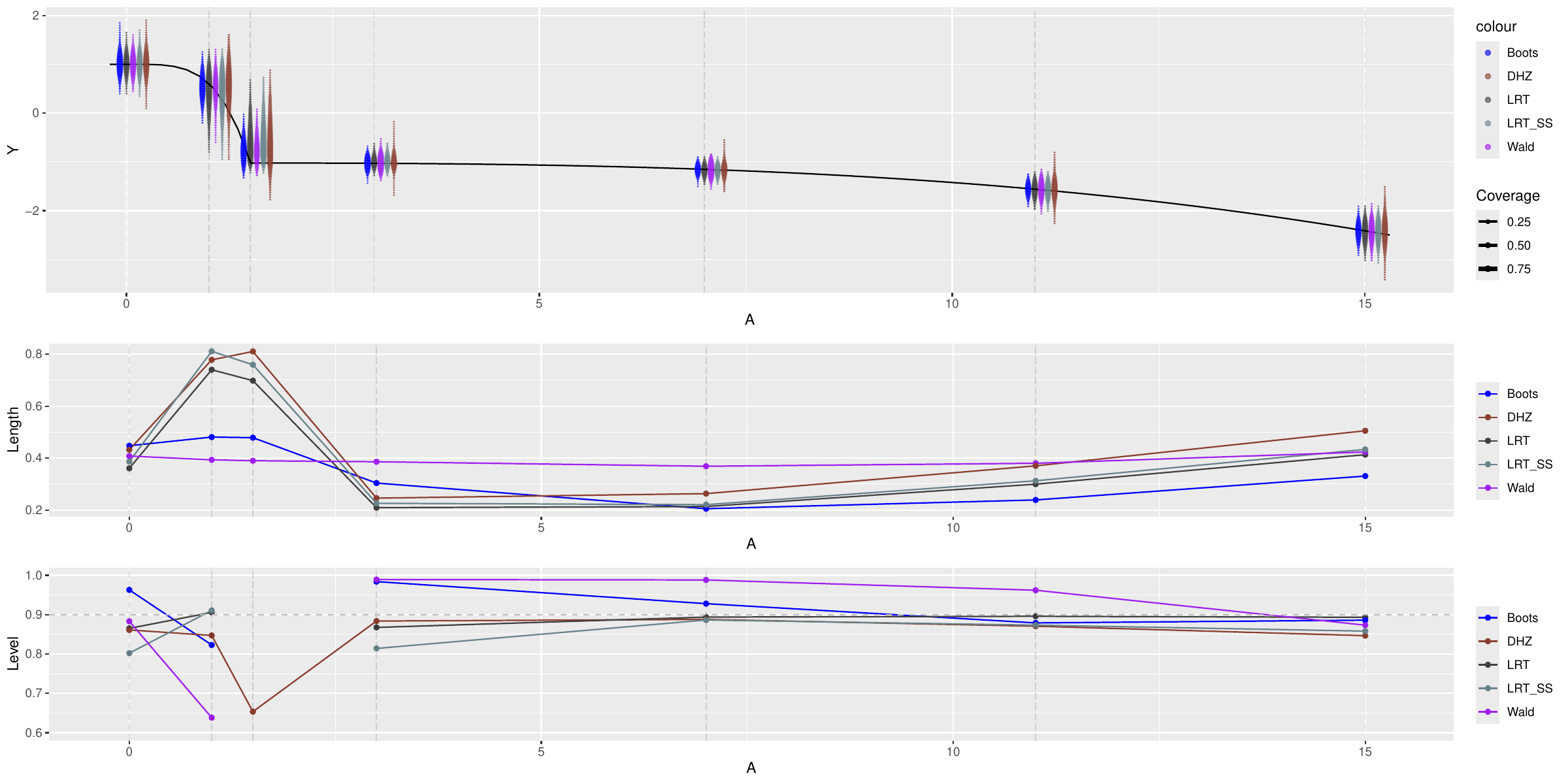}
  \caption{Simulation study ($500$ Monte Carlos) plots with $n=500$, $S = 0.1$,
    and $(\mu, \pi)$   both estimated nonparametrically with SuperLearner.
    \suppplottext \label{fig:sim-plots_confounding1-500-SL}}  
\end{figure}

\begin{figure}%
  \centering  
\includegraphics[width=\linewidth]{\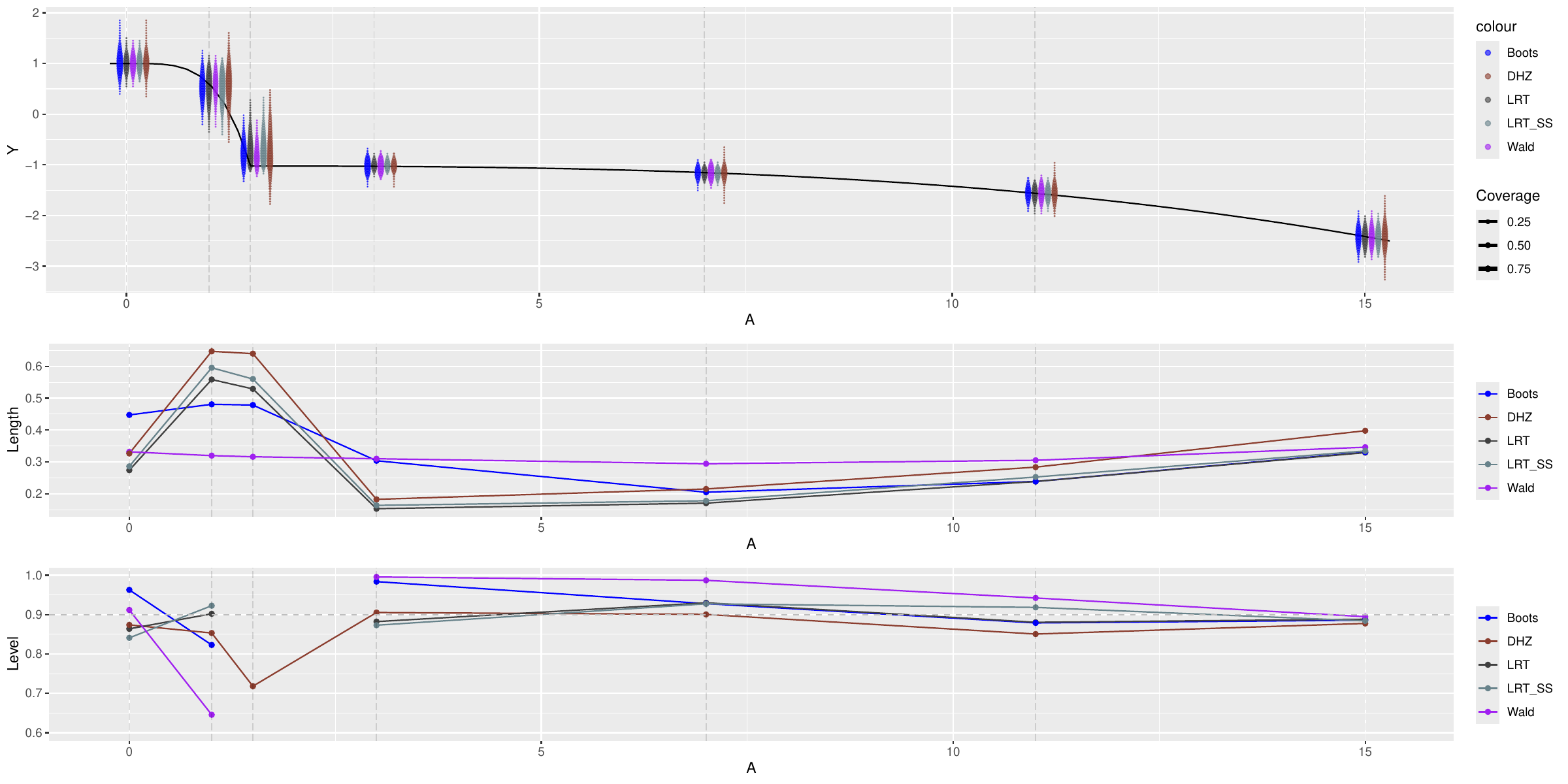}
  \caption{Simulation study ($500$ Monte Carlos) plots with $n=1000$, $S = 0.1$,
    and $(\mu, \pi)$   both estimated nonparametrically with SuperLearner.
    \suppplottext \label{fig:sim-plots_confounding1}}  
\end{figure}

\begin{figure}%
  \centering  
\includegraphics[width=\linewidth]{\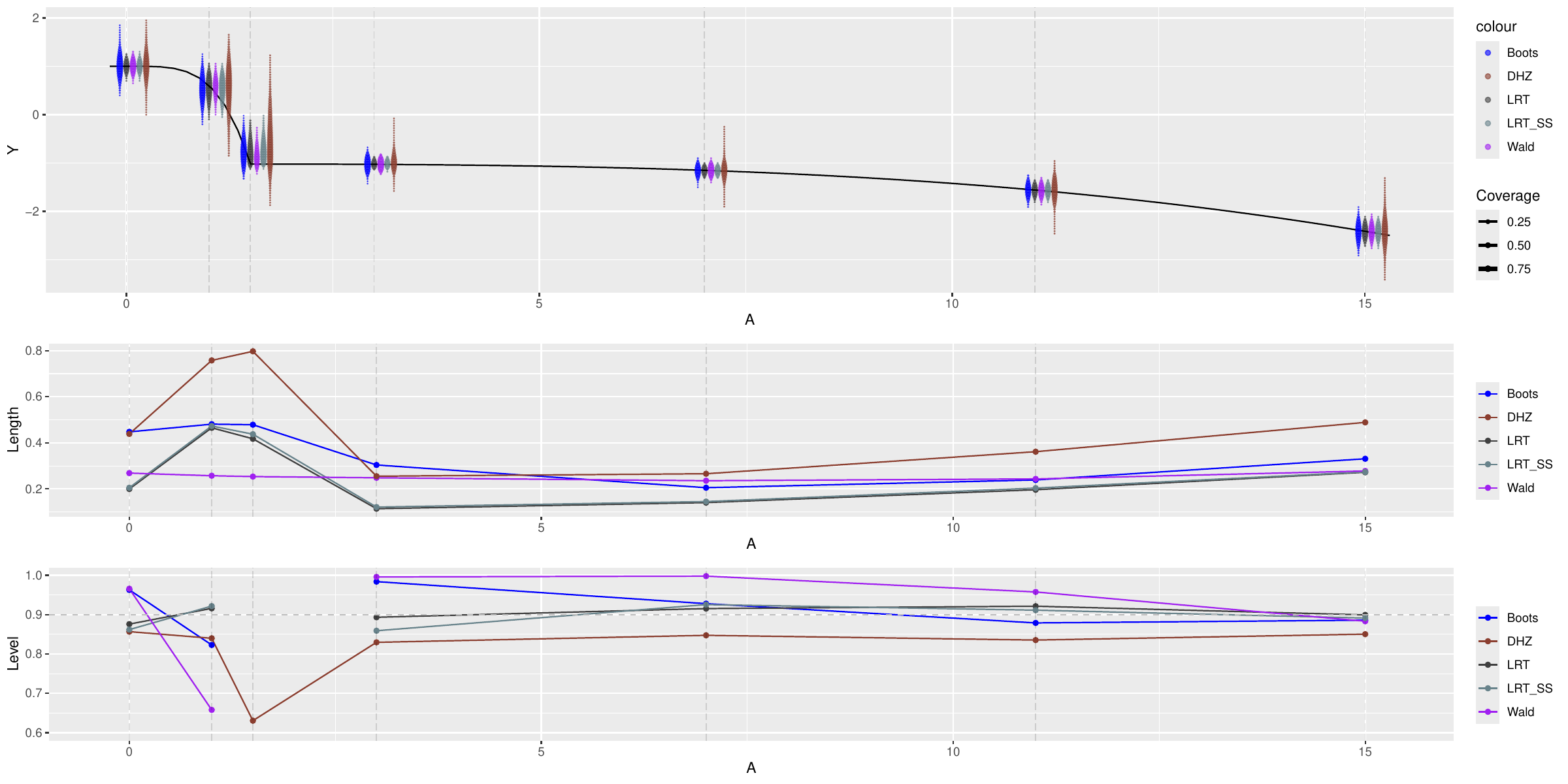}
  \caption{Simulation study ($500$ Monte Carlos) plots with $n=2000$, $S = 0.1$,
    and $(\mu, \pi)$   both estimated nonparametrically with SuperLearner.
    \suppplottext \label{fig:sim-plots_confounding1-2000-SL}}  
\end{figure}

\begin{figure}[ht]
  \centering  \includegraphics[width=\linewidth]{\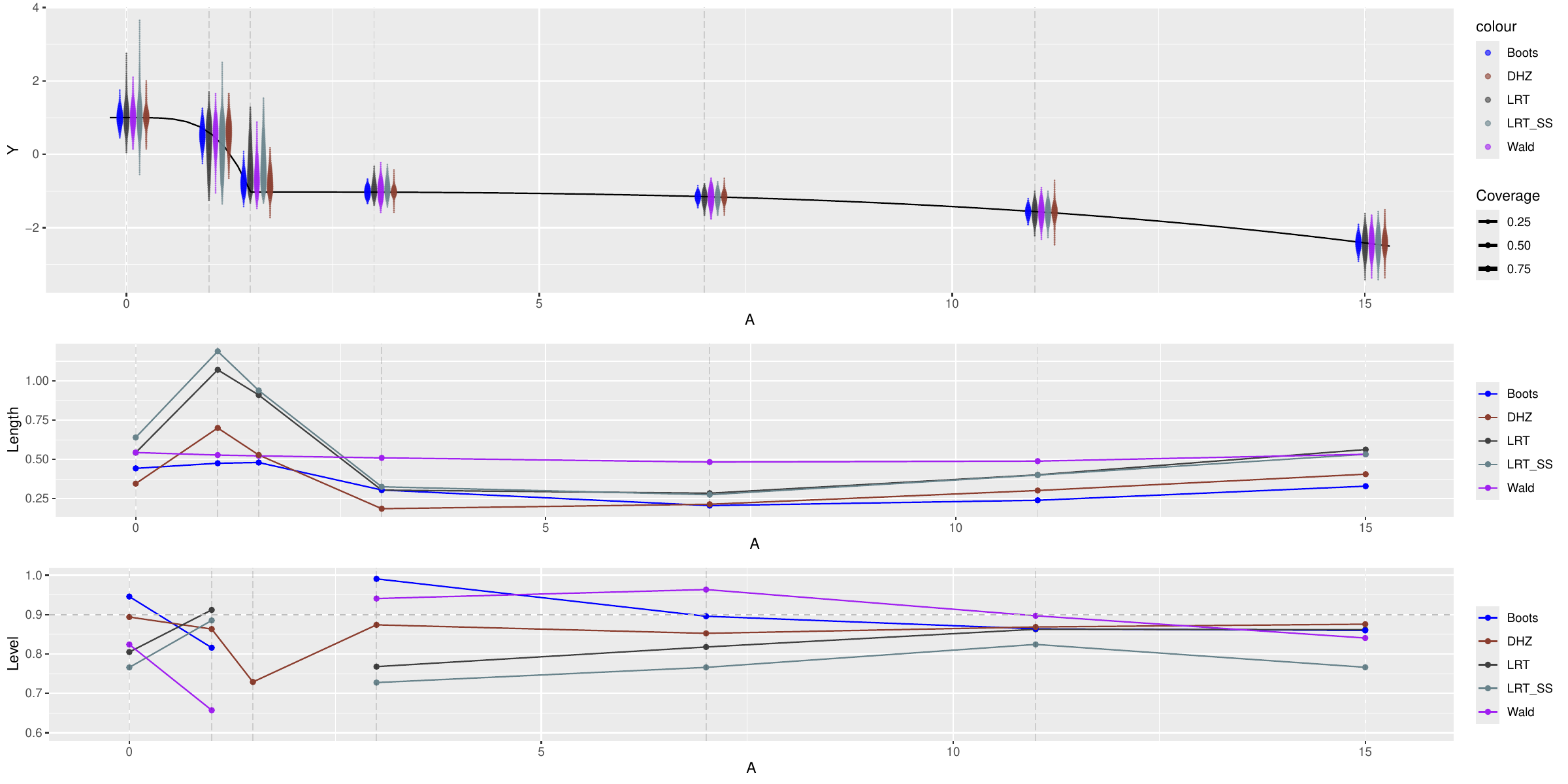}
  \caption{Simulation study ($1000$ Monte Carlos) plots with $n=200$, $S = 0.2$, and $\mu, \pi$ both estimated with well specified (parametric) models. \suppplottext \label{fig:sim-plots_confounding2-200-param}}  
\end{figure}

\begin{figure}[ht]
  \centering  \includegraphics[width=\linewidth]{\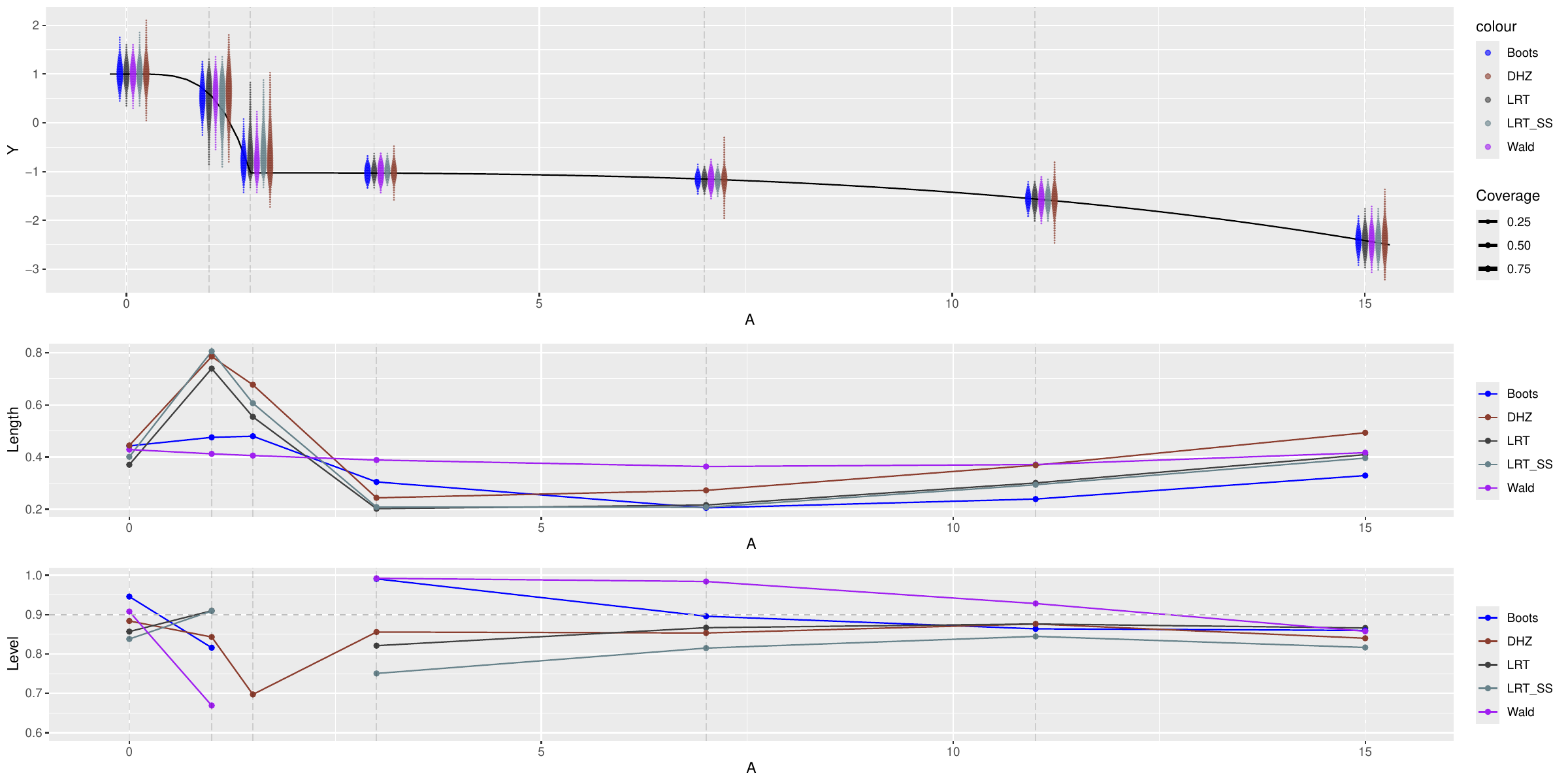}
  \caption{Simulation study ($1000$ Monte Carlos) plots with $n=500$, $S = 0.2$, and $\mu, \pi$ both estimated with well specified (parametric) models. \suppplottext \label{fig:sim-plots_confounding2-500-param}}  
\end{figure}

\begin{figure}[ht]
  \centering  %
  \includegraphics[width=\linewidth]{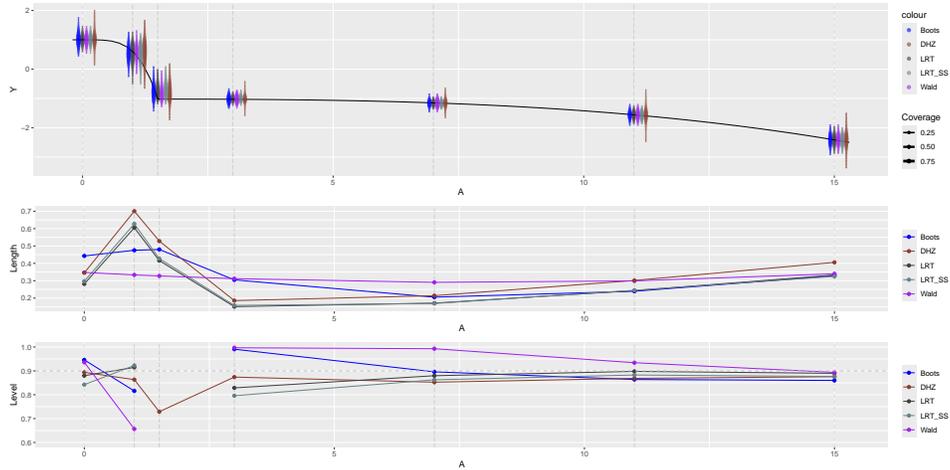}
  \caption{Simulation study ($1000$ Monte Carlos) plots with $n=1000$, $S = 0.2$, and $\mu, \pi$ both estimated with well specified (parametric) models. \suppplottext \label{fig:sim-plots_confounding2-1000-param-bothcorrect}}  
\end{figure}

\begin{figure}[ht]
  \centering  \includegraphics[width=\linewidth]{\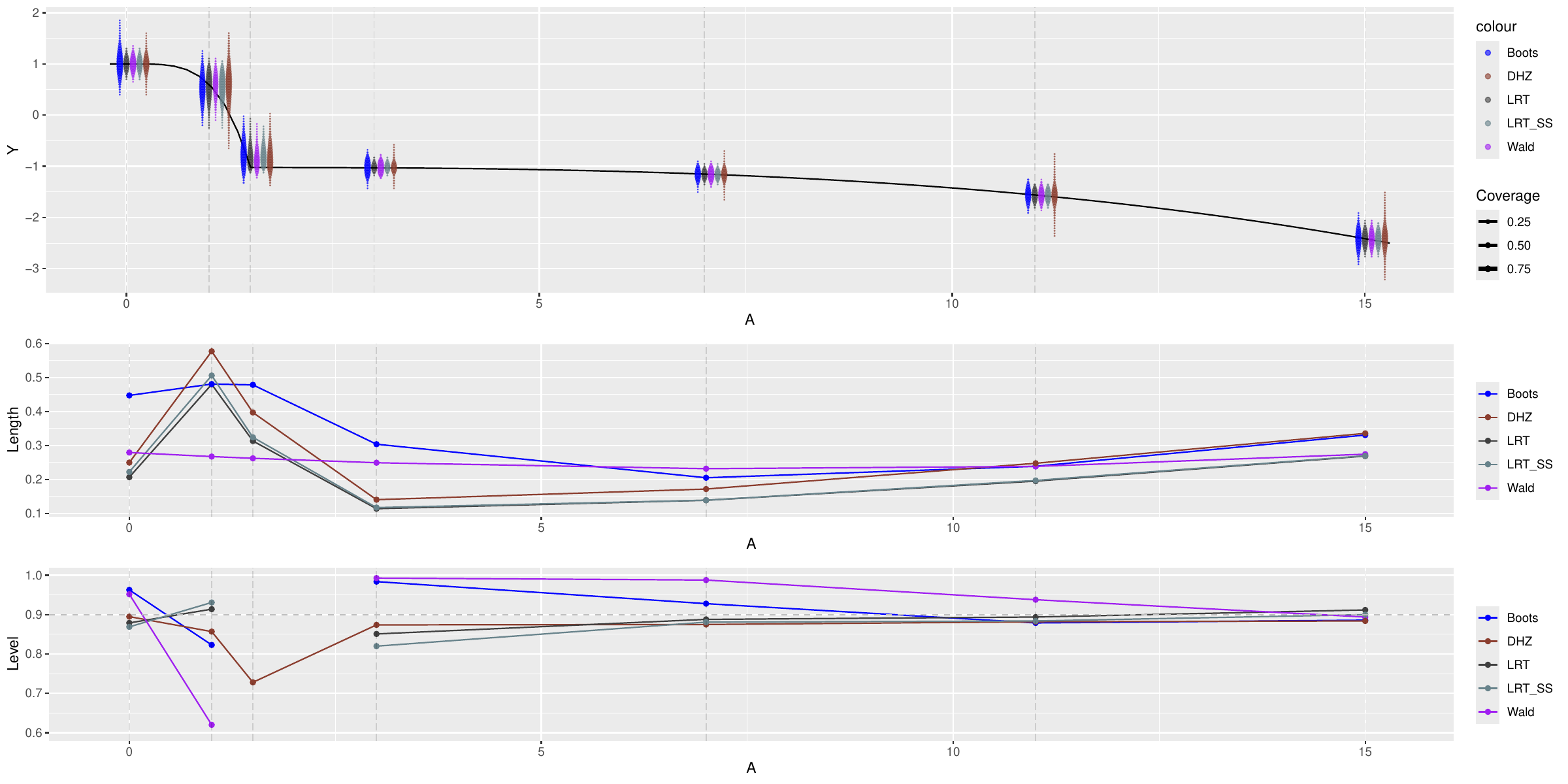} 
  \caption{Simulation study ($1000$ Monte Carlos) plots with $n=2000$, $S = 0.2$, and $\mu, \pi$ both estimated with well specified (parametric) models. \suppplottext \label{fig:sim-plots_confounding2-2000-param-bothcorrect}}  
\end{figure}

\begin{figure}[ht]
  \centering  \includegraphics[width=\linewidth]{\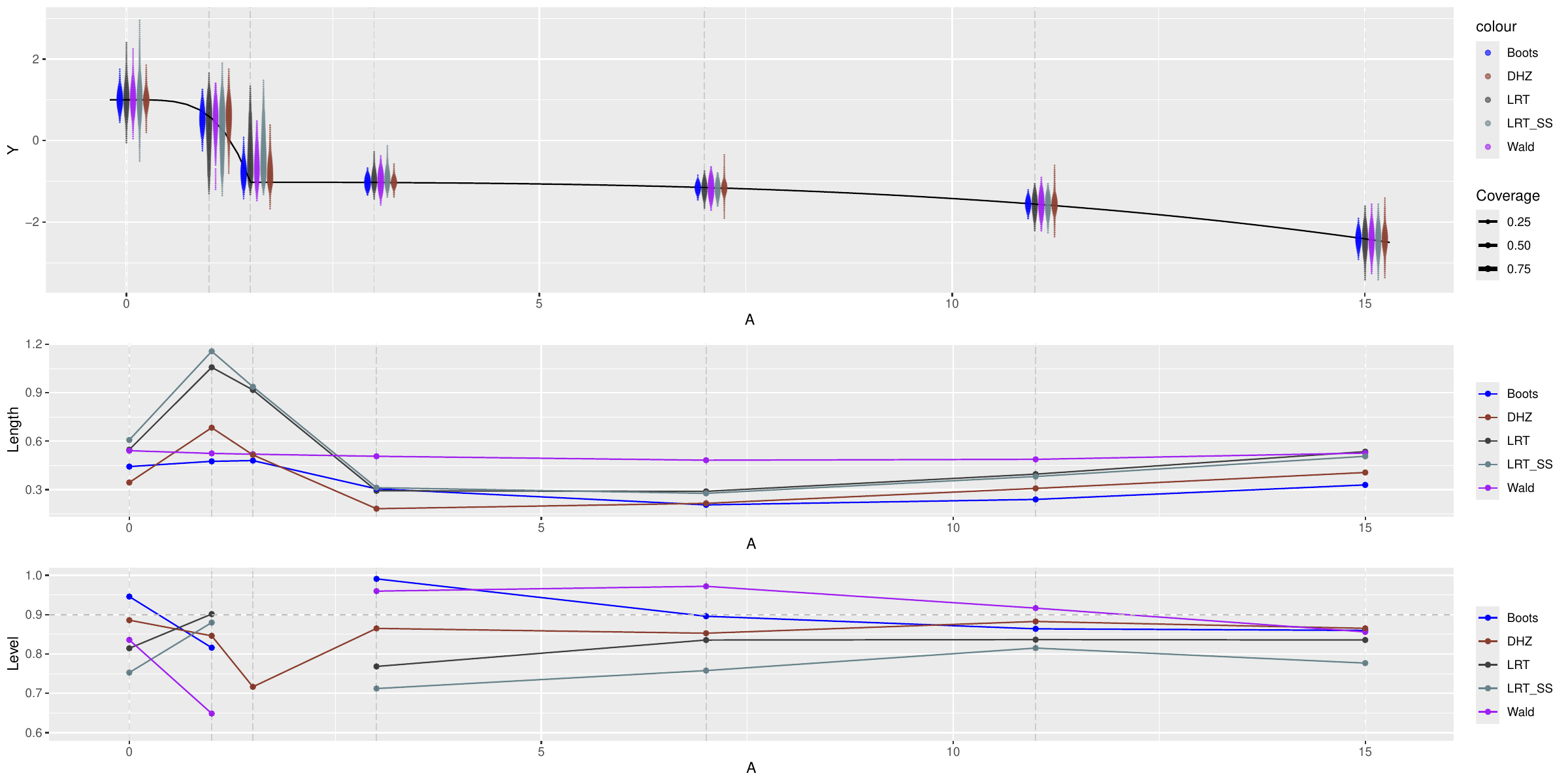}
  \caption{Simulation study ($1000$ Monte Carlos) plots with $n=200$, $S = 0.2$,
    and $(\mu, \pi)$   estimated with (well-, mis-) specified (parametric) models.
    \suppplottext \label{fig:sim-plots_confounding2-200-param-mucorrect}}  
\end{figure}
\begin{figure}[ht]
  \centering  \includegraphics[width=\linewidth]{\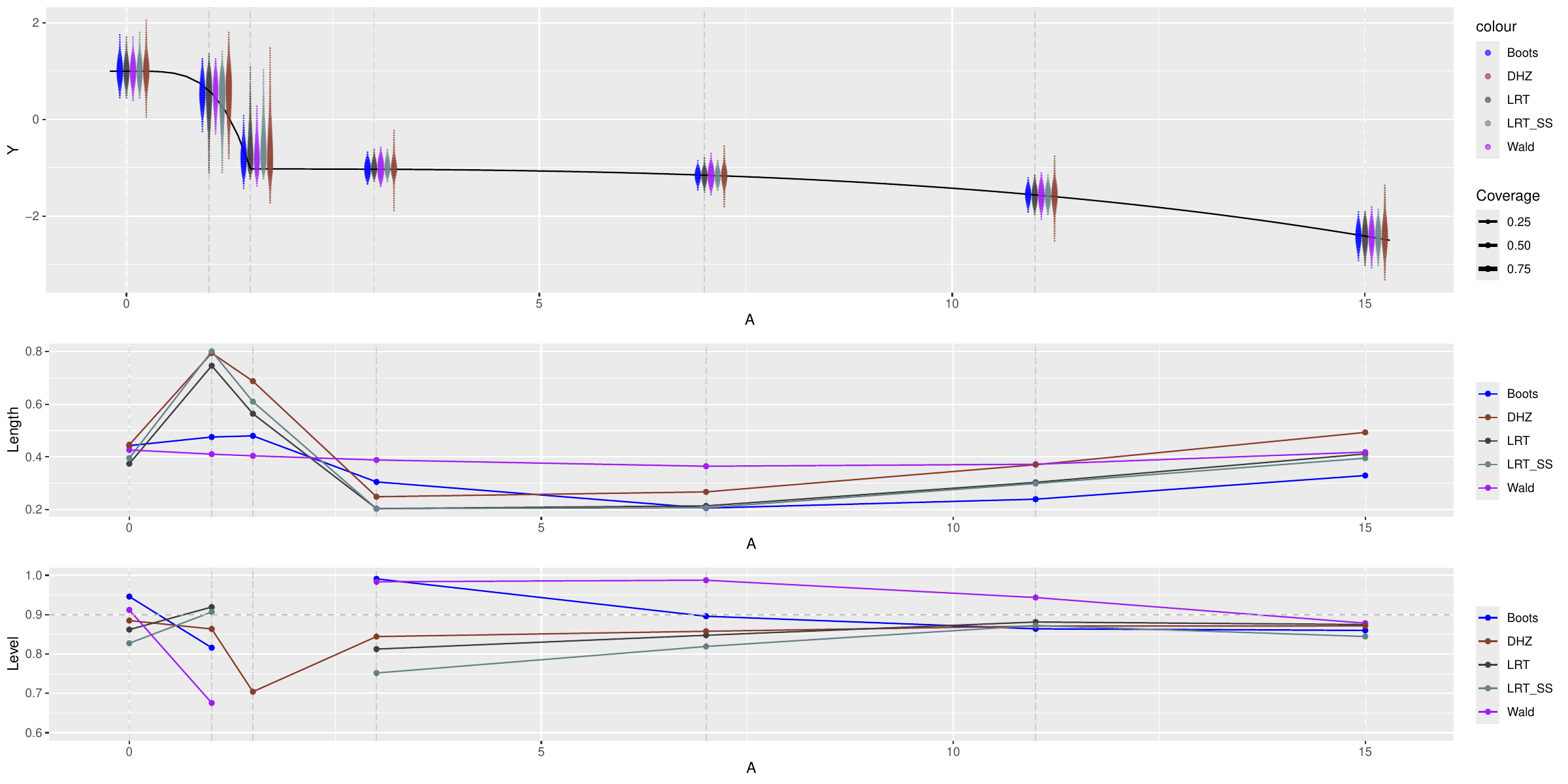}
  \caption{Simulation study ($1000$ Monte Carlos) plots with $n=500$, $S = 0.2$,
    and $(\mu, \pi)$   estimated with (well-, mis-) specified (parametric) models.
    \suppplottext \label{fig:sim-plots_confounding1}}  
\end{figure}

\begin{figure}[ht]
  \centering %
    \includegraphics[width=\linewidth]{\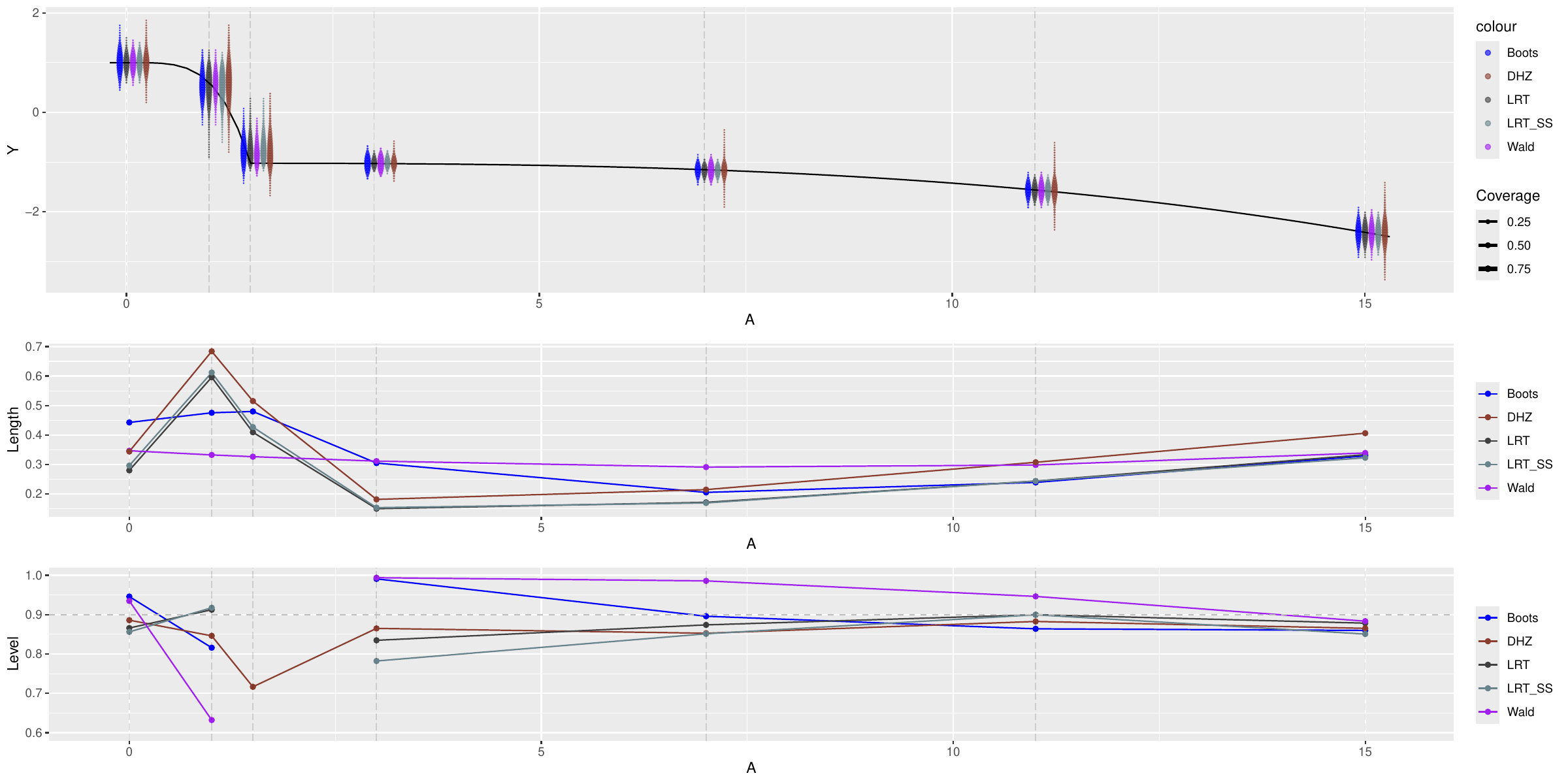}
  \caption{Simulation study ($1000$ Monte Carlos) plots with $n=1000$, $S = 0.2$,
    and $(\mu, \pi)$   estimated with (well-, mis-) specified (parametric) models.
    \suppplottext \label{fig:sim-plots_confounding1}}  
\end{figure}

\begin{figure}[ht]
  \centering  \includegraphics[width=\linewidth]{\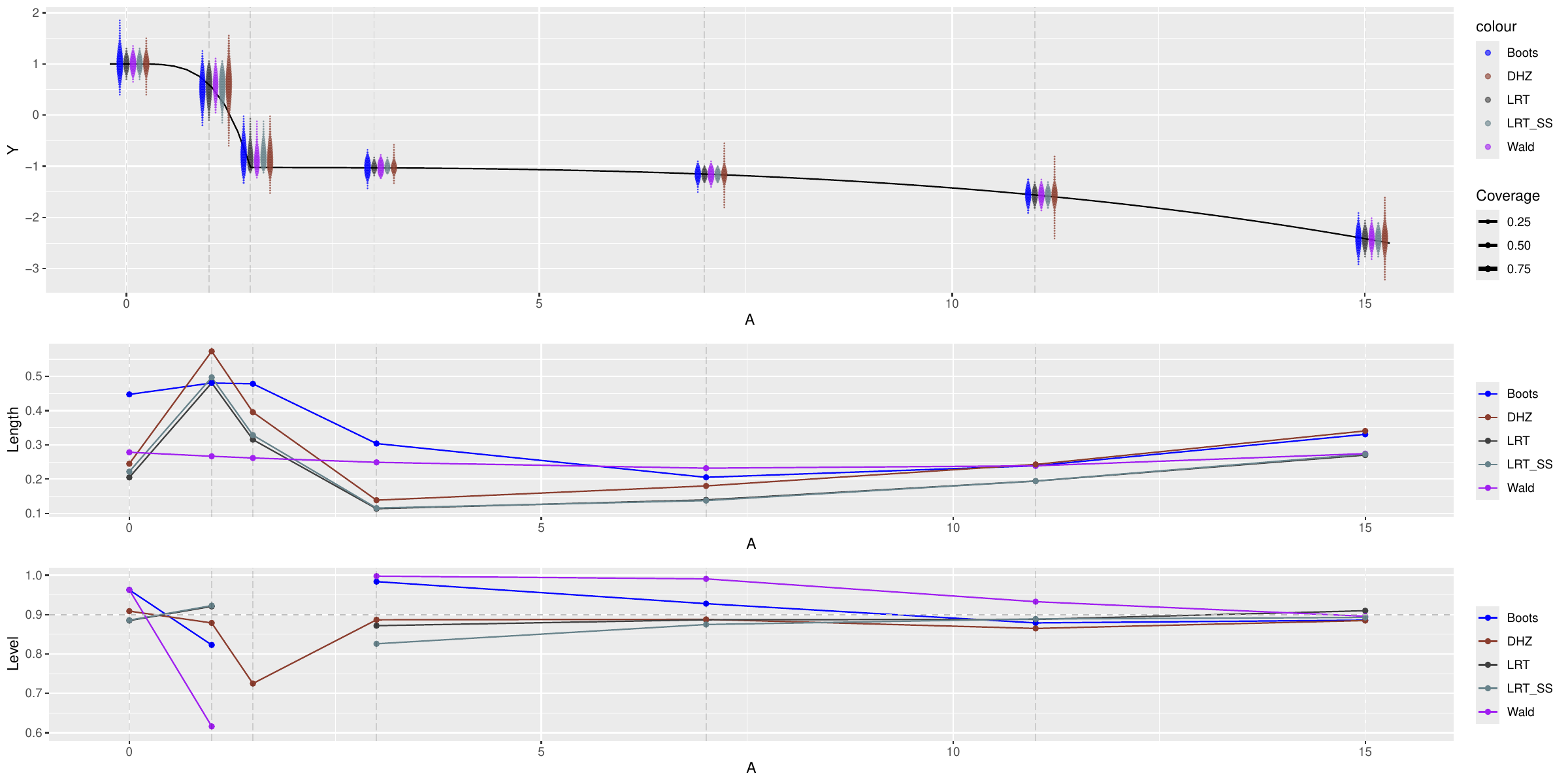} 
  \caption{Simulation study ($1000$ Monte Carlos) plots with $n=2000$, $S = 0.2$,
    and $(\mu, \pi)$   estimated with (well-, mis-) specified (parametric) models.
    \suppplottext \label{fig:sim-plots_confounding1}}  
\end{figure}

\begin{figure}[ht]
  \centering  \includegraphics[width=\linewidth]{\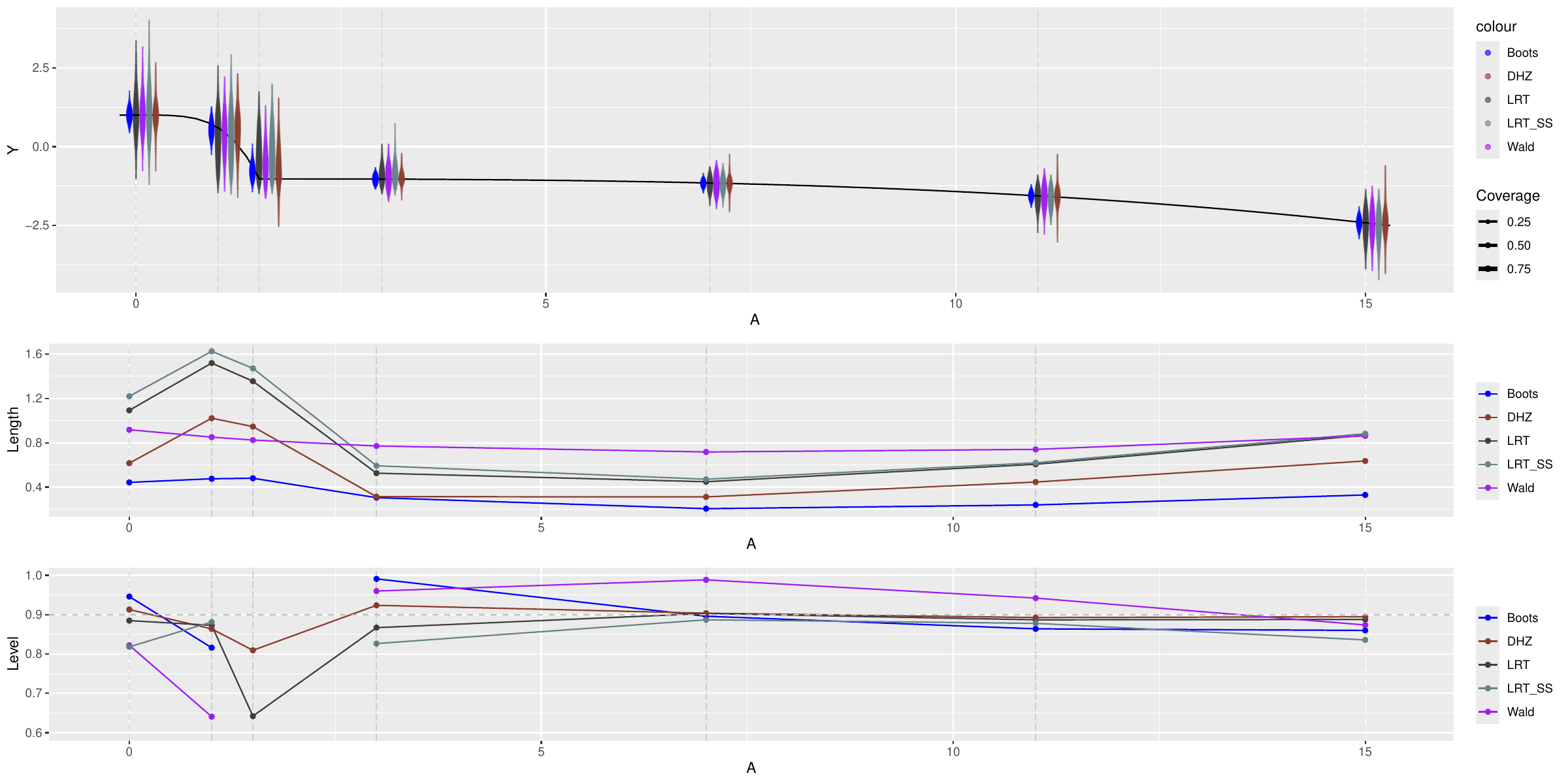}
  \caption{Simulation study ($1000$ Monte Carlos) plots with $n=200$, $S = 0.2$,
    and $(\mu, \pi)$   estimated with (mis-, well-) specified (parametric) models.
    \suppplottext \label{fig:sim-plots_confounding1}}  
\end{figure}
\begin{figure}[ht]
  \centering  \includegraphics[width=\linewidth]{\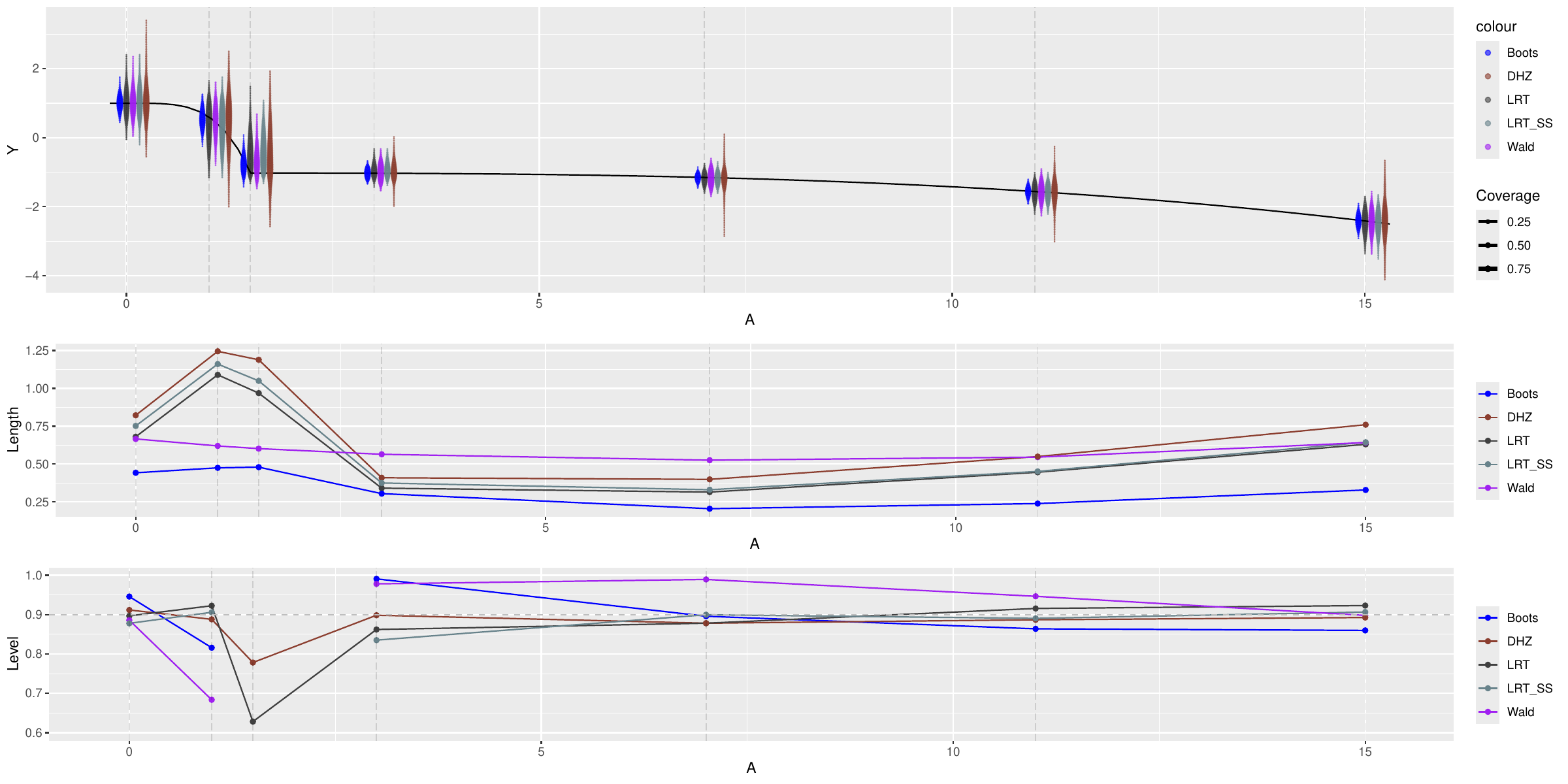}
  \caption{Simulation study ($1000$ Monte Carlos) plots with $n=500$, $S = 0.2$,
    and $(\mu, \pi)$   estimated with (mis-, well-) specified (parametric) models.
    \suppplottext \label{fig:sim-plots_confounding1}}  
\end{figure}

\begin{figure}[ht]
  \centering  \includegraphics[width=\linewidth]{\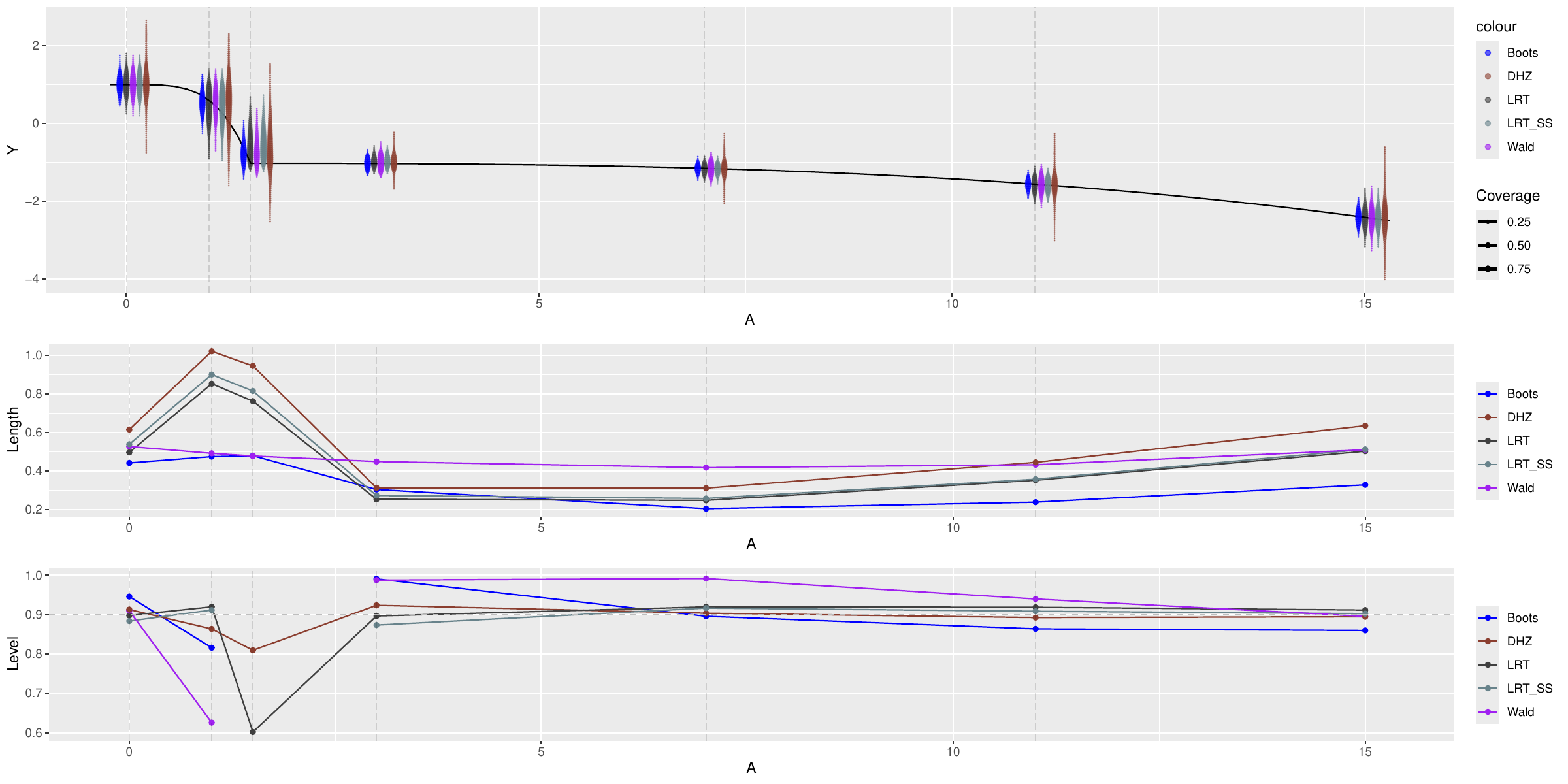}
  \caption{Simulation study ($1000$ Monte Carlos) plots with $n=1000$, $S = 0.2$,
    and $(\mu, \pi)$   estimated with (mis-, well-) specified (parametric) models.
    \suppplottext \label{fig:sim-plots_confounding1}}  
\end{figure}

\begin{figure}[ht]
  \centering  \includegraphics[width=\linewidth]{\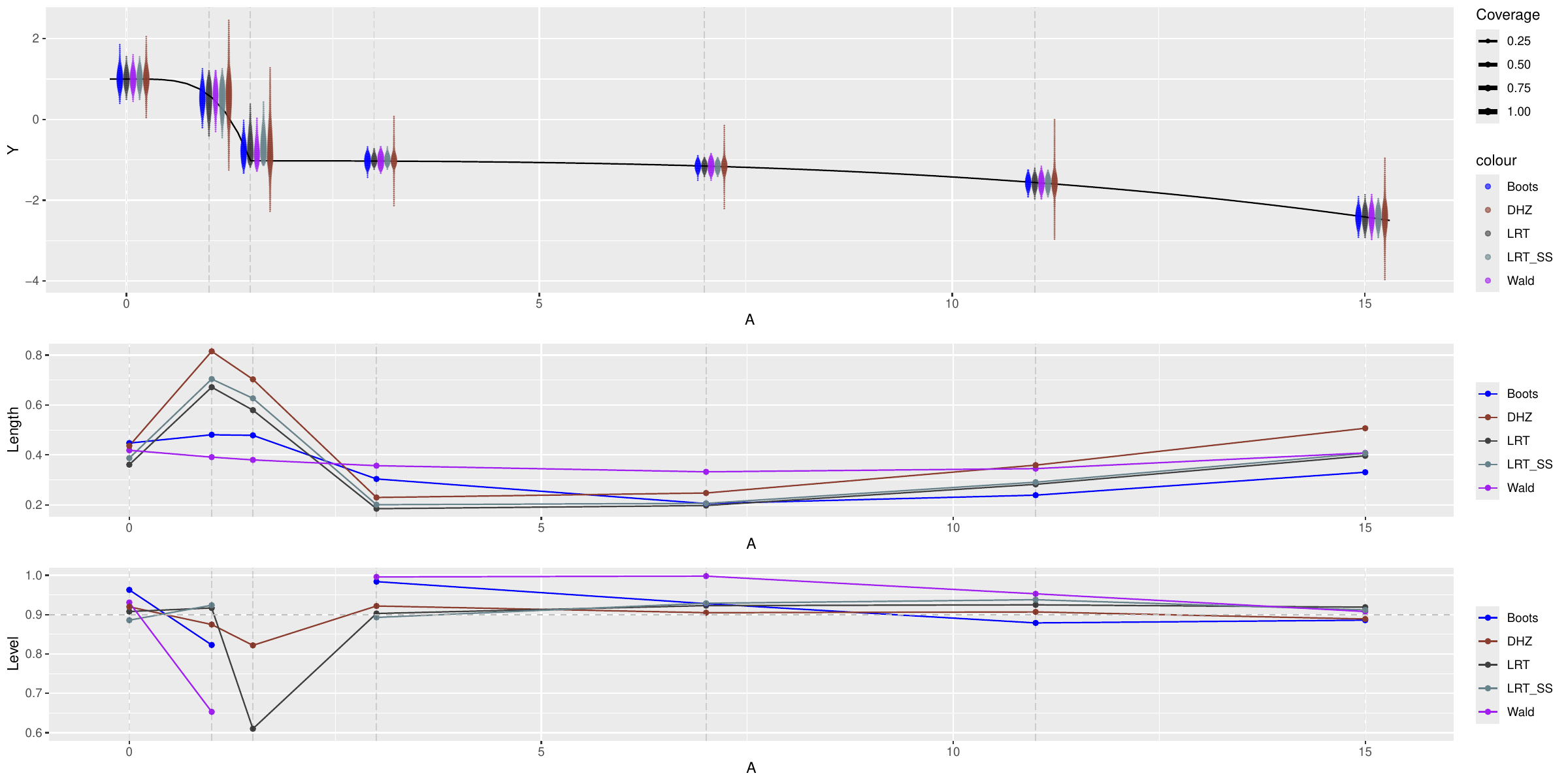} 
  \caption{Simulation study ($1000$ Monte Carlos) plots with $n=2000$, $S = 0.2$,
    and $(\mu, \pi)$   estimated with (mis-, well-) specified (parametric) models.
    \suppplottext \label{fig:sim-plots_confounding2-2000-param-picorrect}}  
\end{figure}

\begin{figure}[ht]
  \centering  \includegraphics[width=\linewidth]{\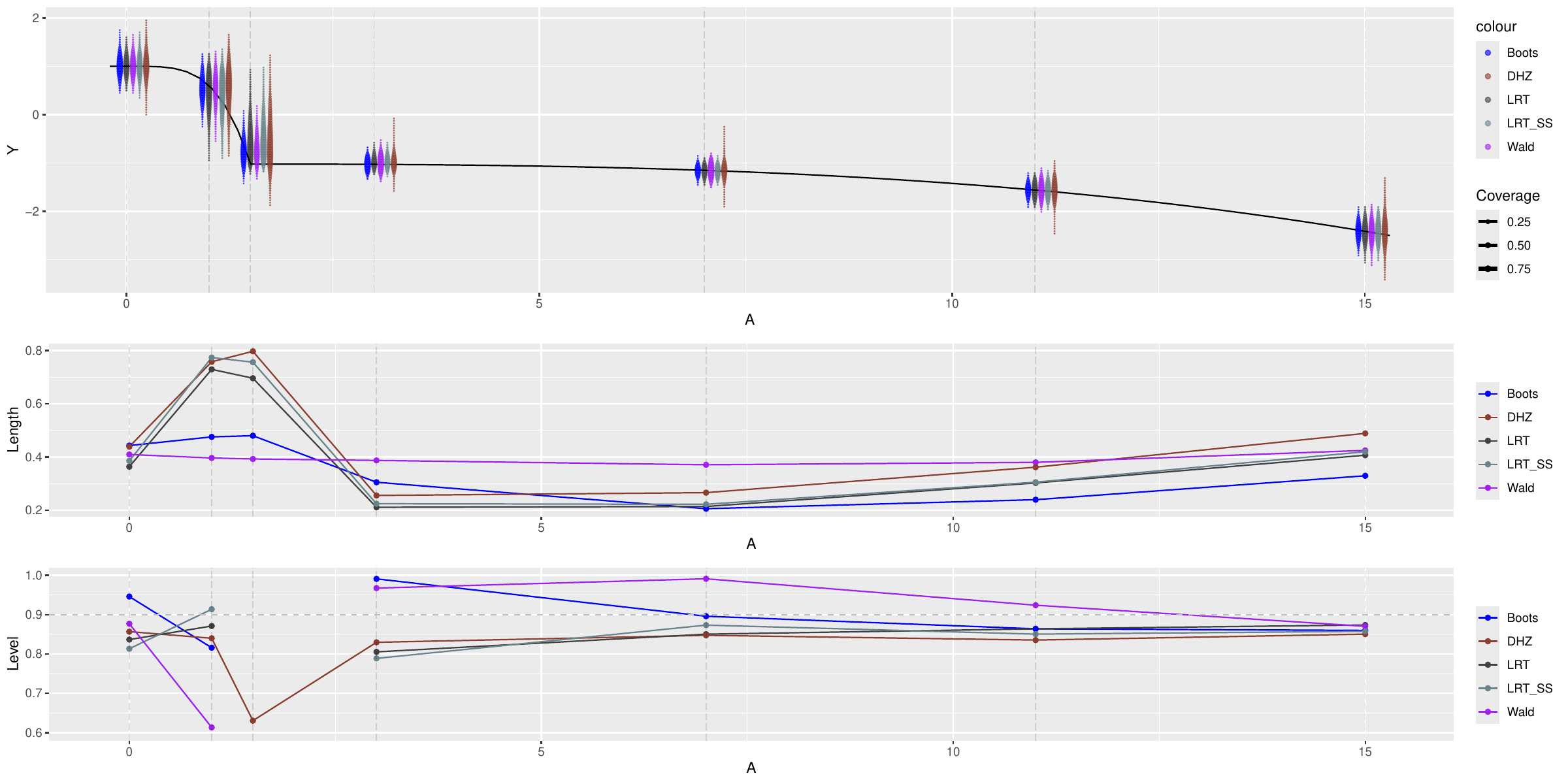}
  \caption{Simulation study ($500$ Monte Carlos) plots with $n=500$, $S = 0.2$,
    and $(\mu, \pi)$   both estimated nonparametrically with SuperLearner.
    \suppplottext \label{fig:sim-plots_confounding2-500-SL}}  
\end{figure}

\begin{figure}[ht]
  \centering
\includegraphics[width=\linewidth]{\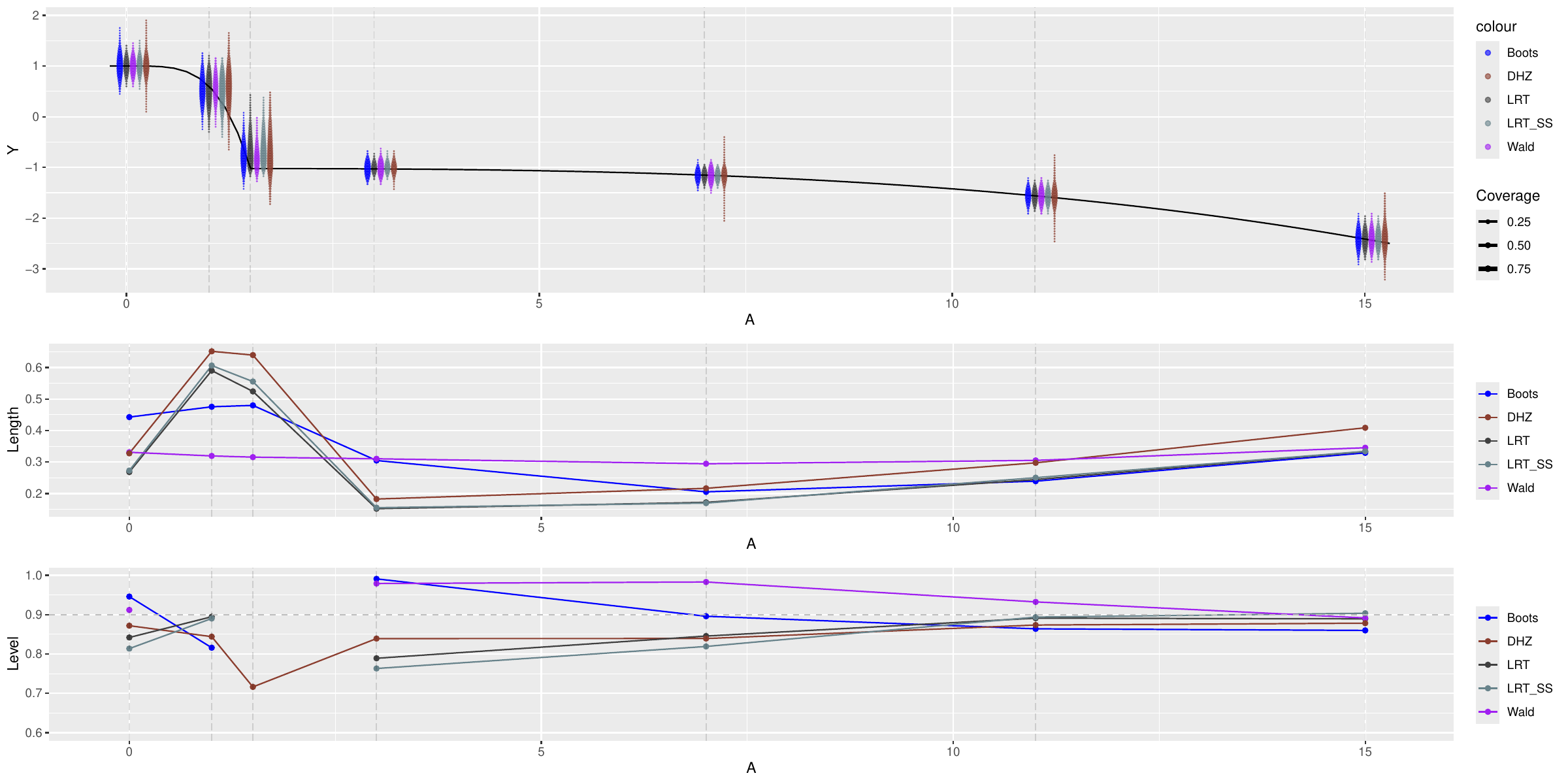}
  \caption{Simulation study ($500$ Monte Carlos) plots with $n=1000$, $S = 0.2$,
    and $(\mu, \pi)$   both estimated nonparametrically with SuperLearner.
    \suppplottext \label{fig:sim-plots_confounding1}}  
\end{figure}

\begin{figure}[!ht]
  \centering
\includegraphics[width=\linewidth]{\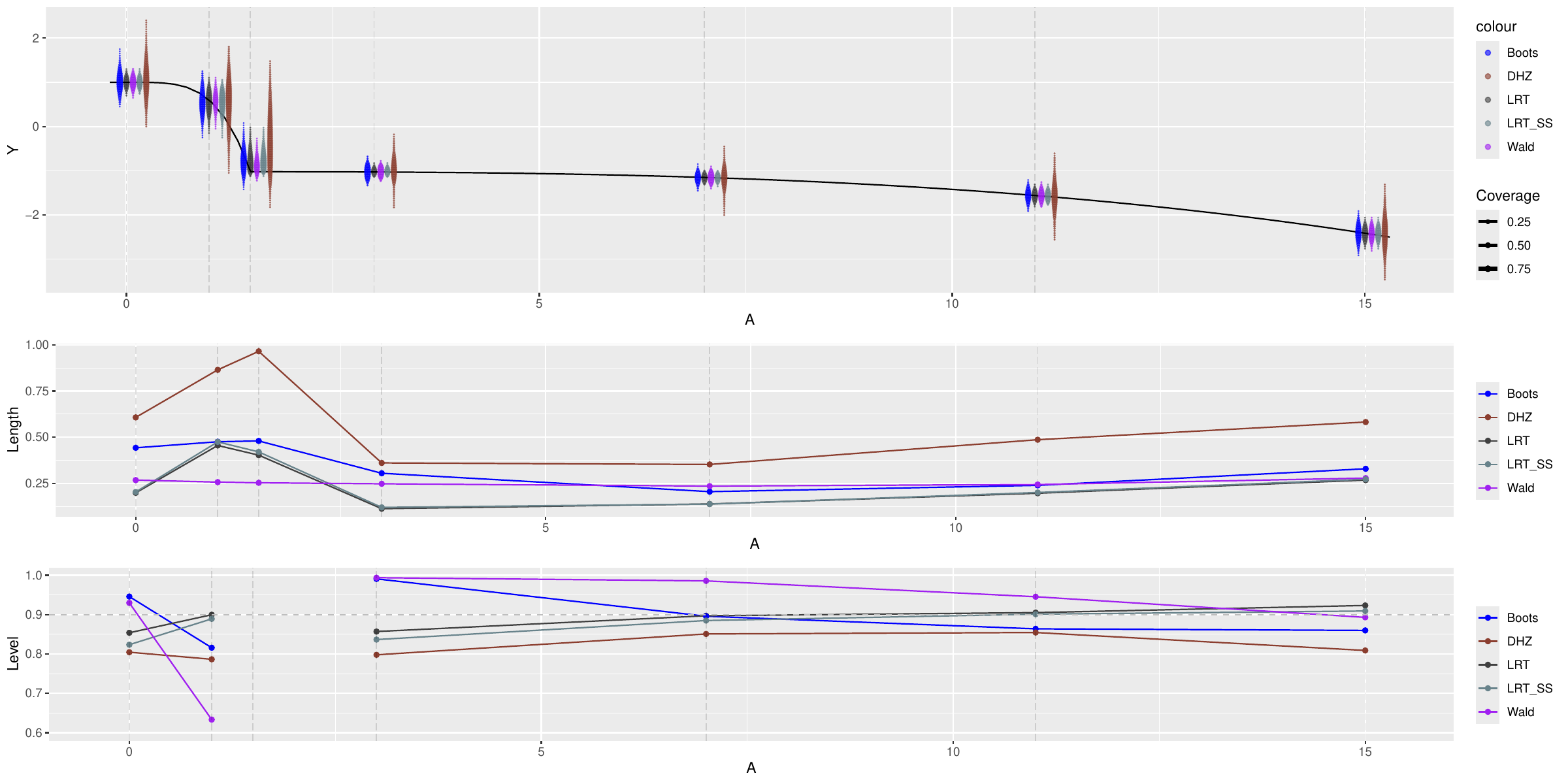}
  \caption{Simulation study ($500$ Monte Carlos) plots with $n=2000$, $S = 0.2$,
    and $(\mu, \pi)$   both estimated nonparametrically with SuperLearner.
    \suppplottext \label{fig:sim-plots_confounding2-2000-SL}}  
\end{figure}

%
%
%
%
%
%

%
%
%
%


\bibliographystyle{alpha}
\bibliography{mybib}

\newcommand{\etalchar}[1]{$^{#1}$}
\begin{thebibliography}{DWW{\etalchar{+}}24b}

\bibitem[And94]{andrews1994empirical}
Donald~WK Andrews.
\newblock Empirical process methods in econometrics.
\newblock {\em Handbook of econometrics}, 4:2247--2294, 1994.

\bibitem[Ban00]{Banerjee:2000wh}
Moulinath Banerjee.
\newblock {Likelihood Ratio Inference in Regular and Non-regular Problems},
  2000.

\bibitem[Ban07]{banerjee2007:LRTmonoresponse}
Moulinath Banerjee.
\newblock {Likelihood based inference for monotone response models}.
\newblock {\em The Annals of Statistics}, 35(3):931 -- 956, 2007.

\bibitem[BCCW18]{belloni2018uniformly}
Alexandre Belloni, Victor Chernozhukov, Denis Chetverikov, and Ying Wei.
\newblock Uniformly valid post-regularization confidence regions for many
  functional parameters in z-estimation framework.
\newblock {\em Annals of statistics}, 46(6B):3643, 2018.

\bibitem[BDS19]{Banerjee:2016uz}
Moulinath Banerjee, Cécile Durot, and Bodhisattva Sen.
\newblock Divide and conquer in nonstandard problems and the super-efficiency
  phenomenon.
\newblock {\em The Annals of Statistics}, 47(2):720–757, 2019.

\bibitem[BR05]{Bang_Robins_2005}
Heejung Bang and James~M. Robins.
\newblock Doubly robust estimation in missing data and causal inference models.
\newblock {\em Biometrics}, 61(4):962–973, 2005.

\bibitem[BW01]{Banerjee:2001jy}
Moulinath Banerjee and Jon~A. Wellner.
\newblock Likelihood ratio tests for monotone functions.
\newblock {\em Ann. Statist.}, 29(6):1699--1731, 2001.

\bibitem[BW05a]{Banerjee:2005gw}
Moulinath Banerjee and Jon~A Wellner.
\newblock {Confidence intervals for current status data}.
\newblock {\em Scand. J. Statist.}, 32(3):405--424, 2005.

\bibitem[BW05b]{Banerjee:2005im}
Moulinath Banerjee and Jon~A. Wellner.
\newblock Score statistics for current status data: comparisons with likelihood
  ratio and {W}ald statistics.
\newblock {\em Int. J. Biostat.}, 1:Art. 3, 29, 2005.

\bibitem[BW07]{Balabdaoui:2007jj}
Fadoua Balabdaoui and Jon~A Wellner.
\newblock {Estimation of a k-monotone Distribution and the Spline Connection}.
\newblock 35(6):2536--2564, December 2007.

\bibitem[CCD{\etalchar{+}}18]{Chernozhukov_Chetverikov_Demirer_Duflo_Hansen_Newey_Robins_2018}
Victor Chernozhukov, Denis Chetverikov, Mert Demirer, Esther Duflo, Christian
  Hansen, Whitney Newey, and James Robins.
\newblock Double/debiased machine learning for treatment and structural
  parameters.
\newblock {\em The Econometrics Journal}, 21(1):C1–C68, 2018.

\bibitem[CJN23]{cattaneo2023bootstrap}
Matias~D Cattaneo, Michael Jansson, and Kenichi Nagasawa.
\newblock Bootstrap-assisted inference for generalized grenander-type
  estimators.
\newblock {\em arXiv preprint arXiv:2303.13598}, 2023.

\bibitem[CL20]{Colangelo:2020tt}
Kyle Colangelo and Ying-Ying Lee.
\newblock Double debiased machine learning nonparametric inference with
  continuous treatments.
\newblock {\em arXiv.org}, April 2020.

\bibitem[CLX13]{Cai_Low_Xia_2013}
T~Tony Cai, Mark~G Low, and Yin Xia.
\newblock Adaptive confidence intervals for regression functions under shape
  constraints.
\newblock {\em The Annals of Statistics}, 41(2):722–750, 2013.

\bibitem[CLY22]{pmlr-v151-chen22c}
Guanhua Chen, Xiaomao Li, and Menggang Yu.
\newblock Policy learning for optimal individualized dose intervals.
\newblock In Gustau Camps-Valls, Francisco J.~R. Ruiz, and Isabel Valera,
  editors, {\em Proceedings of The 25th International Conference on Artificial
  Intelligence and Statistics}, volume 151 of {\em Proceedings of Machine
  Learning Research}, pages 1671--1693. PMLR, 28--30 Mar 2022.

\bibitem[CMP21]{coulombe2021estimating}
Janie Coulombe, Erica~EM Moodie, and Robert~W Platt.
\newblock Estimating the marginal effect of a continuous exposure on an ordinal
  outcome using data subject to covariate-driven treatment and visit processes.
\newblock {\em Statistics in Medicine}, 40(26):5746--5764, 2021.

\bibitem[CZK16]{chen2016personalized}
Guanhua Chen, Donglin Zeng, and Michael~R Kosorok.
\newblock Personalized dose finding using outcome weighted learning.
\newblock {\em Journal of the American Statistical Association},
  111(516):1509--1521, 2016.

\bibitem[DHZ21]{deng2021confidence}
Hang Deng, Qiyang Han, and Cun-Hui Zhang.
\newblock Confidence intervals for multiple isotonic regression and other
  monotone models.
\newblock {\em The Annals of Statistics}, 49(4):2021--2052, 2021.

\bibitem[Dos19]{doss2019concave}
Charles~R Doss.
\newblock Concave regression: value-constrained estimation and likelihood
  ratio-based inference.
\newblock {\em Mathematical Programming}, 174(1):5--39, 2019.

\bibitem[DWW{\etalchar{+}}24a]{drtest-main}
Charles Doss, Guangwei Weng, Lan Wang, Ira Moscovice, and Tongtan Chantarat.
\newblock A nonparametric doubly robust test for a continuous treatment effect.
\newblock {\em (to appear in) The Annals of Statistics}, 2024.

\bibitem[DWW{\etalchar{+}}24b]{drtest-supp}
Charles Doss, Guangwei Weng, Lan Wang, Ira Moscovice, and Tongtan Chantarat.
\newblock Supplementary material for ``{A} nonparametric doubly robust test for
  a continuous treatment effect''.
\newblock 2024.

\bibitem[GJ14]{Groeneboom:2014hk}
Piet Groeneboom and Geurt Jongbloed.
\newblock {\em Nonparametric Estimation under Shape Constraints}, volume~38 of
  {\em Cambridge Series in Statistical and Probabilistic Mathematics}.
\newblock Cambridge University Press, New York, Cambridge, 2014.

\bibitem[GJ15a]{Groeneboom_Jongbloed_2015}
Piet Groeneboom and Geurt Jongbloed.
\newblock Nonparametric confidence intervals for monotone functions.
\newblock {\em The Annals of Statistics}, 43(5):2019–2054, 2015.

\bibitem[GJ15b]{Groeneboom:2015ew}
Piet Groeneboom and Geurt Jongbloed.
\newblock {Nonparametric confidence intervals for monotone functions}.
\newblock 43(5):2019--2054, 2015.

\bibitem[GR01]{gill2001causal}
Richard~D Gill and James~M Robins.
\newblock Causal inference for complex longitudinal data: the continuous case.
\newblock {\em Annals of Statistics}, pages 1785--1811, 2001.

\bibitem[GS18]{guntuboyina2018nonparametric}
Adityanand Guntuboyina and Bodhisattva Sen.
\newblock Nonparametric shape-restricted regression.
\newblock {\em Statistical Science}, 33(4):568--594, 2018.

\bibitem[GW15]{Galvao:2015ju}
Antonio~F Galvao and Liang Wang.
\newblock Uniformly semiparametric efficient estimation of treatment effects
  with a continuous treatment.
\newblock {\em Journal of the American Statistical Association},
  110(512):1528–1542, 2015.

\bibitem[HI04]{Hirano:2004in}
Keisuke Hirano and Guido~W Imbens.
\newblock {\em The propensity score with continuous treatments}, page 73–84.
\newblock Applied Bayesian modeling and causal inference from incomplete-data
  perspectives. Wiley, Chichester, 2004.

\bibitem[Hil11]{Hill:2011bn}
Jennifer~L Hill.
\newblock {Bayesian nonparametric modeling for causal inference}.
\newblock {\em J. Comput. Graph. Statist.}, 20(1):217--240, 2011.

\bibitem[HWD24]{ham2024doubly}
Daeyoung Ham, Ted Westling, and Charles~R Doss.
\newblock Doubly robust estimation and inference for a log-concave
  counterfactual density.
\newblock {\em arXiv preprint arXiv:2403.19917}, 2024.

\bibitem[Imb04]{Imbens:2004cz}
Guido Imbens.
\newblock Nonparametric estimation of average treatment effects under
  exogeneity: A review.
\newblock {\em The Review of Economics and Statistics}, 86(1):4--29, February
  2004.

\bibitem[IvD04]{Imai:2004gd}
Kosuke Imai and David~A van Dyk.
\newblock Causal inference with general treatment regimes.
\newblock {\em Journal of the American Statistical Association},
  99(467):854–866, Jan 2004.

\bibitem[KGDH15]{Kreif:2015cp}
Noémi Kreif, Richard Grieve, Iván Díaz, and David Harrison.
\newblock Evaluation of the effect of a continuous treatment: A machine
  learning approach with an application to treatment for traumatic brain
  injury.
\newblock {\em Health economics}, 24(9):1213–1228, Sep 2015.

\bibitem[KMMS17]{Kennedy:2017cq}
Edward~H Kennedy, Zongming Ma, Matthew~D McHugh, and Dylan~S Small.
\newblock Non-parametric methods for doubly robust estimation of continuous
  treatment effects.
\newblock {\em J. R. Stat. Soc. Ser. B Stat. Methodol.}, 79(4):1229--1245,
  2017.

\bibitem[KP90]{Kim:1990ue}
J.~Kim and D.~Pollard.
\newblock {Cube root asymptotics}.
\newblock pages 191--219, 1990.

\bibitem[KZ18]{Kallus:2018up}
Nathan Kallus and Angela Zhou.
\newblock Policy evaluation and optimization with continuous treatments.
\newblock In {\em International Conference on Artificial Intelligence and
  Statistics}, pages 1243--1251. PMLR, March 2018.

\bibitem[Low97]{Low_1997}
Mark~G. Low.
\newblock On nonparametric confidence intervals.
\newblock {\em The Annals of Statistics}, 25(6):2547–2554, December 1997.

\bibitem[MBS13]{McHugh:2013gn}
Matthew~D McHugh, Julie Berez, and Dylan~S Small.
\newblock Hospitals with higher nurse staffing had lower odds of readmissions
  penalties than hospitals with lower staffing.
\newblock {\em Health Affairs}, 32(10):1740--1747, October 2013.

\bibitem[MS87]{Muller_Stadtmuller_1987}
Hans-Georg Muller and Ulrich Stadtmuller.
\newblock Estimation of heteroscedasticity in regression analysis.
\newblock {\em The Annals of Statistics}, 15(2):610–625, 1987.

\bibitem[NP87]{Nolan:jw}
Deborah Nolan and David Pollard.
\newblock U-processes: rates of convergence.
\newblock {\em The Annals of Statistics}, pages 780--799, 1987.

\bibitem[NvL07]{Neugebauer:2007fg}
Romain Neugebauer and Mark {van der}~Laan.
\newblock Nonparametric causal effects based on marginal structural models.
\newblock {\em Journal of Statistical Planning and Inference},
  137(2):419–434, Feb 2007.

\bibitem[Ric84]{Rice_1984}
John Rice.
\newblock Bandwidth choice for nonparametric regression.
\newblock {\em The Annals of Statistics}, 12(4):1215–1230, 1984.

\bibitem[Rob86]{robins1986new}
James Robins.
\newblock A new approach to causal inference in mortality studies with a
  sustained exposure period—application to control of the healthy worker
  survivor effect.
\newblock {\em Mathematical modelling}, 7(9-12):1393--1512, 1986.

\bibitem[Rob00]{Robins:2000gv}
James~M Robins.
\newblock {\em Marginal structural models versus structural nested models as
  tools for causal inference}, volume 116 of {\em Statistical models in
  epidemiology, the environment, and clinical trials (Minneapolis, MN, 1997)},
  page 95–133.
\newblock Springer, New York, 2000.

\bibitem[RSLGR07]{robins2001comment}
James Robins, Mariela Sued, Quanhong Lei-Gomez, and Andrea Rotnitzky.
\newblock Comment: Performance of double-robust estimators when ``inverse
  probability" weights are highly variable.
\newblock {\em Statistical Science}, 22(4):544--559, 2007.

\bibitem[SC20]{Semenova:2017uc}
Vira Semenova and Victor Chernozhukov.
\newblock {Debiased machine learning of conditional average treatment effects
  and other causal functions}.
\newblock {\em The Econometrics Journal}, 24(2):264--289, 08 2020.

\bibitem[SM21]{schulz2021doubly}
Juliana Schulz and Erica~EM Moodie.
\newblock Doubly robust estimation of optimal dosing strategies.
\newblock {\em Journal of the American Statistical Association},
  116(533):256--268, 2021.

\bibitem[SRR99]{Scharfstein_Rotnitzky_Robins_1999}
Daniel~O. Scharfstein, Andrea Rotnitzky, and James~M. Robins.
\newblock Adjusting for nonignorable drop-out using semiparametric nonresponse
  models.
\newblock {\em Journal of the American Statistical Association},
  94:1096–1120, 1999.

\bibitem[SUZ19]{su2019non}
Liangjun Su, Takuya Ura, and Yichong Zhang.
\newblock Non-separable models with high-dimensional data.
\newblock {\em Journal of Econometrics}, 212(2):646--677, 2019.

\bibitem[TW24]{takatsu2024debiased}
Kenta Takatsu and Ted Westling.
\newblock Debiased inference for a covariate-adjusted regression function.
\newblock {\em Journal of the Royal Statistical Society Series B: Statistical
  Methodology}, 2024.

\bibitem[vdLD03]{VanDerLaan:2003uu}
M~J van~der Laan and S~Dudoit.
\newblock {"Unified Cross-Validation Methodology For Selection Among Estimators
  an" by Mark J. van der Laan and Sandrine Dudoit}.
\newblock 2003.

\bibitem[VdLPH07]{van2007super}
Mark~J Van~der Laan, Eric~C Polley, and Alan~E Hubbard.
\newblock Super learner.
\newblock {\em Statistical applications in genetics and molecular biology},
  6(1), 2007.

\bibitem[vdVW96]{vanderVaart:1996tf}
Aad~W van~der Vaart and Jon~A Wellner.
\newblock {\em Weak Convergence and Empirical Processes}.
\newblock Springer Series in Statistics. Springer-Verlag, New York, first
  edition, 1996.

\bibitem[WC20]{Westling_Carone_2020}
Ted Westling and Marco Carone.
\newblock A unified study of nonparametric inference for monotone functions.
\newblock {\em The Annals of Statistics}, 48(2):1001–1024, 2020.

\bibitem[WGC20a]{Westling_Gilbert_Carone_2020}
Ted Westling, Peter Gilbert, and Marco Carone.
\newblock Causal isotonic regression.
\newblock {\em Journal of the Royal Statistical Society. Series B. Statistical
  Methodology}, 82(3):719–747, 2020.

\bibitem[WGC20b]{Westling_Gilbert_Carone_2020_supp}
Ted Westling, Peter Gilbert, and Marco Carone.
\newblock Supplementary material for ``causal isotonic regression".
\newblock {\em Journal of the Royal Statistical Society. Series B. Statistical
  Methodology}, 82(3):1–23, 2020.

\bibitem[Wri81]{Wright_1981}
F~T Wright.
\newblock The asymptotic behavior of monotone regression estimates.
\newblock {\em The Annals of Statistics}, 9(2):443–448, 1981.

\end{thebibliography}


\end{document}